\documentclass[preprint,number,times,sort&compress,12pt,3p]{elsarticle}

\usepackage{amssymb}
\usepackage{amsmath}

\usepackage{lineno}

\usepackage{subcaption}
\usepackage{float}
\usepackage{multirow}
\usepackage{color,soul}

\usepackage{tikz}
\usetikzlibrary{shapes.geometric, arrows}
\tikzstyle{startstop} = [rectangle, 
rounded corners, 
minimum width=3cm, 
minimum height=0.5cm,
text centered, 
draw=black] 

\tikzstyle{io} = [trapezium, 
trapezium stretches=true, 
trapezium left angle=70, 
trapezium right angle=110, 
minimum width=3cm, 
minimum height=1cm, 
text centered, 
draw=black]

\tikzstyle{process} = [rectangle, 
minimum width=5.1cm, 
minimum height=1cm,
text centered, 
text width=5.1cm, 
draw=black]

\tikzstyle{decision} = [diamond, 
minimum width=3cm, 
minimum height=1cm, 
text centered, 
draw=black,
aspect=3,
text width=14.5em, 
inner sep=-4pt,
align=center]

\tikzstyle{arrow} = [thick,->,>=stealth]

\usepackage{hyperref}
\hypersetup{
    colorlinks=true,
    linkcolor=blue,
    filecolor=blue,
    urlcolor=blue,
    citecolor=blue}

\journal{Journal of Fluids and Structures}

\begin{document}

\begin{frontmatter}



\title{Stable fluid-rigid body interaction algorithm using the direct-forcing immersed boundary method (DF-IBM)}

\author[UTC,CU]{Elias FARAH\corref{cor1}}
\ead{elias.farah@cranfield.ac.uk}
\author[UTC]{Abdellatif OUAHSINE}
\author[CU]{Patrick G. VERDIN}
\author[UTC]{Badr KAOUI}

\affiliation[UTC]{organization={Centre de Recherche Royallieu, UMR 7337 Roberval, Université de Technologie de Compiègne, Alliance Sorbonne Université},
            addressline={Rue du Dr Schweitzer},
            city={Compiègne},
            postcode={60200},
            country={France}}

\affiliation[CU]{organization={Faculty of Engineering and Applied Sciences, Cranfield University},
            addressline={College Rd},
            city={Cranfield, Wharley End},
            postcode={MK43 0AL},
            country={United Kingdom}}

\cortext[cor1]{Corresponding author}



\begin{abstract}
The direct-forcing immersed boundary method (DF-IBM) algorithm previously developed by the authors is extended by coupling the Navier-Stokes equations with the Newton-Euler equations for rigid body dynamics within the DF-IBM framework. This coupling broadens the applicability of the previous development, from stationary or prescribed motion to flow-induced (free) motion cases. To address fluid-rigid body interactions under a partitioned approach, an implicit coupling algorithm is developed to handle strongly coupled interface conditions. Stability and convergence issues, particularly stemming from critical solid-fluid density ratios and from the rigid body approximation of internal mass effects in rotational dynamics, are mitigated using a fixed relaxation technique for the rigid body kinematics to ensure numerical robustness. Additionally, the proposed algorithm leverages the previously developed DF-IBM formulation and the predictor-corrector strategy of the pressure implicit with splitting of operators (PISO) algorithm by omitting the momentum predictor step and the costly corrector loops from the implicit iterations. The method is validated against several benchmark cases, demonstrating robustness, stability, and efficiency in capturing complex fluid-rigid body interactions across a range of challenging scenarios.
\end{abstract}


\begin{highlights}
\item The Navier-Stokes equations were coupled to the Newton-Euler equations under the direct-forcing immersed boundary framework.
\item An implicit coupling algorithm was developed under the partitioned approach for fluid-rigid body interaction problems.
\item Numerical challenges stemming from the internal mass effect were mitigated using a fixed relaxation technique.
\end{highlights}


\end{frontmatter}



\section{Introduction} \label{sec: introduction}
Fluid-structure interaction (FSI) problems in the context of the immersed boundary method (IBM) typically involve two types of interactions. The first concerns the interaction between fluid dynamics and a prescribed solid structure kinematics previously discussed by the authors~\cite{eliasIBM}. The second type, being the focus of the present work, pertains to the dynamic interaction of the solid structure in response to fluid forces. In the first category, the position, velocity, and acceleration of the structure are known a priori, resulting in a prescribed or forced motion. In contrast, the second category considers the motion of the solid structure to be free, induced by the fluid forces interacting with the solid boundary.

The primary goal of any FSI strategy is to ensure interface matching between the kinematic and dynamic properties of the fluid and structural domains. In the body conforming realm, the interface is inherently aligned with the meshes of both domains, facilitating a matched interface; as seen in Arbitrary Lagrangian-Eulerian (ALE)~\cite{ALE2000} and overset (Chimera)~\cite{chimeraMethod2} methods. On the other hand, for body non-conformal methods like IBM, a kinematic interface matching must be enforced in the numerical method through a force term applied at the interface. This approach introduces an additional kinematic constraint on the fluid-solid system, particularly for incompressible fluid flow, which already has the incompressibility constraint that will be enforced inherently by the pressure Poisson equation (PPE).

Solution procedures for the numerical simulation of a multiphysics FSI problem are broadly classified into monolithic (fully coupled) and partitioned (segregated) methods. Monolithic methods solve fluid and solid subproblems within a single system, offering high accuracy and stability but demanding substantial development effort. Different solution techniques are described to solve such a system, block-iterative, quasi-direct, and direct coupling techniques as described in{~\cite{mono_FSI}}. Partitioned methods treat each subproblem independently using separate solvers, providing flexibility and making them particularly suitable for coupling an existing fluid solver like OpenFOAM with an in-house developed solid solver, as in the present work. The main challenge of partitioned coupling lies in handling interface conditions between the independently solved subproblems. Extensive research has focused on developing and optimizing such coupling algorithms{~\cite{fsiReview1, fsiReview3, matthiesFSI2006, kassiotisFSI1}}.

Significant research has been conducted on fluid-structure coupling algorithms within the IBM framework, with numerous studies examining various methods of coupling using the partitioned strategy{~\cite{IBMreview2014, IBMreview2019, IBMreview2019_2, IBMreview2020, additionalIBMFSIArticle}}. Several numerical challenges arise when addressing FSI problems within the IBM framework~\cite{FSI_IBM1, FSI_IBM2, timeLaggedIBPM_1}. These issues are mainly related to the internal mass effect (IME). The IME represents the contribution of the internal fluid enclosed within the immersed boundary of the rigid body. In translational and rotational dynamics, the IME is equivalent to the rate of change of the linear and angular momentum of the internal fluid (or mass), respectively. Instabilities and convergence problems stem from improper handling of the IME. Therefore, different schemes for addressing the IME term have been proposed in the literature. 

The first scheme is set by ignoring the IME in the Newton-Euler equations, as in~\cite{shangguiPhd2016}. This simplification is valid for low translational solid Reynolds numbers, which are defined based on the maximum linear velocity ($Re_{solid}^{trans} < 1$) and low solid rotational Reynolds numbers based on the maximum angular velocity experienced by the rigid body ($Re_{solid}^{rot} < 1$). However, this scheme leads to significant inaccuracies when either $Re_{solid}^{trans}$ or $Re_{solid}^{rot} \gg 1$~\cite{suzuki_IME}. The second scheme, introduced by Uhlmann~\cite{uhlmann2005} in the DF-IBM context, discretizes the IME implicitly using the first-order backward Euler scheme. Despite its simplicity, it introduces a singularity in the discretized equations when the solid to fluid density ratio $\rho_s/\rho_f$ is equal to $1$, corresponding to a neutrally buoyant rigid body, constraining stability. Nevertheless, Uhlmann~\cite{uhlmann2005} achieved stable computations for 2D rigid bodies when $\rho_s/\rho_f \ge 1.05$, though the stability limits depend on the problem, numerical method, and mesh resolution~\cite{suzuki_IME,kempe&frohlich2012}. Additionally, this scheme was found to yield inaccurate results when $Re_{solid}^{rot} \gg 10$~\cite{suzuki_IME}. To address the density ratio limitation, Feng and Michaelides~\cite{feng_IME} proposed an explicit scheme using the first-order forward Euler method, which requires few lines of code, allowing simulations of neutrally buoyant rigid bodies ($\rho_s/\rho_f \approx 1$). This scheme enables stable computations with reasonable accuracy but introduces phase discrepancies in the rotational dynamics quantities when $Re_{solid}^{rot} \gg 10$~\cite{suzuki_IME}. Suzuki et al.~\cite{suzuki_IME} introduced the Lagrangian markers approximation by computing the rate of change of the linear and angular momentum using uniformly distributed internal Lagrangian markers that move with the rigid body. This method achieves good computations even for $Re_{solid}^{rot} \gg 10$ while also enabling stable computations without imposing any restrictions on the density ratio. Nevertheless, this approach significantly increases the computational effort of the algorithm, especially when the number of internal Lagrangian markers becomes large. Additionally, for rigid bodies with complex shapes, the process of uniformly distributing internal Lagrangian markers within the solid becomes a challenging and time-consuming task. A different method to alleviate the density ratio limitation was presented by Kempe and Fröhlich~\cite{kempe&frohlich2012}, named the volume fraction method. This method evaluates numerically the IME volume integrals by employing the second-order midpoint quadrature rule, with the aid of a signed-distance function and a level-set function defined at the fluid-solid interface. This approach ensures accuracy and stability even for low-density ratios as low as $\rho_s/\rho_f \approx 0.3$~\cite{kempe&frohlich2012}. However, the increased computational cost associated with computing the volume fraction as the number of Lagrangian markers gets large enough makes it less attractive compared to the explicit IME method of Feng and Michaelides~\cite{feng_IME}. Finally, it is important to emphasize that, in terms of implementation complexity, the explicit scheme is considerably simpler than both the Lagrangian Markers Approximation and the Volume Fraction Method.

All of these methods were implemented within a weakly (explicit) coupled framework. A comparison between weak and strong (implicit) coupling strategies for various FSI problems, presented in{~\cite{FSI_IBM1}}, demonstrated that weak coupling becomes unstable when the solid to fluid mass ratio falls below a certain threshold. Strong coupling schemes exhibit similar challenges; however, the introduction of a relaxation scheme has been shown to address these instabilities. The introduction of a relaxation technique has been shown to improve stability, but at the cost of increased computational effort. In other words, enhancing stability through stronger relaxation inevitably leads to slower convergence and higher computational cost. It is also worth noting that the aforementioned explicit methods were primarily developed and validated for geometrically simple bodies such as spheres or cylinders. For more complex, non-symmetric shapes where the fluid forces acting on the solid can be significantly larger and vary sharply with orientation or position, these explicit schemes often become unstable, as reported in{~\cite{timeLaggedIBPM_1}}. For such cases, stability can only be maintained for relatively dense rigid bodies. Therefore, employing an implicit coupling scheme enables stable simulations for systems with lower solid to fluid density ratios. Moreover, by appropriately tuning the fixed relaxation parameter, a practical balance between numerical stability and computational efficiency can be achieved.

Another challenge related to the IME is the rigid body approximation used to evaluate the hydrodynamic force and torque. This approximation assumes that the internal fluid inside the immersed boundary follows the rigid body motion, which can cause stability issues, especially in rotational dynamics. It was proved theoretically by Uhlmann~\cite{uhlmann2003} that the rate of change of the linear momentum of the internal fluid is exactly equal to the rate of change of the linear momentum of the rigid body, despite internal flow development. However, the same principle does not apply analogously to the rate of change of angular momentum, unless the internal fluid follows the rigid body motion, as demonstrated by Uhlmann. Hence, it becomes necessary to impose an additional approximation by assuming that the internal fluid is forced to follow the rigid body motion regardless of the true nature of the internal fluid motion.

In the present work, the improved implicit DF-IBM algorithm developed in~\cite{eliasIBM} is further extended to efficiently simulate flow-induced rigid body motions. To achieve this, an efficient implicit coupling algorithm for a strongly coupled interface conditions is proposed. The Navier-Stokes equations governing the fluid are coupled with the Newton-Euler equations of motion within the DF-IBM framework, ensuring an accurate representation of fluid-rigid body interactions. The dual constraints of the system, defined by the divergence-free condition and interface matching, are sequentially satisfied within the PISO predictor-corrector algorithm. To enhance stability, a fixed-point strategy is employed alongside a fixed relaxation technique, which improves convergence while enforcing the interface condition with high fidelity at each time-step. This relaxation approach effectively mitigates challenges associated with critical density ratios and the rigid body approximation of the IME in rotational dynamics.

The current approach could revolutionize the way FSI problems are addressed. Conventional body conformal mesh techniques used in FSI problems often require complex and time-consuming body-fitted meshing strategies, especially for moving rigid bodies. IBM offers exact geometry representation, allowing the use of simple fixed Cartesian grids, which drastically reduce computational costs.

This paper is organized as follows: the fluid-rigid body governing equations are presented first in Section~\ref{sec: Fluid-Rigid Body Coupling}, followed by the numerical discretization of the rigid body dynamics equations. Additionally, the challenges encountered by the IME are thoroughly discussed. A partitioned coupling algorithm using the predictor-corrector strategy of the PISO solution algorithm for strongly coupled interface conditions is proposed in Section~\ref{sec: Coupling Algorithm}. The algorithm is implemented within the open-source Computational Fluid Dynamics (CFD) code OpenFOAM, version 7.0 of the OpenFOAM Foundation variant. In Section~\ref{sec: results}, numerical validations are carried out for the proposed algorithm against published data found in the literature. Finally, conclusions are drawn in Section~\ref{sec: conclusions} and potential future research avenues are proposed.

\section{Fluid-Rigid Body Coupling} \label{sec: Fluid-Rigid Body Coupling}
Fluid variables are expressed in the Eulerian domain and represented using the subscript $f$, whereas all solid variables described in the Lagrangian domain and computed about the rigid body's center of mass are denoted with the subscript $s$. Lagrangian variables computed at the fluid-solid interface, i.e., the rigid immersed boundary, are represented with the subscript $n$, where $n$ corresponds to the Lagrangian marker under consideration.

\subsection{Governing Equations}
\begin{figure}[!h]
\centering
\includegraphics[width=.5\linewidth]{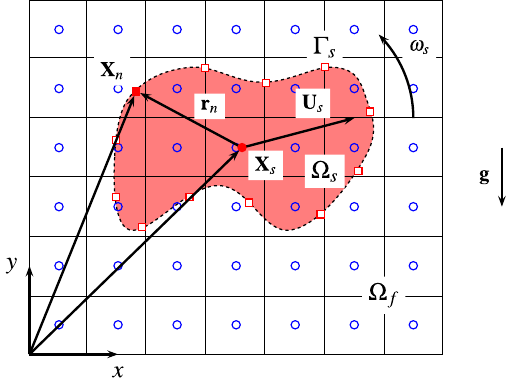}
\caption{Schematic of the fluid-rigid body domain used in DF-IBM framework to simulate FSI problems.}
\label{fig: Schematic of the fluid-rigid body domain used in DF-IBM framework to simulate FSI problems.}
\end{figure}

In a general fluid-rigid body interaction problem, an arbitrarily shaped rigid body occupies the domain $\Omega_s$ and undergoes general planar motion within a viscous incompressible fluid, $\Omega_f$. Within the context of the IBM, the rigid body is modeled through its immersed boundary, $\Gamma_s$, with the fluid occupying its interior. The interaction occurs at the fluid-solid interface, which coincides with the immersed boundary, $\Gamma_s = \Omega_f \cap \Omega_s$. The fluid motion is governed by the Navier-Stokes equations, while the rigid body follows the rigid body dynamics equations, expressed through the Newton-Euler equations of motion. A schematic illustrating the physical domain of the fluid-rigid body system is provided in Fig.~(\ref{fig: Schematic of the fluid-rigid body domain used in DF-IBM framework to simulate FSI problems.}). The system is subjected to Earth's gravitational acceleration.

The mathematical formulation of the fluid-rigid body system under the DF-IBM framework is derived as follows:
\begin{subequations}
\allowdisplaybreaks
\begin{align}
&\nabla \cdot \mathbf{u}_f = 0, & \quad &\text{in } \Omega_f, \label{eq: fsi_mass_cons} \\
&\dfrac{\partial \mathbf{u}_f}{\partial t} 
+ \nabla \cdot (\mathbf{u}_f \mathbf{u}_f) 
= 
\nabla \cdot \boldsymbol{\sigma}_f 
+ \mathbf{f}, & \quad &\text{in } \Omega_f, \label{eq: fsi_momentum_cons} \\ 
&\mathbf{u}_f(\mathbf{x}_{i,j}, t) = \mathbf{U}_{s}(t) + \omega_s (t) \times \mathbf{r}_{n}(\mathbf{X}_{n}, t), & \quad &\text{on } \Gamma_s, \label{eq: fsi_noslip_condition} \\[2mm]
&\mathbf{U}_{n} (\mathbf{X}_{n}, t) = \mathcal{I}[\mathbf{u}_f (\mathbf{x}_{i,j}, t)], & \quad &\text{from } \Omega_f \text{ to } \Gamma_s, \label{eq: fsi_ibm_interpolation} \\[2mm]
&\mathbf{f} (\mathbf{x}_{i,j}, t) = \mathcal{S}[\mathbf{F}_{n} (\mathbf{X}_{n}, t)], & \quad &\text{from } \Gamma_s \text{ to } \Omega_f, \label{eq: fsi_ibm_spreading} \\
&m_s \dfrac{d \mathbf{U}_{s}}{dt} = \sum \mathbf{F}_{ext}, & \quad &\text{on } \Omega_s, \label{eq: fsi_newton_law_motion} \\ 
&I_s \dfrac{d \omega_s}{dt} = \sum \mathbf{T}_{ext}, & \quad &\text{on } \Omega_s, \label{eq: fsi_euler_law_motion}
\end{align}
\end{subequations}
where,
\begin{itemize}
    \item Eqs.~(\ref{eq: fsi_mass_cons}) and~(\ref{eq: fsi_momentum_cons}) are the fluid governing equations for a dimensional incompressible flow, with $\mathbf{u}_f$ being the fluid Eulerian velocity vector, $\boldsymbol{\sigma}_f$ is the fluid Cauchy stress tensor (or total stress tensor) divided by the fluid density $\rho_f$, and $\mathbf{f}$ is the boundary force, added to account for the presence of $\Gamma_s$ within $\Omega_f$.

    \item Eqs.~(\ref{eq: fsi_newton_law_motion}) and~(\ref{eq: fsi_euler_law_motion}) are the Newton-Euler equations of motion. They describe a rigid body's combined translational (Newton) and rotational (Euler) dynamics. For a general 3D motion, a system with a 6-degrees-of-freedom (6-DOF) is formed. However, for 2D cases, i.e., for a general planar motion the system reduces to a 3-DOF system only. $\mathbf{F}_{ext}$ and $\mathbf{T}_{ext}$ represent the external forces and torques respectively, $\mathbf{U}_{s} = (U_s,~ V_s)$ is the linear velocity, $\omega_s$ is the angular velocity (about the z-axis), $m_s = \rho_s V$ is the rigid body mass, $V = \int_{\Omega_s}dV$ is the rigid body volume, $\rho_s$ is the rigid body density, $I_s = \rho_s \int_{V} \mathbf{r}_{n}^2~dV$ is the mass moment of inertia, and $\mathbf{r}_{n} = \mathbf{X}_{n} - \mathbf{X}_s$ is the relative position vector of any solid Lagrangian marker $\mathbf{X}_{n}$ situated on the rigid body's immersed surface $\Gamma_s$ with respect to the rigid body center of mass $\mathbf{X}_s$.

    \item Eq.~(\ref{eq: fsi_noslip_condition}) is the fluid-solid interface matching and no-slip velocity boundary condition.
    
    \item Eqs.~(\ref{eq: fsi_ibm_interpolation}) and~(\ref{eq: fsi_ibm_spreading}) are the IBM-related linear operators used to map the data between the Lagrangian and Eulerian domains, with $\mathbf{U}_n$ and $\mathbf{F}_n$ being the desired velocity and singular boundary force of the Lagrangian markers $\mathbf{X}_n$, respectively.
\end{itemize}

\subsection{External Forces and Torques}
A rigid body undergoing a general planar motion is influenced by external forces and torques that alter its position, velocity, and acceleration. In this study, the external forces and torques are limited to:
\begin{subequations}
\begin{equation}
\sum \mathbf{F}_{ext} = \mathbf{F}_{g} + \mathbf{F}_{b} + \mathbf{F}_{h} + \mathbf{F}_c,
\end{equation}
\begin{equation}
\sum \mathbf{T}_{ext} = \mathbf{T}_{h},
\end{equation}
\end{subequations}
with
\begin{itemize}
    \item $\mathbf{F}_{g} = m_s \mathbf{g} = \rho_s V \mathbf{g}$, the downward gravitational force.
    
    \item $\mathbf{F}_b = -\rho_f V \mathbf{g}$, the upward buoyancy force.
    
    \item $\mathbf{F}_{h} = \rho_f \oint_{\Gamma_s} \boldsymbol{\sigma}_f \cdot \mathbf{n} ~dS = -\rho_f \sum_{\forall \mathbf{X}_{n} \in \Gamma_s} \mathbf{F}_{n}(\mathbf{X}_{n}) W_{n} + \mathbf{F}_{IME}$, the hydrodynamic force with $\mathbf{F}_{IME} = \rho_f V \dfrac{d \mathbf{U}_{s}}{dt}$, being the contribution of the internal mass effect in translational dynamics.
    
    \item $\mathbf{F}_c$, the collision force.

    \item $\mathbf{T}_{h} = \rho_f \oint_{\Gamma_s} \mathbf{r}_{i,j} \times (\boldsymbol{\sigma}_f \cdot \mathbf{n}) ~dS = -\rho_f \sum_{\forall \mathbf{X}_{n} \in \Gamma_s} \mathbf{r}_{n}(\mathbf{X}_{n}) \times \mathbf{F}_{n}(\mathbf{X}_{n}) W_{n} + \mathbf{T}_{IME}$, the hydrodynamic torque having $\mathbf{T}_{IME} \approx \rho_f \dfrac{I_s}{\rho_s} \dfrac{d \omega_s}{dt}$ as the contribution of the internal mass effect in rotational dynamics.
\end{itemize}

A detailed explanation behind the hydrodynamic force and torque formulations is presented in~\ref{sec: Appendix A}, and the repulsive potential model (RPM) used to compute the collision force is shown in~\ref{sec: Appendix B}.

\subsection{Internal Mass Effect (IME) Challenges}
FSI problems involving incompressible viscous fluids are particularly challenging due to the IME, which becomes prominent when the fluid density closely matches the solid density~\cite{addedMassFSI1, addedMassFSI2, addedMassFSI3, addedMassFSI4}, i.e., for $\rho_s/\rho_f \approx 1$. The IME represents the inertia added to the system from the fluid's acceleration around a moving body and manifests as numerical difficulties in the FSI realm, especially in incompressible flows.

\begin{figure}[!h]
\centering
\includegraphics[width=.5\linewidth]{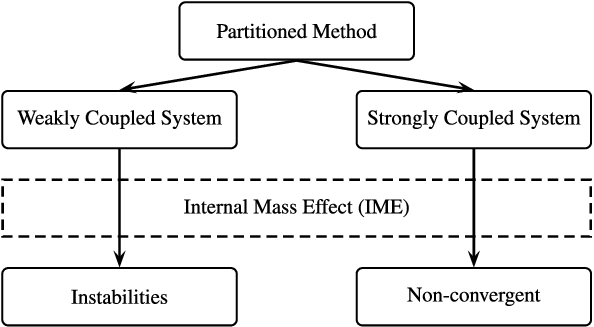}
\caption{Numerical challenges induced by the IME in the FSI partitioned coupling methods.}
\label{fig: Numerical difficulties induced by the IME in the FSI partitioned coupling methods}
\end{figure}

Explicit coupling methods, typically used for weakly coupled systems, encounter numerical instabilities in the presence of strong IME because they fail to enforce the interface conditions accurately~\cite{addedMassFSI1, addedMassFSI2, addedMassFSI4}. This failure leads to spurious numerical power at the fluid-solid interface, resulting in an unstable coupling~\cite{addedMassFSI2}. In contrast, implicit coupling methods, which are designed to preserve energy balance, are more stable for strongly coupled systems. They effectively suppress the spurious power on the fluid-solid interface by enforcing the interface condition more accurately. Nevertheless, implicit methods can encounter convergence issues during the iterative process under significant IME~\cite{addedMassFSI4}. Fig.~(\ref{fig: Numerical difficulties induced by the IME in the FSI partitioned coupling methods}) highlights the difficulties associated with partitioned coupling for FSI problems in the presence of IME.

These challenges are particularly critical in biomechanics and biomedical applications, such as blood flow simulations, where the IME is unavoidable due to the similar densities of the fluid and solid. Furthermore, reducing the time-step fails to resolve these instabilities~\cite{addedMassFSI2}, highlighting the fundamental nature of the issue. In contrast, for compressible flows where the IME is less significant, explicit coupling methods are more stable, as seen in aeroelastic applications where interface conditions are less stringent~\cite{addedMassFSI2, addedMassFSI4}. 

For fluid-rigid body interaction and in the context of IBM, it was found that these difficulties are further exacerbated by the rigid body approximation for the IME when the rigid body is undergoing rotational dynamics, which forces the internal fluid to follow the rigid body motion of the immersed boundary, regardless of the true induced motion of the internal flow field. This behavior, depicted by the IME contribution in rotational dynamics in~\ref{sec: Appendix A}, highlights a key source of instability when simulating such interactions.

\subsection{Direct-Forcing Immersed Boundary Method (DF-IBM)}
The IBM-related linear operators in the DF-IBM framework are divided between a velocity interpolation operator $\mathcal{I}$, written as:
\begin{align}
\mathbf{U}_n(\mathbf{X}_n, t) &= \mathcal{I}[\mathbf{u}_f(\mathbf{x}_{i,j}, t)] = \sum_{\mathbf{x}_{i,j} \in \mathcal{D}_{\mathbf{X}_n}} \mathbf{u}_f(\mathbf{x}_{i,j}, t) \delta_h(\mathbf{x}_{i,j} - \mathbf{X}_n) \Delta V_{i,j},
\label{eq: velocity_interpolation_eq}
\end{align}
with $\Delta V_{i,j} = h^2$ the Eulerian weight, $h$ the fluid mesh spacing, and $\left( \mathbf{x}_{i,j} \in \mathcal{D}_{\mathbf{X}_n} \right)$ the collection of all neighboring Eulerian grid points that fall inside the support domain $\mathcal{D}_{\mathbf{X}_n}$ of the Lagrangian marker $\mathbf{X}_n$. In addition to, a force spreading operator $\mathcal{S}$ formulated as:
\begin{align}
\mathbf{f}(\mathbf{x}_{i,j}, t) &= \mathcal{S}[\mathbf{F}_{n}(\mathbf{X}_n, t)] = \sum_{\mathbf{X}_n \in \mathcal{D}_{\mathbf{x}_{i,j}}} \mathbf{F}_{n}(\mathbf{X}_n, t) \delta_h(\mathbf{x}_{i,j} - \mathbf{X}_n) W_n,
\label{eq: force_spreading_eq}
\end{align}
where $(\mathbf{X}_n \in \mathcal{D}_{\mathbf{x}_{i,j}})$ refers to a loop over all the Lagrangian markers whose support domain $\mathcal{D}_{\mathbf{X}_n}$ contains the same Eulerian grid point $\mathbf{x}_{i,j}$ and $W_n$ is the Lagrangian weight and it is calculated according to~\cite{julienFavier2023}:
\begin{equation}
W_n = \dfrac{1}{\sum_{\mathbf{x}_{i,j} \in \mathcal{D}_{\mathbf{X}_n}} \sum_{\mathbf{X}_m \in \mathcal{D}_{\mathbf{x}_{i,j}}} 
\delta_h (\mathbf{x}_{i,j} - \mathbf{X}_{n}) \delta_h (\mathbf{x}_{i,j} - \mathbf{X}_{m}) h^2}.
\end{equation}

The interpolation function $\delta_h$ used for the Lagrangian-Eulerian mapping is constructed using:
\begin{equation}
\delta_{h} (\mathbf{x}_{i,j} - \mathbf{X}_{n}) = \dfrac{1}{h^2} \phi \left(\dfrac{x_{i,j} - X_{n}}{h}\right) \phi \left(\dfrac{y_{i,j} - Y_{n}}{h}\right).
\end{equation}
The kernel function $\phi$ adopted in this work is the three-point-width function developed in~\cite{roma&peskin1999}.

\subsection{Rigid Body Kinematics}
The kinematics of a rigid body undergoing a general planar motion are used to evaluate the position, velocity, and acceleration due to both translational and rotational dynamics. The absolute velocity $\mathbf{U}_{n}$ of a Lagrangian marker $\mathbf{X}_{n}$ undergoing a general planar motion is equal to the combined translational and rotational contributions of $\mathbf{X}_{n}$:
\begin{align}
\mathbf{U}_{n}(\mathbf{X}_{n}, t)
&= \mathbf{U}_{s}(t) + \omega_s (t) \times \mathbf{r}_{n}(\mathbf{X}_{n}, t).
\label{eq: rigid_body_abs_velocity}
\end{align}

The center of mass $\mathbf{X}_s$ undergoing translational motion is computed as:
\begin{equation}
\dfrac{d \mathbf{X}_s(t)}{dt} = \mathbf{U}_{s}(t).
\label{eq: linear_position_update_equation}
\end{equation}

The overall angular position $\theta_s$ of the rigid body undergoing rotational motion about its center of mass is computed as:
\begin{equation}
\dfrac{d \theta_{s}(t)}{dt} = \omega_s(t).
\label{eq: angular_position_update_equation}
\end{equation}

\subsection{Numerical Discretization}
The fluid governing equations, Eqs.~(\ref{eq: fsi_mass_cons}) and~(\ref{eq: fsi_momentum_cons}), are described in the OpenFOAM environment, which is based on the collocated grid arrangement and employs a cell-centered finite volume method (FVM) discretization. The detailed discretization procedure adopted in~\cite{eliasIBM} is used here. The obtained linear algebraic equation is written as:
\begin{equation}
a_{P} \mathbf{u}_{f_P}^{n+1}
+ \sum_{N} a_N \mathbf{u}_{f_N}^{n+1} 
= 
\mathbf{RHS}\left(\mathbf{u}_{f_P}^{n}, \mathbf{u}_{f_P}^{n-1}\right)
- \left(\nabla p\right)^{n+1}
+ \mathbf{f}^{n+1},
\label{eq: NSE_linear_algebraic_eq}
\end{equation}

The solid governing equations, Eqs.~(\ref{eq: fsi_newton_law_motion}) and~(\ref{eq: fsi_euler_law_motion}), are advanced in time using the first-order implicit backward Euler. Hence, the linear and angular accelerations are discretized as follows:
\begin{subequations}
\begin{equation}
m_s \dfrac{d \mathbf{U}_{s}}{dt} = \rho_s V \dfrac{\mathbf{U}_{s}^{n+1} - \mathbf{U}_{s}^{n}}{\Delta t},
\end{equation}
\begin{equation}
I_{s} \dfrac{d \omega_s}{dt} = I_{s} \dfrac{\omega_s^{n+1} - \omega_s^{n}}{\Delta t}.
\end{equation}
\end{subequations}

The collision force term $\mathbf{F}_c$ will be kept as it is for simplicity. To counter the density ratio limitation, the explicit scheme proposed by Feng and Michaelides~\cite{feng_IME} is applied to the IME. Theoretically speaking, this allows the simulation of neutrally buoyant solids $\rho_s / \rho_f = 1$, non-buoyant (dense) solids $\rho_s / \rho_f > 1$, and buoyant (light) solids $\rho_s / \rho_f < 1$. Therefore, the first-order forward Euler scheme is chosen to discretize the IME:
\begin{subequations}
\begin{equation}
\mathbf{F}_{IME} = \rho_f V \dfrac{\mathbf{U}_{s}^{n} - \mathbf{U}_{s}^{n-1}}{\Delta t},
\label{eq: explicit_IME_force}
\end{equation}
\begin{equation}
\mathbf{T}_{IME} \approx \rho_f \dfrac{I_s}{\rho_s} \dfrac{\omega_s^{n} - \omega_s^{n-1}}{\Delta t}.
\label{eq: explicit_IME_torque}
\end{equation}
\end{subequations}

Adding all external forces and torque, and substituting Eqs.~(\ref{eq: explicit_IME_force}) and~(\ref{eq: explicit_IME_torque}) into the Newton-Euler Eqs.~(\ref{eq: fsi_newton_law_motion}) and~(\ref{eq: fsi_euler_law_motion}), while treating the interface variables implicitly as unknowns to be solved for, yields the final discretized form of the Newton-Euler equations:
\begin{subequations}
\begin{align} 
\mathbf{U}_{s}^{n+1} = \Delta t \left( \dfrac{\rho_s - \rho_f}{\rho_s} \mathbf{g} - \dfrac{\rho_f}{\rho_s V} \sum_{\forall \mathbf{X}_{n} \in \Gamma_s} \mathbf{F}_{n}^{n+1}(\mathbf{X}_{n}^{n+1}) W_{n} + \dfrac{1}{\rho_s V} \mathbf{F}_c^{n+1}(\mathbf{X}_{n}^{n+1}) \right) + \dfrac{\rho_s + \rho_f}{\rho_s}\mathbf{U}_{s}^{n} - \dfrac{\rho_f}{\rho_s} \mathbf{U}_{s}^{n-1},
\label{eq: explicit_IME_newton_equations}
\end{align}
\begin{align}
\omega_s^{n+1} = \Delta t \left( -\dfrac{\rho_f}{I_s} \sum_{\forall \mathbf{X}_{n} \in \Gamma_s} \mathbf{r}_{n}^{n+1}(\mathbf{X}_{n}^{n+1}) \times \mathbf{F}_{n}^{n+1}(\mathbf{X}_{n}^{n+1}) W_{n} \right) + \dfrac{\rho_s + \rho_f}{\rho_s} \omega_s^{n} - \dfrac{\rho_f}{\rho_s} \omega_s^{n-1}.
\label{eq: explicit_IME_euler_equations}
\end{align}
\end{subequations}

The aforementioned discretization, i.e., Eq.~(\ref{eq: NSE_linear_algebraic_eq}) is linearly implicit in $\mathbf{u}_f^{n+1}$, with two additional unknowns: the pressure $p^{n+1}$ and the boundary force $\mathbf{f}^{n+1}$. While Eqs.~(\ref{eq: explicit_IME_newton_equations}) and~(\ref{eq: explicit_IME_euler_equations}) are linearly explicit in $\mathbf{U}_s^{n+1}$ and $\omega_s^{n+1}$, respectively. This introduces three additional unknowns: $\mathbf{X}_s^{n+1}$, $\mathbf{X}_n^{n+1}$, and $\mathbf{F}_n^{n+1}$. Bearing in mind that $\mathbf{f}^{n+1} = \mathcal{S}[\mathbf{F}_n^{n+1}]$.

\section{Partitioned Coupling Algorithm} \label{sec: Coupling Algorithm}
In this section, an efficient implicit coupling algorithm is proposed under the partitioned method category for a strongly coupled interface condition. The proposed coupling method leverages the improved implicit DF-IBM algorithm already developed in~\cite{eliasIBM}, comprised of three new major steps, listed as: (1) momentum predictor, (2) fluid-rigid body coupling, and (3) PISO corrector loops, aimed at improving the overall efficiency of the FSI coupling methods. The two subproblems and the interface condition formulated for the fluid-rigid body coupling partition method are elaborated in Table~\ref{tab: Physical domains in a fluid-rigid body coupling algorithm.}. The two subproblems are coupled through the fluid-solid interface $\Gamma_s$ condition. Fluid unknowns are stored at the fluid cell centroids, solid unknowns at the rigid body center of mass, and fluid-solid interface variables at each Lagrangian marker.

\begin{table}[!h]
\centering
\caption{Physical domains in a fluid-rigid body coupling algorithm.}
\label{tab: Physical domains in a fluid-rigid body coupling algorithm.}
\resizebox{\columnwidth}{!}{%
\begin{tabular}{lccc}
\hline
Physical Domain     & \textit{subproblem 1} & \textit{interface}  & \textit{subproblem 2} \\ \hline
Solver              & Fluid        & Interface Condition (DF-IBM) & Rigid Body   \\
Domain              & $\Omega_f$   & $\Gamma_s$                   & $\Omega_s$   \\
Governing Equations & Eqs.~(\ref{eq: fsi_mass_cons}) and~(\ref{eq: fsi_momentum_cons}) & Eqs.~(\ref{eq: fsi_noslip_condition}),~(\ref{eq: fsi_ibm_interpolation}), and~(\ref{eq: fsi_ibm_spreading}) & Eqs.~(\ref{eq: fsi_newton_law_motion}) and~(\ref{eq: fsi_euler_law_motion}) \\
Unknowns            & $(\mathbf{u}_f^{n+1}, p^{n+1})$ & $(\mathbf{X}_{n}^{n+1}, \mathbf{U}_{n}^{n+1}, \mathbf{F}_{n}^{n+1})$ & $(\mathbf{X}_s^{n+1}, \mathbf{U}_s^{n+1}, \omega_s^{n+1})$ \\ \hline
\end{tabular}%
}
\end{table}

The boundary force term is partially derived from the variables associated with the rigid body and partially from the neighboring fluid Eulerian cells, i.e., $\mathbf{F}_{n} = (\mathbf{U}_n - \mathcal{I}[{\mathbf{u}_f^{*}}]) / \Delta t$ (refer to direct-forcing step in~\cite{eliasIBM}), with $\mathbf{U}_n$ computed here using Eq.~(\ref{eq: rigid_body_abs_velocity}). This approach contrasts with the traditional coupling found in body conforming mesh methods, such as the ALE method, where the interface coupling is achieved through the direct interaction of the pressure field, $p$, across the matched interface. 

This highlights two important findings: (1) The momentum predictor can be omitted because the governing equations are solved on a fixed Eulerian mesh without the boundary force term. The PISO corrector loops are also omitted since the modified PPE (mPPE)~\cite{eliasIBM} inherently accounts for the boundary force where the no-slip condition is already enforced. This optimization reduces computational cost, particularly by avoiding the mPPE iteration as in conventional implicit methods (e.g. those used in the ALE formulation), which is typically the most demanding step. (2) Since the Lagrangian domain is much smaller than the Eulerian domain $(N_{Lag} \in \Gamma_s) \ll (N_{Eul} \in \Omega_f)$, with $N_{Lag}$ and $N_{Eul}$ being the total number of Lagrangian markers and Eulerian grid points, respectively, updating and constructing the Lagrangian support domain for each marker and its associated variables during each FSI iteration is computationally efficient. This process is significantly faster compared to updating the ALE mesh.

In a similar fashion to the DF-IBM algorithm developed in~\cite{eliasIBM}, the fluid-rigid body system dual constraints problem formed for an incompressible viscous flow interacting with a rigid body is satisfied sequentially in the solution process of the PISO predictor-corrector algorithm:
\begin{itemize}
    \item[(1)] Divergence-free condition on the fluid domain: $\nabla \cdot \mathbf{u}_f^{n+1} = 0$ in $\Omega_f$,

    \item[(2)] Fluid-solid interface no-slip velocity boundary condition based on Eq.~(\ref{eq: fsi_noslip_condition}) and Eq.~(\ref{eq: fsi_ibm_interpolation}) in the DF-IBM framework: $\mathbf{U}_{n}^{n+1} = \mathcal{I}[\Tilde{\mathbf{u}}_{f}]$ on $\Gamma_s$, having $\mathbf{U}_{n} = \mathbf{U}_s + \omega_s \times \mathbf{r}_{n}$ and $\Tilde{\mathbf{u}}_{f}$ the updated (corrected) Eulerian velocity field.
\end{itemize}

The second fluid-solid interface condition has two physical meanings: first, to enforce the DF-IBM no-slip velocity boundary condition, and second, to ensure that the fluid-solid interface accurately follows the rigid body motion. The interface condition can be either weakly coupled, as in the case of explicit coupling, and is expressed as:
\begin{align}
\mathbf{U}_{n}^{n+1}
&= \mathbf{U}_s^{n} + \omega_s^{n} \times \mathbf{r}_{n}^{n} \notag \\
&= \mathcal{I}[\Tilde{\mathbf{u}}_{f}],
\label{eq: weakly_coupled_interface_condition}
\end{align}
where the rigid body variables are used from the previous time level $t^n$, or the same condition can be accurately enforced as in the case of implicit coupling for the strongly coupled schemes, therefore:
\begin{align}
\mathbf{U}_{n}^{n+1}
&= \mathbf{U}_s^{n+1} + \omega_s^{n+1} \times \mathbf{r}_{n}^{n+1} \notag \\
&= \mathcal{I}[\Tilde{\mathbf{u}}_{f}],
\label{eq: strongly_coupled_interface_condition}
\end{align}
with the rigid body variables taken from the current time-step $t^{n+1}$.

\subsection{Implicit Coupling: Strongly Coupled Interface Condition}
A fixed-point strategy is employed to enforce iteratively the fluid-solid interface condition with high accuracy at each time-step. To mitigate the two main challenges leading to non-convergence in implicit FSI iterations: (1) issues arising with neutrally buoyant and buoyant solids, and (2) difficulties related to the rigid body approximation of the IME in rotational dynamics, a fixed relaxation technique was used to stabilize the algorithm. This method reduces high-velocity fluctuation rates, improving overall stability and convergence rate as suggested in~\cite{improvedPerformanceFSI1, fixedPointRelaxation1, fixedPointRelaxation2}. Additionally, the same relaxation technique when applied to the angular velocity in rotational motion, helps to overcome the convergence issues introduced by the rigid body approximation in rotational dynamics. The proposed algorithm was observed to produce stable solution even for solid to fluid density ratios as low as 0.3.

The following FSI procedure is performed at each time-step, where the superscript $n$ and $n+1$ denote the past and current time levels, respectively:
\begin{enumerate}
    \item[1.] Momentum Predictor: \\
    Prediction of the velocity field $\mathbf{u}_f^{*}$:
    \begin{equation}
    a_P \mathbf{u}_{f_P}^{*} 
    + \sum_{N} a_N \mathbf{u}_{f_N}^{*} 
    = \mathbf{RHS}\left(\mathbf{u}_{f}^{n}, \mathbf{u}_{f}^{n-1}\right)
    - \left(\nabla p\right)^{n}.
    \label{eq: predictor_step_algorithm}
    \end{equation}
    
    \item[2.] Fluid-Rigid Body Coupling: \\
    Outer implicit FSI iterations start for $k = 1$, with $\mathbf{X}_{n}^{(0)} = \mathbf{X}_{n}^{n}$, $\mathbf{X}_{s}^{(0)} = \mathbf{X}_{s}^{n}$, $\mathbf{U}_s^{(0)} = \mathbf{U}_s^{n}$, and $\omega_s^{(0)} = \omega_s^{n}$. The implicit coupling of the fluid-rigid body interaction is carried out using the following steps:
    \begin{enumerate}
        \item[(a)] Determine desired Lagrangian velocity $\mathbf{U}_{n}$ on $\Gamma_s$, using Eq.~(\ref{eq: strongly_coupled_interface_condition}):
        \begin{align}
        \mathbf{U}_{n}^{(k-1)} \bigl(\mathbf{X}_{n}^{(k-1)}\bigl) 
        &= \mathbf{U}_s^{(k-1)} + \omega_s^{(k-1)} \times \mathbf{r}_{n}^{(k-1)} \bigl(\mathbf{X}_{n}^{(k-1)}\bigl) \notag \\
        &= \mathbf{U}_s^{(k-1)} + \omega_s^{(k-1)} \times \bigl(\mathbf{X}_{n}^{(k-1)} - \mathbf{X}_{s}^{(k-1)}\bigl).
        \label{eq: implicit_coupling_desired_lagrangian_velocity}
        \end{align}
        
        \item[(b)] Inner implicit DF-IBM iterations start for $m = 1$, with $\Tilde{\mathbf{u}}_f^{~(k=0,m=0)} = \mathbf{u}_f^{*}$ and $\mathbf{F}_{n}^{(k,m=0)} \bigl(\mathbf{X}_{n}^{(k-1)}\bigl) = \mathbf{F}_{n}^{n} \bigl(\mathbf{X}_{n}^{n}\bigl)$. The iterative procedure is executed as follows:
        \begin{enumerate}
            \item[(i)] Eulerian Velocity Interpolation:
            \begin{equation}
            \mathbf{U}^{(k,m-1)}\bigl(\mathbf{X}_{n}^{(k-1)}\bigl) = \mathcal{I}[\Tilde{\mathbf{u}}_f^{~(k,m-1)}],
            \label{eq: interpolation_step_algorithm_impfsi}
            \end{equation}
            
            \item[(ii)] Direct-Forcing:
            \begin{equation}
            \mathbf{F}_{n}^{(k,m)}\bigl(\mathbf{X}_{n}^{(k-1)}\bigl) = \dfrac{\mathbf{U}_{n}^{(k-1)} \bigl(\mathbf{X}_{n}^{(k-1)}\bigl) - \mathbf{U}^{(k,m-1)}\bigl(\mathbf{X}_{n}^{(k-1)}\bigl)}{\Delta t} 
            + \mathbf{F}_{n}^{(k,m-1)}\bigl(\mathbf{X}_{n}^{(k-1)}\bigl),
            \label{eq: lag_boundary_force_init_impfsi}
            \end{equation}
            
            \item[(iii)] Lagrangian Force Spreading:
            \begin{equation}
            {\mathbf{f}}^{(k,m)} = \mathcal{S}[\mathbf{F}_{n}^{(k,m)}\bigl(\mathbf{X}_{n}^{(k-1)}\bigl)],
            \label{eq: spreading_step_algorithm_impfsi}
            \end{equation}
            
            \item[(iv)] Update Eulerian Velocity:
            \begin{equation}
            a_P \Tilde{\mathbf{u}}_{f_P}^{~(k,m)} 
            + \sum_{N} a_N \Tilde{\mathbf{u}}_{f_N}^{~(k,m)} 
            = \mathbf{RHS}\left(\mathbf{u}_{f}^{n}, \mathbf{u}_{f}^{n-1}\right)
            - \left(\nabla p\right)^{n}
            + {\mathbf{f}}^{(k,m)},
            \label{eq: update_velocity_algorithm_impfsi}
            \end{equation}
            
            \item[(v)] Fluid-solid interface no-slip velocity boundary condition:
            \begin{equation}
            \bigl\| E_{no-slip} \bigl\|_{2} = 
            \bigl\| E_{\Tilde{\mathbf{u}}_f^{~(k,m)}} \bigl\|_{2} - 
            \bigl\| E_{\Tilde{\mathbf{u}}_f^{~(k,m-1)}} \bigl\|_{2} < \boldsymbol{\epsilon}_{IBM},
            \label{eq: noslip_error_criterion_impfsi}
            \end{equation}
            with
            \begin{align*}
            \bigl\| E_{\Tilde{\mathbf{u}}_f^{~(k,m)}} \bigl\|_{2} = 
            \bigl\| \mathbf{U}_{n}^{(k-1)} \bigl(\mathbf{X}_{n}^{(k-1)}\bigl) - \mathbf{U}^{(k,m)}\bigl(\mathbf{X}_{n}^{(k-1)}\bigl) \bigl\|_{2},
            \end{align*}
            and
            \begin{align*}
            \bigl\| E_{\Tilde{\mathbf{u}}_f^{~(k,m-1)}} \bigl\|_{2} = 
            \bigl\| \mathbf{U}_{n}^{(k-1)} \bigl(\mathbf{X}_{n}^{(k-1)}\bigl) - \mathbf{U}^{(k,m-1)}\bigl(\mathbf{X}_{n}^{(k-1)}\bigl) \bigl\|_{2},
            \end{align*}
            where $\boldsymbol{\epsilon}_{IBM}$ is a user-specified tolerance for the implicit DF-IBM iterations.

            \item[(vi)] If the defined convergence criterion is reached in \textit{Step 2 (v)}, then
            \begin{equation}
            \mathbf{f}^{(k)} = \mathbf{f}^{(k,m)}, ~\Tilde{\mathbf{u}}_f^{~(k)} = \Tilde{\mathbf{u}}_f^{~(k,m)}, ~\text{and}~\mathbf{F}_{n}^{(k)} \bigl(\mathbf{X}_{n}^{(k-1)}\bigl) = \mathbf{F}_{n}^{(k,m)}\bigl(\mathbf{X}_{n}^{(k-1)}\bigl),
            \end{equation}
            otherwise, return to \textit{Step 2 (i)}, and let $m = m + 1$. \\
            It can be noted that after this step, the DF-IBM no-slip velocity boundary condition is fulfilled, $\mathbf{U}^{(k,m)}\bigl(\mathbf{X}_{n}^{(k-1)}\bigl) = \mathcal{I}[\Tilde{\mathbf{u}}_f^{~(k,m)}] = \mathbf{U}_{n}^{(k-1)} \bigl(\mathbf{X}_{n}^{(k-1)}\bigl)$.
        \end{enumerate}
        
        \item[(c)] Solve the Newton-Euler equations of motion in $\Omega_s$, based on Eqs.~(\ref{eq: explicit_IME_newton_equations}) and~(\ref{eq: explicit_IME_euler_equations}):
        \begin{subequations}
        \begin{multline}
        \mathbf{U}_{s}^{(k), *} = \Delta t \Biggl( \dfrac{\rho_s - \rho_f}{\rho_s} \mathbf{g} - \dfrac{\rho_f}{\rho_s V} \sum_{\forall \mathbf{X}_{n} \in \Gamma_s} \mathbf{F}_{n}^{(k)} \bigl(\mathbf{X}_{n}^{(k-1)}\bigl) W_{n} + \dfrac{1}{\rho_s V} \mathbf{F}_c^{(k)} \bigl(\mathbf{X}_{n}^{(k-1)}\bigl) \Biggl) \\ + \dfrac{\rho_s + \rho_f}{\rho_s}\mathbf{U}_{s}^{n} - \dfrac{\rho_f}{\rho_s} \mathbf{U}_{s}^{n-1},
        \label{eq: implicit_coupling_newton_equation}
        \end{multline}
        \begin{multline}
        \omega_s^{(k), *} = \Delta t \left( -\dfrac{\rho_f}{I_s} \sum_{\forall \mathbf{X}_{n} \in \Gamma_s} \mathbf{r}_{n}^{(k-1)} \bigl(\mathbf{X}_{n}^{(k-1)}\bigl) \times \mathbf{F}_{n}^{(k)} \bigl(\mathbf{X}_{n}^{(k-1)}\bigl) W_{n} \right) \\ + \dfrac{\rho_s + \rho_f}{\rho_s} \omega_s^{n} - \dfrac{\rho_f}{\rho_s} \omega_s^{n-1}.
        \label{eq: implicit_coupling_euler_equation}
        \end{multline}
        \end{subequations}

        \item[(d)] Fixed relaxation of the linear and angular velocities:
        \begin{subequations}
        \begin{equation}
        \mathbf{U}_{s}^{(k)} = \alpha_r \mathbf{U}_{s}^{(k), *} + (1 - \alpha_r) \mathbf{U}_{s}^{n},
        \label{eq: relax_linear_velocity_implicit_algorithm}
        \end{equation}
        \begin{equation}
        \omega_{s}^{(k)} = \alpha_r \omega_{s}^{(k), *} + (1 - \alpha_r) \omega_{s}^{n},
        \label{eq: relax_angular_velocity_implicit_algorithm}
        \end{equation}
        \end{subequations}
        where $\alpha_r$ is the fixed relaxation parameter, with $0 < \alpha_r \leq 1$. For $\alpha_r = 1$, no relaxation occurs, i.e., $(\cdot)_s^{(k)} = (\cdot)_s^{(k), *}$.
        
        \item[(e)] Update the position of all Lagrangian markers $\mathbf{X}_{n}^{(k)}$ and rigid body center of mass $\mathbf{X}_s^{(k)}$.

        \item[(f)] Check iterations residuals to exit implicit coupling step:
        \begin{equation}
        \mathbf{R}_1 =\dfrac{\bigl\| \mathbf{U}_s^{(k)} - \mathbf{U}_s^{(k-1)} \bigl\|}{\bigl\| \mathbf{U}_s^{(k)} \bigl\|}< \boldsymbol{\epsilon}_{FSI} \quad \text{and} \quad \mathbf{R}_2 = \dfrac{\bigl\| \omega_s^{(k)} - \omega_s^{(k-1)} \bigl\|}{\bigl\| \omega_s^{(k)} \bigl\|} < \boldsymbol{\epsilon}_{FSI},
        \end{equation}
        where $\boldsymbol{\epsilon}_{FSI}$ is a user-specified tolerance for the FSI iterations. \\
        If the convergence criteria are met, then:
        \begin{subequations}
        \begin{equation}
        \mathbf{X}_{s}^{n+1} = \mathbf{X}_{s}^{(k)}, 
        ~\mathbf{U}_s^{n+1} = \mathbf{U}_s^{(k)}, ~\text{and} 
        ~\omega_s^{n+1} = \omega_s^{(k)},
        \end{equation}
        \begin{equation}
        ~\mathbf{X}_{n}^{n+1} = \mathbf{X}_{n}^{(k)}, 
        ~\mathbf{U}_{n}^{n+1} = \mathbf{U}_{s}^{n+1} + \omega_s^{n+1} \times \mathbf{r}_n^{n+1}, ~\text{and} 
        ~\mathbf{F}_{n}^{n+1} = \mathbf{F}_{n}^{(k)}.
        \end{equation}
        \end{subequations}
        Otherwise, return to \textit{Step 2 (a)} and let $k = k + 1$.
    \end{enumerate}
    
    \item[3.] PISO Corrector Loops: \\
    Inner corrector loops start for $m = 1$, where $1 \leq m \leq m_f$, with $m_f = 3$ being the fixed number of loops. For $m=1$, $\hat{\mathbf{u}}_f^{(0)} = \Tilde{\mathbf{u}}_{f}^{(k)}$:
    \begin{enumerate}
        \item[(a)] Solve the mPPE for a new pressure field, to enforce the divergence-free constraint:
        \begin{align}
        \nabla \cdot \left( \dfrac{1}{a_P} \left(\nabla p\right)^{(m)} \right) 
        = \nabla \cdot \left( \dfrac{\mathbf{H} \left(\hat{\mathbf{u}}_f^{(m-1)}, \mathbf{u}_f^{n}, \mathbf{u}_f^{n-1}\right)}{a_P} \right) 
        + \nabla \cdot \left( \dfrac{1}{a_P} {\mathbf{f}}^{n+1} \right),
        \label{eq: mppe_solving_algorithm}
        \end{align}
        
        \item[(b)] Correct the velocity field using the former pressure solution $p^{(m)}$:
        \begin{equation}
        \hat{\mathbf{u}}_f^{(m)}
        = \dfrac{\mathbf{H} \left(\hat{\mathbf{u}}_f^{(m-1)}, \mathbf{u}_f^{n}, \mathbf{u}_f^{n-1}\right)}{a_P} 
        - \dfrac{1}{a_P} \left(\nabla p\right)^{(m)} 
        + \dfrac{1}{a_P} {\mathbf{f}}^{n+1},
        \label{eq: explicit_velocity_correction_algorithm}
        \end{equation}
        
        \item[(c)] At $m = m_f$, update velocity and pressure fields for $t^{n+1}$:
        \begin{equation}
        \mathbf{u}_f^{n+1} = \hat{\mathbf{u}}_f^{(m_f)},
        \end{equation}
        \begin{equation}
        p^{n+1} = p^{(m_f)}.
        \end{equation}
        The divergence-free constraint is satisfied at this stage, $\nabla \cdot \mathbf{u}_f^{n+1} = 0$.
    \end{enumerate}
\end{enumerate}

\begin{figure}[!h]
\centering
{
\scalebox{0.75}{
\begin{tikzpicture}[node distance=1.5cm]
\node (start) [startstop] {Start at $t^{n}$};

\node (in1) [io, below of=start, yshift=0.2cm] {\textbf{Momentum Predictor}};
\node (pro1) [process, below of=in1] {Predict velocity $\mathbf{u}^{*}_f$};

\node (in2) [io, below of=pro1] {\textbf{Fluid-Rigid Body Coupling}};
\node (pro20) [process, below of=in2, minimum width=6.5cm, text width=6.5cm] {Desired Lagrangian velocity $\mathbf{U}_n^{(k-1)}$};

\node (pro21) [process, below of=pro20, xshift=-8.5cm] {Velocity interpolation $\mathbf{U}^{(k,m-1)}$};
\node (pro22) [process, below of=pro20, xshift=-2.83cm] {Direct-forcing $\mathbf{F}_{n}^{(k,m)}$};
\node (pro23) [process, below of=pro20, xshift=2.83cm] {Force spreading ${\mathbf{f}}^{(k,m)}$};
\node (pro24) [process, below of=pro20, xshift=8.5cm] {Update velocity $\Tilde{\mathbf{u}}_f^{(k,m)}$};
\node (dec25) [decision, below of=pro20, yshift=-2.2cm] {Interface matching? \\ $\mathbf{U}_{n}^{(k-1)} \stackrel{?}{=} \mathbf{U}^{(k,m)}$};

\node (pro31) [process, below of=dec25, yshift=-1cm, xshift=-6.75cm, minimum width=6cm, text width=6cm] {Newton-Euler $\mathbf{U}_s^{(k),*}$ and $\omega_s^{(k),*}$};
\node (pro32) [process, below of=dec25, yshift=-1cm, minimum width=6cm, text width=6cm] {Relaxation $\mathbf{U}_s^{(k)}$ and $\omega_s^{(k)}$};
\node (pro33) [process, below of=dec25, yshift=-1cm, xshift=6.75cm, minimum width=6cm, text width=6cm] {Position update $\mathbf{X}_{s}^{(k)}$ and $\mathbf{X}_n^{(k)}$};
\node (dec34) [decision, below of=pro32, yshift=-1cm] {Residuals converged? \\ $\mathbf{R}_1,~\mathbf{R}_2 < \boldsymbol{\epsilon}_{FSI}$};

\node (in3) [io, below of=dec34, yshift=-1cm] {\textbf{PISO Corrector Loops}};
\node (pro41) [process, below of=in3, xshift=-4.5cm] {Solve mPPE for $p^{(m)}$}; 
\node (pro42) [process, below of=in3, xshift=4.5cm] {Correct velocity field $\hat{\mathbf{u}}_f^{(m)}$};
\node (dec43) [decision, below of=pro42, xshift=-4.5cm, yshift=-0.75cm] {Max. number of \\ correctors $m_f$ reached?};
\node (pro44) [process, below of=dec43, yshift=-0.75cm, minimum width=6.5cm, text width=6.5cm] {Update $\mathbf{u}_f^{n+1} = \hat{\mathbf{u}}_f^{(m_f)}$ and $p^{n+1} = p^{(m_f)}$};

\node (stop) [startstop, below of=pro44, yshift=0.2cm] {Proceed to $t^{n+1}$};

\draw [arrow] (start) -- (in1);
\draw [arrow] (in1) -- (pro1);
\draw [arrow] (pro1) -- (in2);
\draw [arrow] (in2) -- (pro20);
\draw [arrow] (pro20.west) -| (pro21.north);
\draw [arrow] (pro21) -- (pro22);
\draw [arrow] (pro22) -- (pro23);
\draw [arrow] (pro23) -- (pro24);
\draw [arrow] (pro24.south) |- (dec25.east);
\draw [arrow] (dec25.west) --++ (-25pt,0) node[anchor=south] {\textbf{No}} -| (pro21.south);
\draw [arrow] (dec25.south) --++ (0,-7.5pt) node[anchor=south, xshift=-35pt] {\textbf{Yes}} -| (pro31.north);
\draw [arrow] (pro31) -- (pro32);
\draw [arrow] (pro32) -- (pro33);
\draw [arrow] (pro33.south) --++ (0,-7.5pt) -| (dec34.north);
\draw [arrow] (dec34.east) --++ (25pt,0) node[anchor=south] {\textbf{No}} --++ (200pt,0) |- (pro20.east);
\draw [arrow] (dec34.south) --++ (0,-10pt) node[anchor=west] {\textbf{Yes}} -- (in3.north);
\draw [arrow] (in3.south) --++ (0,-7.5pt) -| (pro41.north);
\draw [arrow] (pro41) -- (pro42);
\draw [arrow] (pro42.south) |- (dec43.east);
\draw [arrow] (dec43.west) --++ (-7.5pt,0) node[anchor=south] {\textbf{No}} -| (pro41.south);
\draw [arrow] (dec43.south) --++ (0,-6.5pt) node[anchor=west] {\textbf{Yes}} -- (pro44.north);
\draw [arrow] (pro44) -- (stop);
\end{tikzpicture}
}
}
\caption{Flowchart of the proposed fluid-rigid body implicit coupling algorithm for one simulation time-step.}
\label{fig: Flowchart of the proposed fluid-rigid body implicit coupling algorithm for one simulation time-step.}
\end{figure}
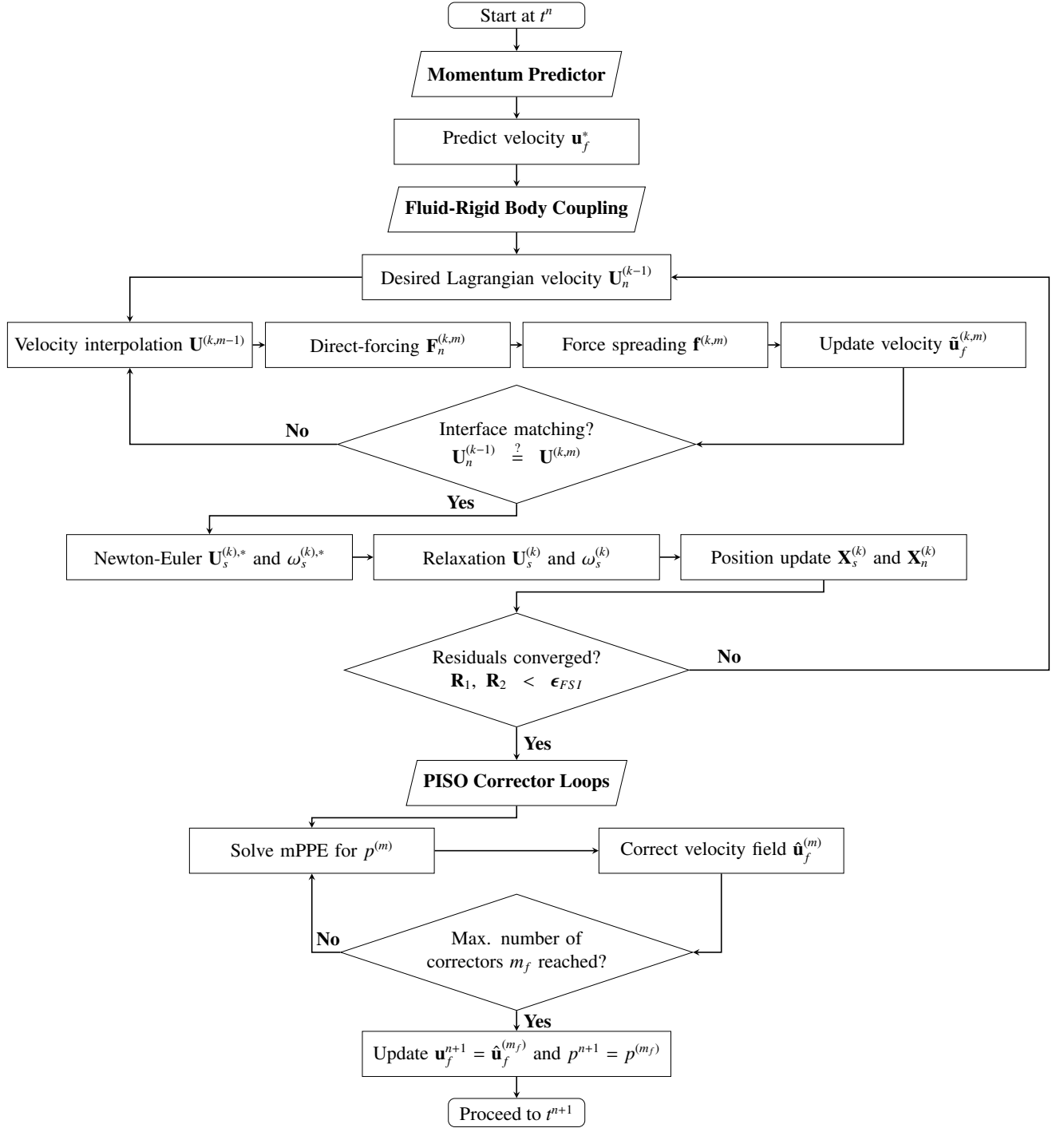

A flowchart summarizing the proposed algorithm is presented in Fig.~(\ref{fig: Flowchart of the proposed fluid-rigid body implicit coupling algorithm for one simulation time-step.}). The choice of relaxation parameter $\alpha_r$ in Eqs.{~(\ref{eq: relax_linear_velocity_implicit_algorithm})} and{~(\ref{eq: relax_angular_velocity_implicit_algorithm})} is user-defined. Smaller values improve stability by damping velocity overshoots but slow the convergence of $(\cdot)_s^{k}$, whereas larger values accelerate convergence at the expense of reduced stability. A practical compromise is $\alpha_r = 0.5$, which provides a balanced trade-off between stability and convergence and will be adopted throughout the remainder of this paper. Unlike the coupling algorithms developed in~\cite{timeLaggedIBPM_1, timeLaggedIBPM_2}, where the velocity interpolation and force spreading operators are performed on Lagrangian markers from the previous time-step, $t^n$, i.e., $\mathcal{I}[\mathbf{X}_{n}^{n}]$ and $\mathcal{S}[\mathbf{X}_{n}^{n}]$, these methods introduce time-lagged operators that theoretically reduce the overall time accuracy to first-order. In contrast, for the present approach and from a theoretical standpoint, the error introduced by the time-lagged operators~\cite{timeLaggedIBPM_1, timeLaggedIBPM_2} diminishes. Owing to the strong coupling, the fluid-solid interface is accurately enforced using the rigid body variables at the new time level $t^{n+1}$. Moreover, in the final iteration of the FSI coupling, the collision force $\mathbf{F}_c$ and the position vector $\mathbf{r}_{n}$ are computed based on the Lagrangian markers and the rigid body center of mass from the new time level $t^{n+1}$.

\section{Results} \label{sec: results}
\subsection{Sedimentation of a 2D Disk Inside a Confined Channel}
\label{subsec: Sedimentation of a 2D Disk Inside a Confined Channel}
\begin{figure}[!h]
\centering
\includegraphics[width=.4\linewidth]{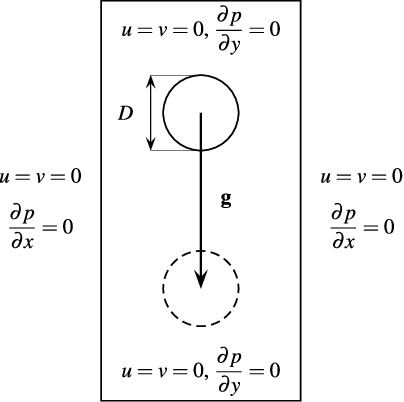}
\caption{Computational domain and boundary conditions for the sedimenting 2D disk inside a confined channel.}
\label{fig: Computational domain and boundary conditions for the sedimenting 2D circular disk inside a confined channel.}
\end{figure}

The sedimentation of a 2D disk in a quiescent fluid is first examined to test the coupling during pure translational motion for non-buoyant rigid bodies with a density ratio $\rho_s / \rho_f = 1.25$. The disk is placed inside a confined channel with the following dimensions $\Omega_f = [0,~0.02] \times [0,~0.06]~m^2$. The disk, with a diameter $D = 2.5 \times 10^{-3}~m$ and density $\rho_s = 1250~kg/m^3$, is initially at rest, with its center of mass located at $(X_s,~Y_s) = (0.01~m,~0.04~m)$. The disk is allowed to fall freely inside the fluid due to gravity, with a density $\rho_f = 1000~kg/m^3$ and dynamic viscosity $\mu_f = 10^{-2}~kg/m \cdot s$. Once the disk reaches the bottom of the fluid channel, the wall collision model is activated automatically. The boundary conditions applied on all sides of the computational domain are illustrated in Fig.~(\ref{fig: Computational domain and boundary conditions for the sedimenting 2D circular disk inside a confined channel.}). A computational setup similar to the work of Wan and Turek~\cite{WanAndTurek2006} was chosen to allow comparison. Simulations are conducted on two different uniform mesh sizes: a coarse mesh with $h = 1/4800~m$ and a fine mesh with $h = 1/9600~m$. Additionally, two different time-steps are tested: $\Delta t = 0.001~s$ and $\Delta t = 0.0005~s$.

\begin{figure}[!h]
\centering
\begin{subfigure}[b]{.4\textwidth}
\centering
\includegraphics[width=0.95\linewidth]{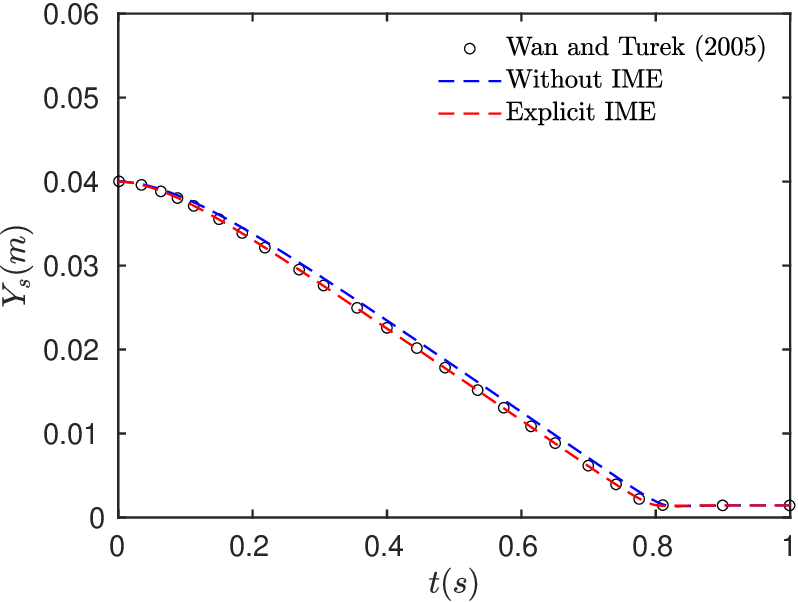}
\subcaption{}
\end{subfigure}%
\begin{subfigure}[b]{.4\textwidth}
\centering
\includegraphics[width=0.95\linewidth]{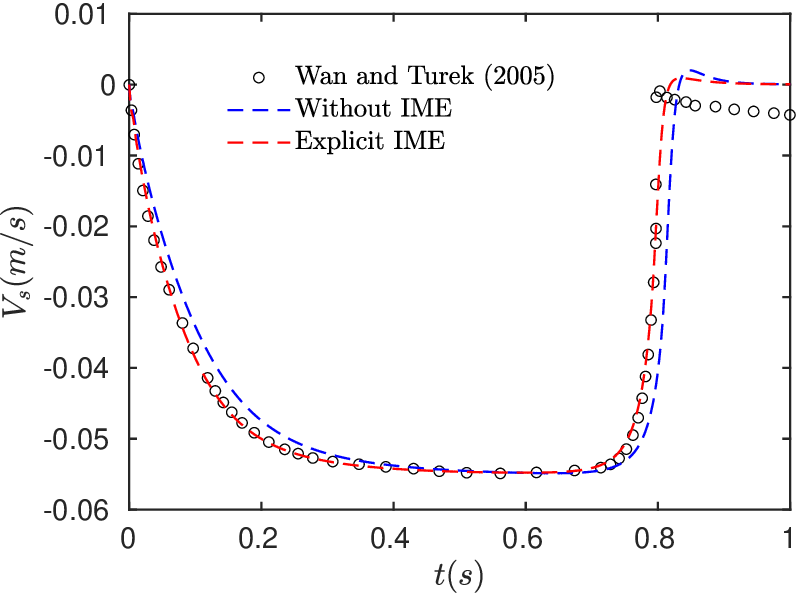}
\subcaption{}
\end{subfigure}
\begin{subfigure}[b]{.4\textwidth}
\centering
\includegraphics[width=0.95\linewidth]{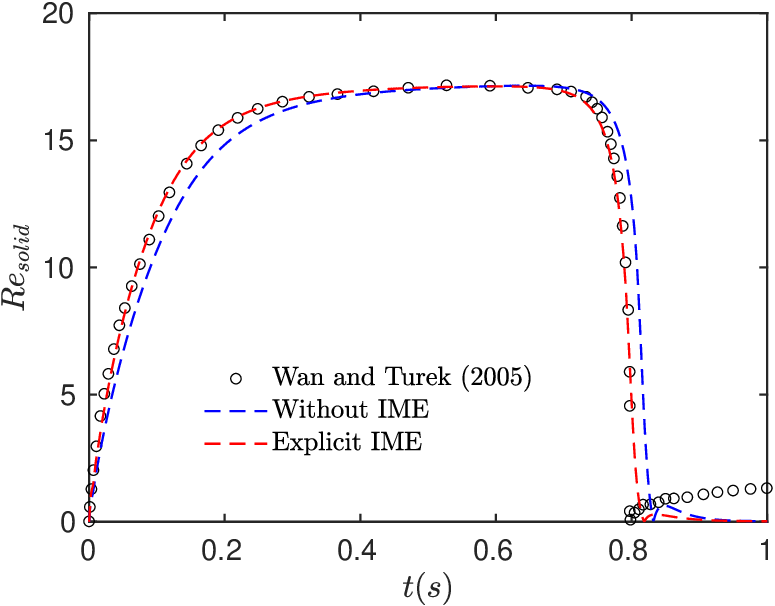}
\subcaption{}
\end{subfigure}%
\begin{subfigure}[b]{.4\textwidth}
\centering
\includegraphics[width=0.95\linewidth]{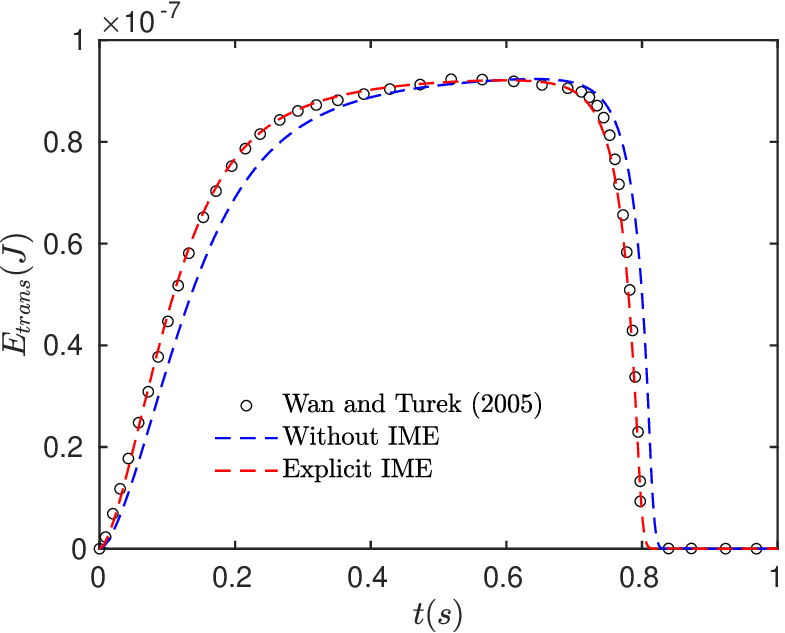}
\subcaption{}
\end{subfigure}
\caption{Time history of the sedimenting 2D disk for different IME treatment using $h = 1/4800~m$ and $\Delta t = 0.001~s$. (a) $y$-coordinate of the disk center of mass position $Y_s$; (b) v-component of the disk velocity $V_s$; (c) solid Reynolds number of the disk $Re_{solid}$; and (d) translational kinetic energy of the disk $E_{trans}$.}
\label{fig: Time history of the sedimenting 2D circular disk for different FSI coupling strategies (explicit/implicit) and IME (without/explicit).}
\end{figure}

The first investigation focuses on examining the influence of explicit IME on the computed results. The solid disk parameters of interest include the y-coordinate of the disk's center of mass $Y_s$ in $(m)$, the vertical component of the disk's translational velocity $V_s$ in $(m/s)$, the solid Reynolds number based on the disk's translational velocity, $Re_{solid} = \rho_s D \sqrt{U_s^2 + V_s^2}/ {\mu_f}$, and the disk's translational kinetic energy, defined as $E_T = \rho_s V ({U_s}^2 + {V_s}^2) / 2$ in $(J)$. Fig.~(\ref{fig: Time history of the sedimenting 2D circular disk for different FSI coupling strategies (explicit/implicit) and IME (without/explicit).}) presents these four quantities as functions of the simulated time $t$. For this analysis, the coarse mesh with $h = 1/4800~m$ and a time-step of $\Delta t = 0.001~s$ is used for all simulations. It is observed that, the sedimenting disk achieves a terminal velocity, $V_t = 0.055~m/s$ corresponding to $Re_{solid} \approx 17.18$, after approximately $t \approx 0.3~s$, before hitting the bottom of the channel at $t \approx 0.8~s$. However, excluding the IME shifts all quantities slightly to the right as the sedimentation is delayed, aligning with findings in~\cite{suzuki_IME}, which state that for $Re_{solid} > 1$, ignoring the IME leads to inaccuracies. Once the IME is included, the results are found to be in great agreement with the numerical data reported by Wan and Turek~\cite{WanAndTurek2006}, demonstrating the importance of the IME in fluid-rigid body interactions. After colliding with the bottom wall, the disk's behavior differs from Wan and Turek's results. The present study shows a slight upward rebound with a positive velocity, which is more physically realistic, whereas Wan and Turek's data incorrectly show a downward velocity after impact.

\begin{figure}[!h]
\centering
\begin{subfigure}[b]{.4\textwidth}
\centering
\includegraphics[width=0.95\linewidth]{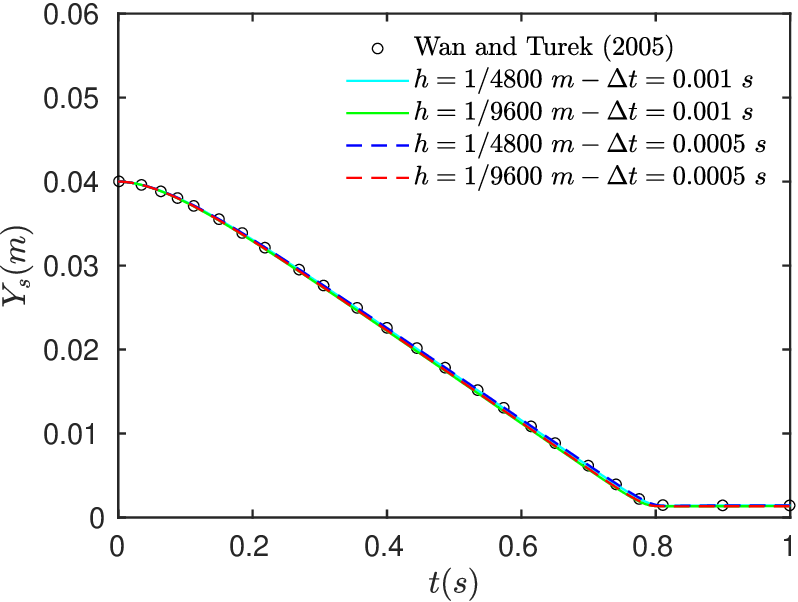}
\subcaption{}
\end{subfigure}%
\begin{subfigure}[b]{.4\textwidth}
\centering
\includegraphics[width=0.95\linewidth]{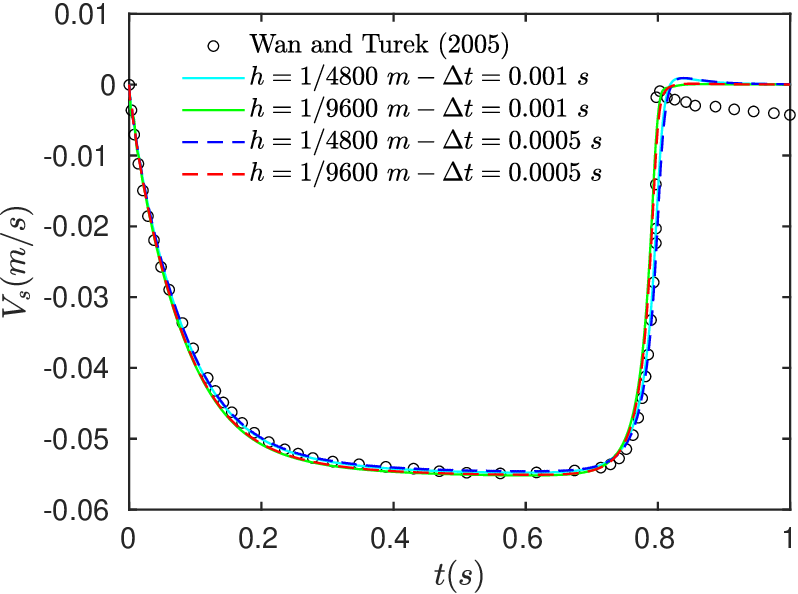}
\subcaption{}
\end{subfigure}
\begin{subfigure}[b]{.4\textwidth}
\centering
\includegraphics[width=0.95\linewidth]{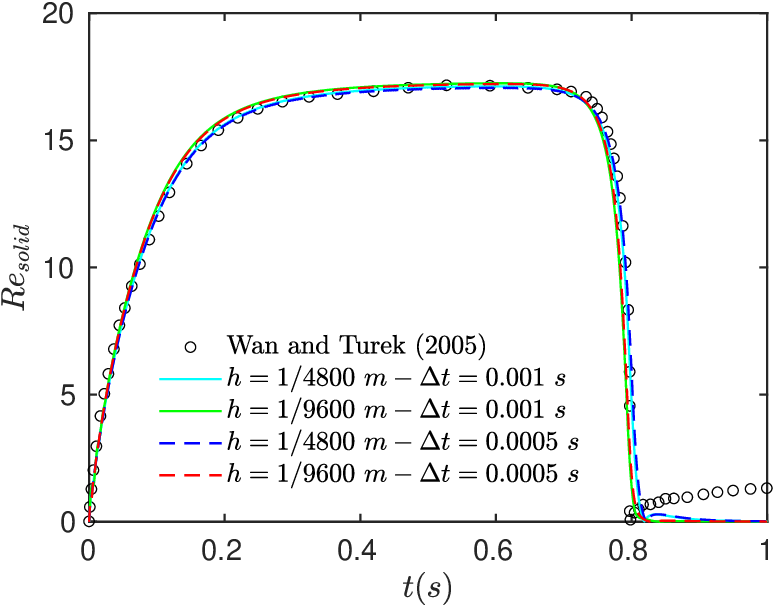}
\subcaption{}
\end{subfigure}%
\begin{subfigure}[b]{.4\textwidth}
\centering
\includegraphics[width=0.95\linewidth]{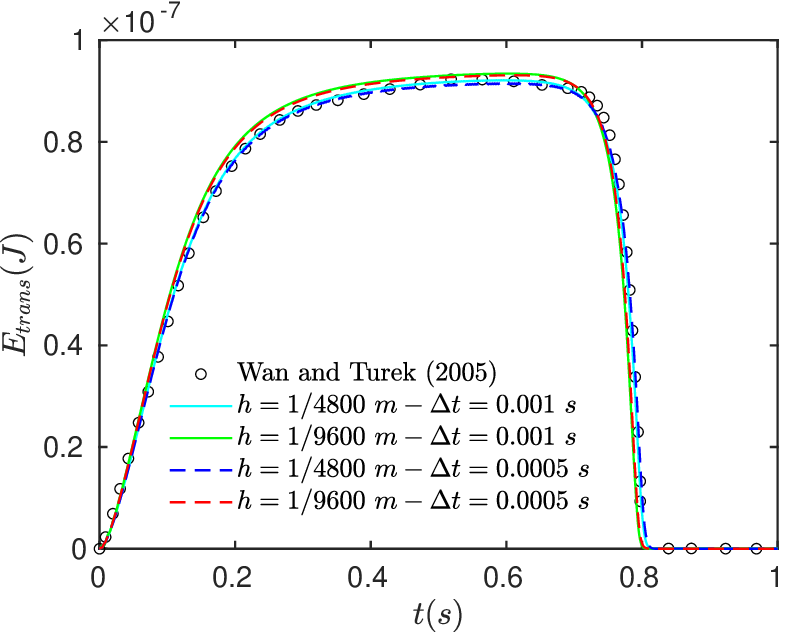}
\subcaption{}
\end{subfigure}
\caption{Time history of the sedimenting 2D disk for different mesh sizes $(h = 1/4800~m, ~h=1/9600~m)$ and time-steps $(\Delta t = 0.001~s, ~\Delta t = 0.0005~s)$ using explicit IME. (a) $y$-coordinate of the disk center of mass position $Y_s$; (b) v-component of the disk velocity $V_s$; (c) solid Reynolds number of the disk $Re_{solid}$; and (d) translational kinetic energy of the disk $E_{trans}$.}
\label{fig: Time history of the sedimenting 2D circular disk for two different mesh sizes and time-steps.}
\end{figure}

The second investigation assesses the sensitivity of the coupling to different mesh sizes and time-steps. For this analysis, the explicit IME is used for all simulations. Fig.~(\ref{fig: Time history of the sedimenting 2D circular disk for two different mesh sizes and time-steps.}) illustrates the four quantities of interest for all $h$ and $\Delta t$. It can be concluded that reducing the time-step has minimal influence on the results, whereas refining the mesh size slightly overpredicts the quantities once the sedimenting disk reaches its terminal velocity. Despite this overprediction, the results remain in good agreement with Wan and Turek~\cite{WanAndTurek2006}, further validating the implementation of translational dynamics in the present work.

\begin{figure}[!h]
\centering
\begin{subfigure}[b]{.2\textwidth}
\centering
\includegraphics[width=0.975\linewidth]{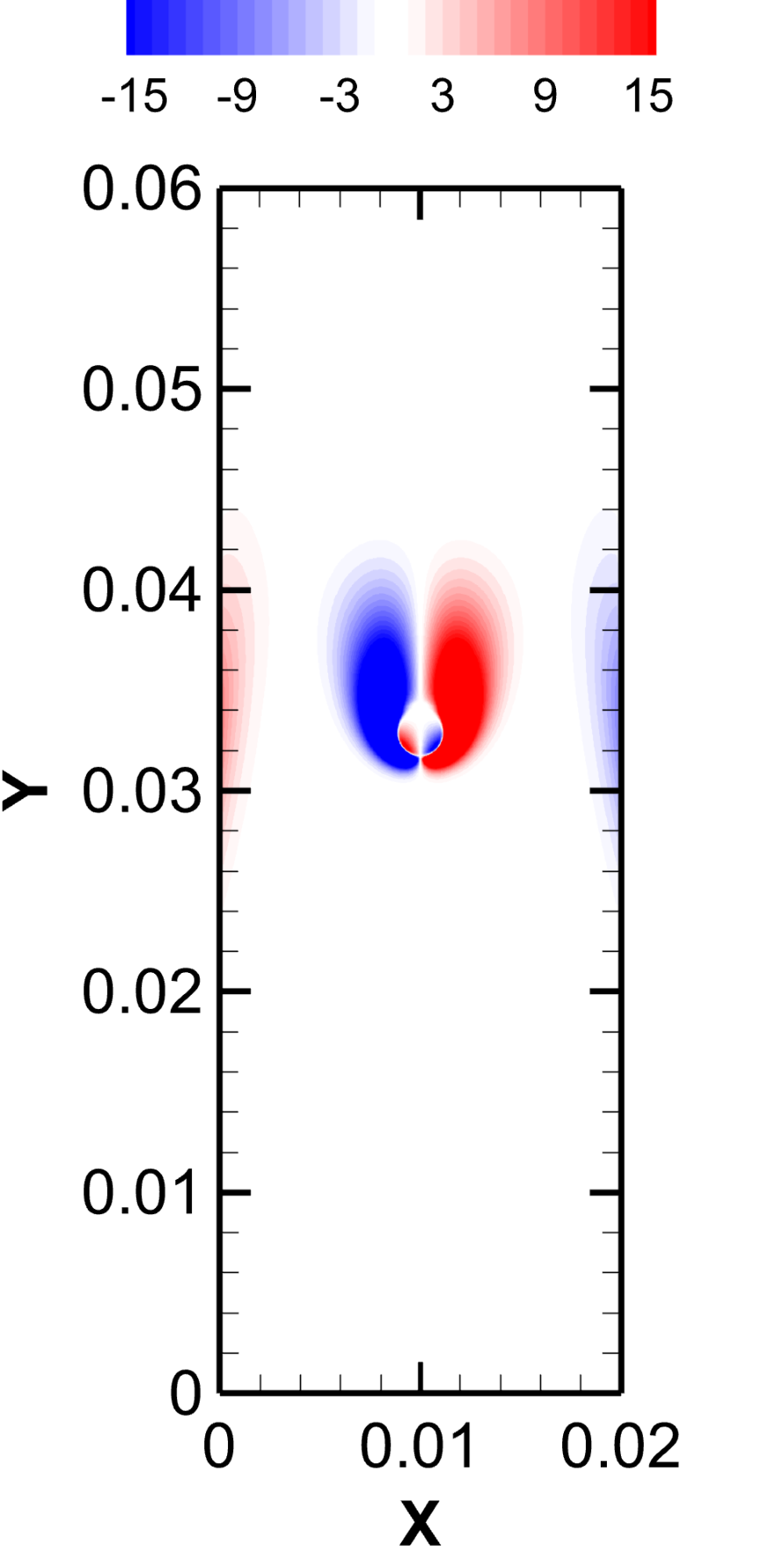}
\subcaption{$t = 0.2~s$}
\end{subfigure}%
\begin{subfigure}[b]{.2\textwidth}
\centering
\includegraphics[width=0.975\linewidth]{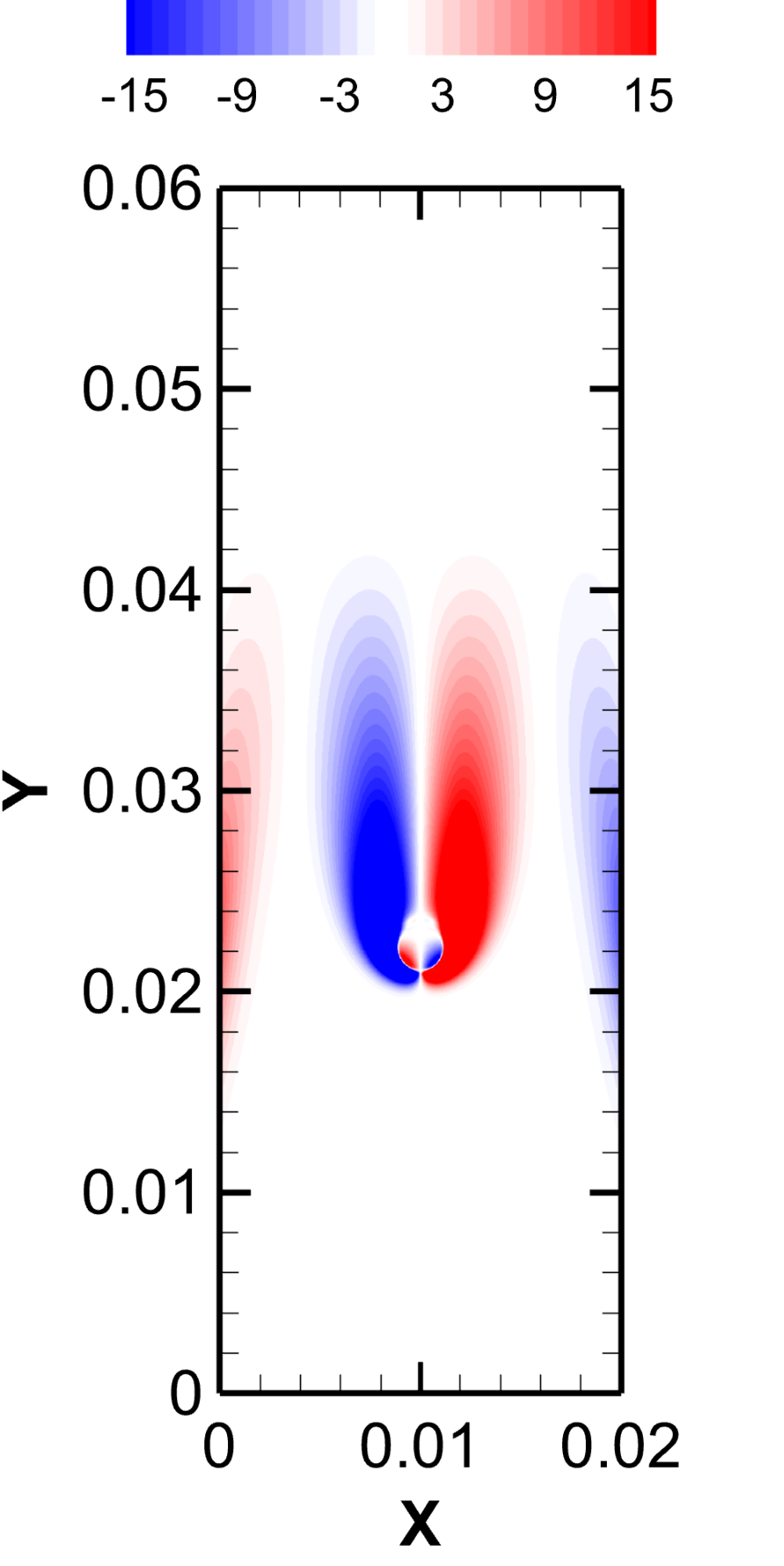}
\subcaption{$t = 0.4~s$}
\end{subfigure}%
\begin{subfigure}[b]{.2\textwidth}
\centering
\includegraphics[width=0.975\linewidth]{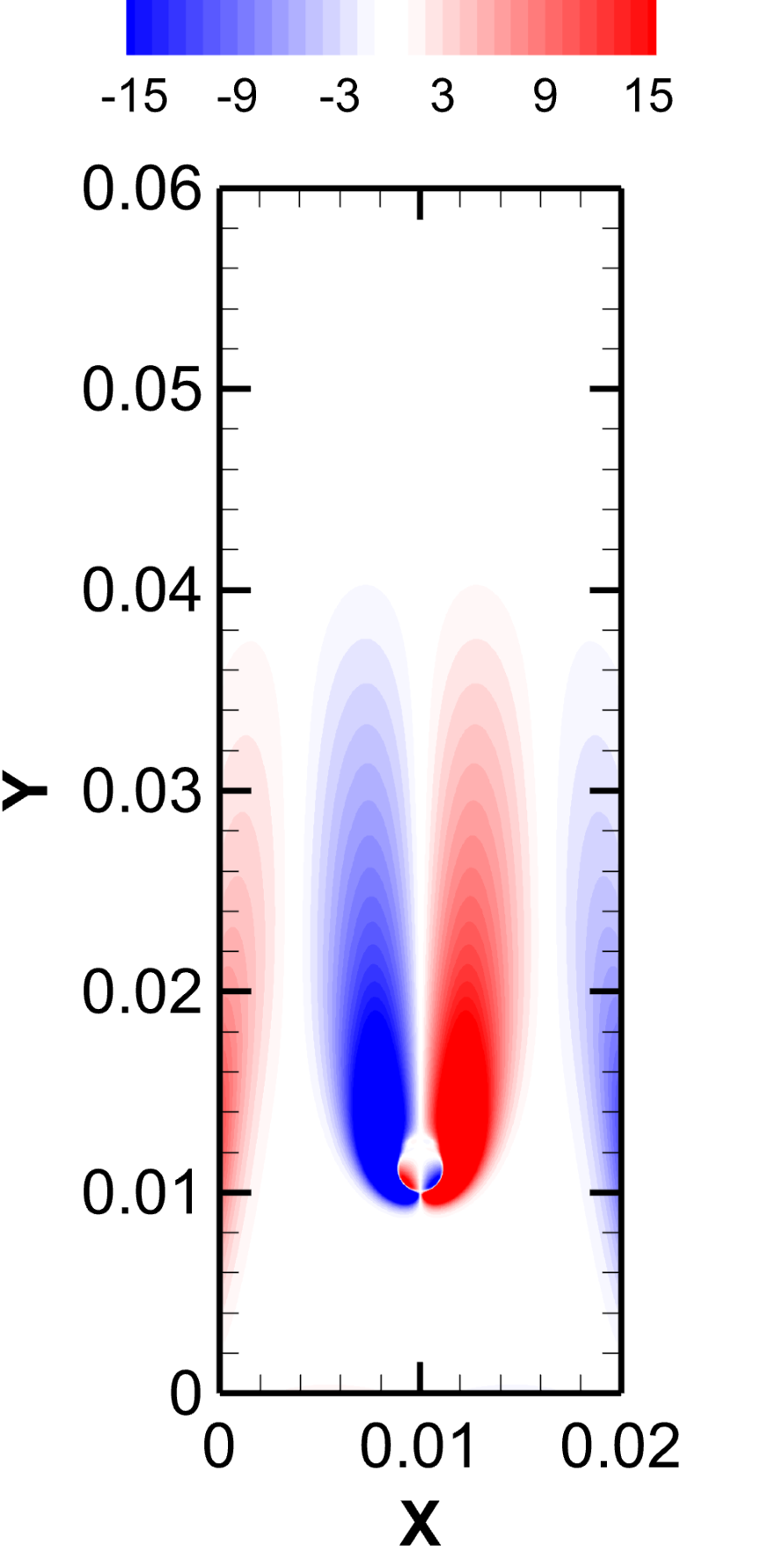}
\subcaption{$t = 0.6~s$}
\end{subfigure}%
\begin{subfigure}[b]{.2\textwidth}
\centering
\includegraphics[width=0.975\linewidth]{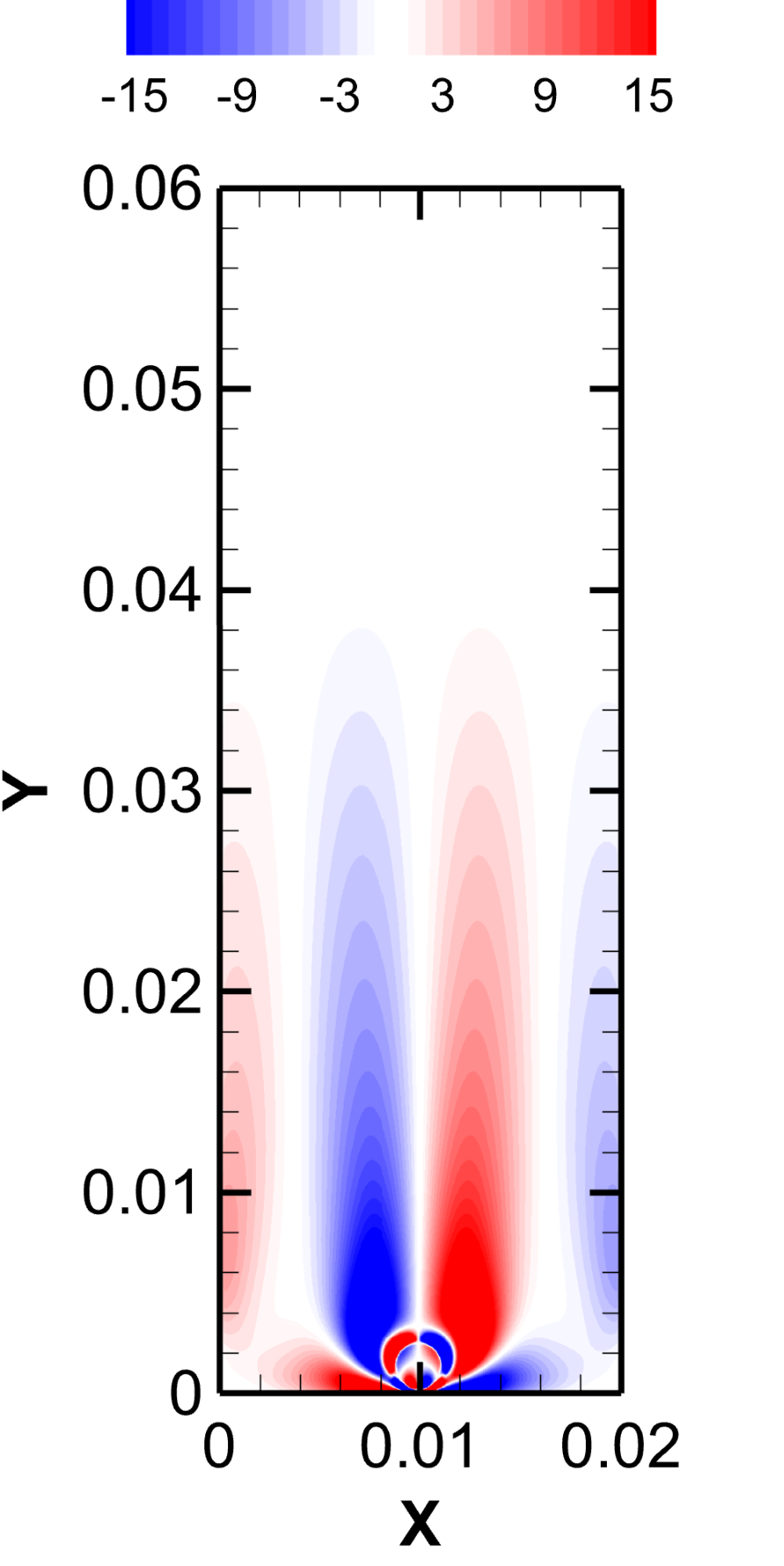}
\subcaption{$t = 0.8~s$}
\end{subfigure}%
\begin{subfigure}[b]{.2\textwidth}
\centering
\includegraphics[width=0.975\linewidth]{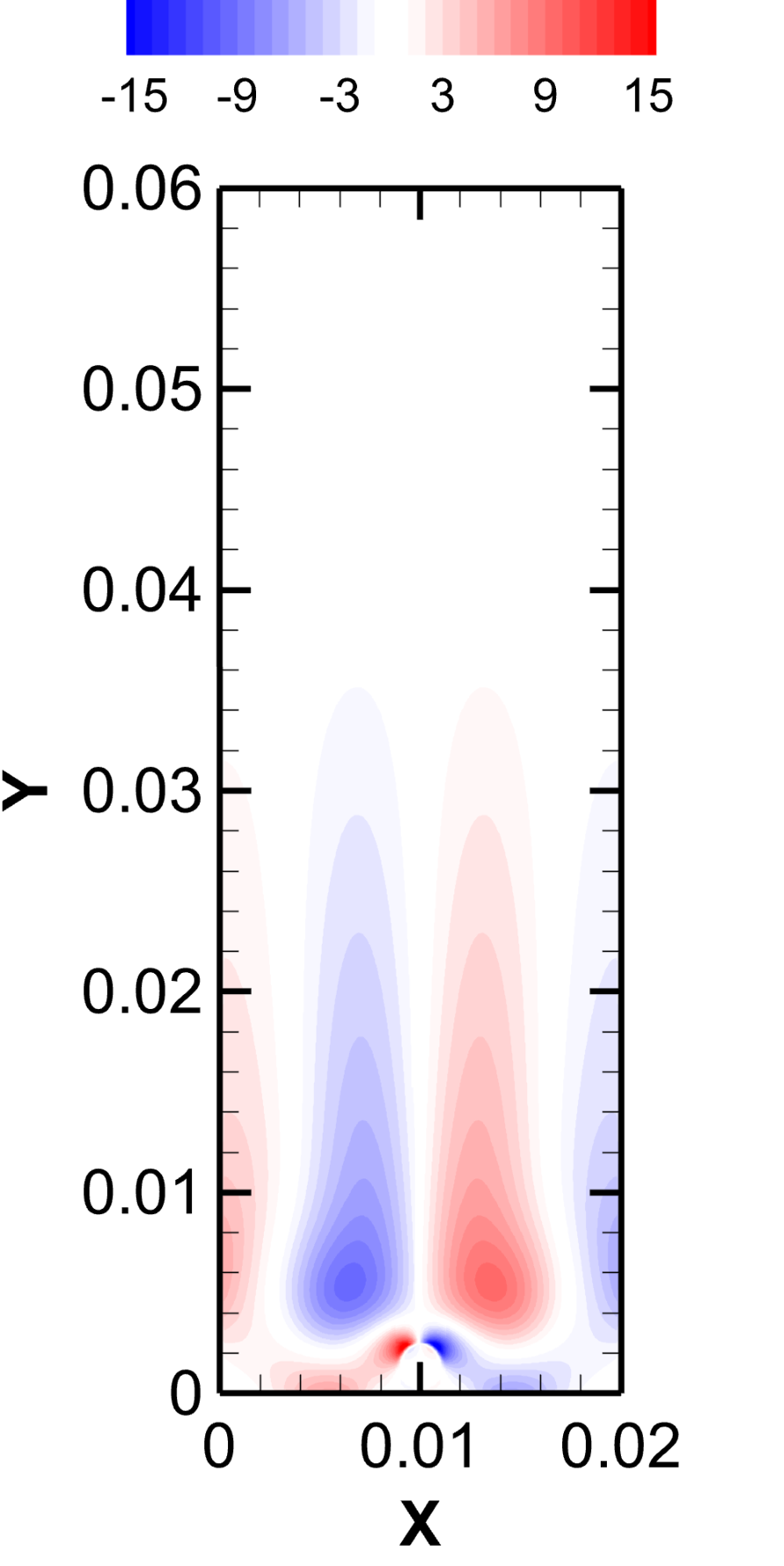}
\subcaption{$t = 1.0~s$}
\end{subfigure}
\caption{Vorticity contours at different time $t$ instances of the sedimenting 2D circular disk.}
\label{fig: Instantaneous vorticity contours at different time t instances of the sedimenting 2D circular disk.}
\end{figure}

The instantaneous vorticity contours generated by the sedimenting disk are shown in Fig.~(\ref{fig: Instantaneous vorticity contours at different time t instances of the sedimenting 2D circular disk.}) for five simulation times: $t = 0.2,~0.4,~0.6,~0.8,~\text{and}~1.0~s$. The vorticity field dissipates at $t = 1.0~s$, after the disk reaches the bottom of the channel approximately at $t \approx 0.8~s$.

\subsection{Shear Flow-Induced Rotation of a Neutrally Buoyant 2D Disk}
\begin{figure}[!h]
\centering
\includegraphics[width=.6\linewidth]{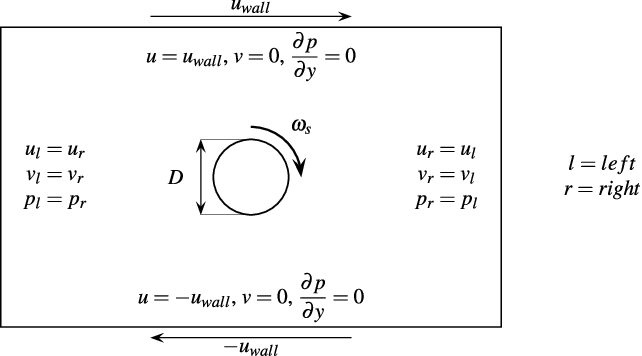}
\caption{Computational domain and boundary conditions for the shear flow-induced rotation of a 2D disk.}
\label{fig: Computational domain and boundary conditions for the free rotating 2D circular disk due to a linear shear flow.}
\end{figure}

The shear flow-induced rotation of a 2D disk is examined to evaluate the coupling during pure rotational motion of neutrally buoyant rigid bodies. In this problem, the disk is placed inside a fluid domain defined as $\Omega_f = [-0.03,~0.03] \times [-0.02,~0.02]~m^2$. The domain is discretized uniformly with a resolution of $h = 1/9600~m$. The disk is positioned at the center of the computational domain. The fluid density is set to $\rho_f = 1000~kg/m^3$, and the kinematic viscosity is $\nu_f = 10^{-6}~m^2/s$. Three disk diameters are considered for this problem, $D = 0.004,~0.008,~\text{and}~0.02~m$. The boundary conditions of the fluid domain are presented in Fig.~(\ref{fig: Computational domain and boundary conditions for the free rotating 2D circular disk due to a linear shear flow.}). The top and bottom walls move at constant velocities in opposite directions, with $u_{wall} = \pm 2 \times 10^{-4}~m/s$. The simulations are run until the non-dimensional time, $t u_{wall}/D$, reaches a value of $600$.  

Since the center of mass of the disk is located at the center of the computational domain, the disk will exhibit no translational motion. Instead, only rotational dynamics are expected about its center of mass. For a linear shear flow without any solid immersed ($D \rightarrow 0$) in the fluid domain, the flow is expected to reach a uniform state, characterized by a constant angular velocity given by $\omega_s = -u_{wall}/H = -0.005~rad/s$, where $H$ is the distance between the top and bottom walls ($H = 0.04~m$). However, as the disk diameter increases, the angular velocity is expected to decrease due to the disruption of the linear shear flow caused by the presence of the disk. The same fluid and solid Reynolds numbers as those in Wan and Turek~\cite{WanAndTurek2006} and Cai~\cite{shangguiPhd2016} are used to enable comparison with data from the literature. These values are $Re_{fluid} = 2 u_{wall} H / \nu_f = 16$ and $Re_{solid}^{rot} = \omega_s D^2 / 2\nu_f = 0.04,~0.16,~\text{and}~1$, corresponding to all disk diameters, respectively.

\begin{figure}[!h]
\centering
\includegraphics[width=0.6\linewidth]{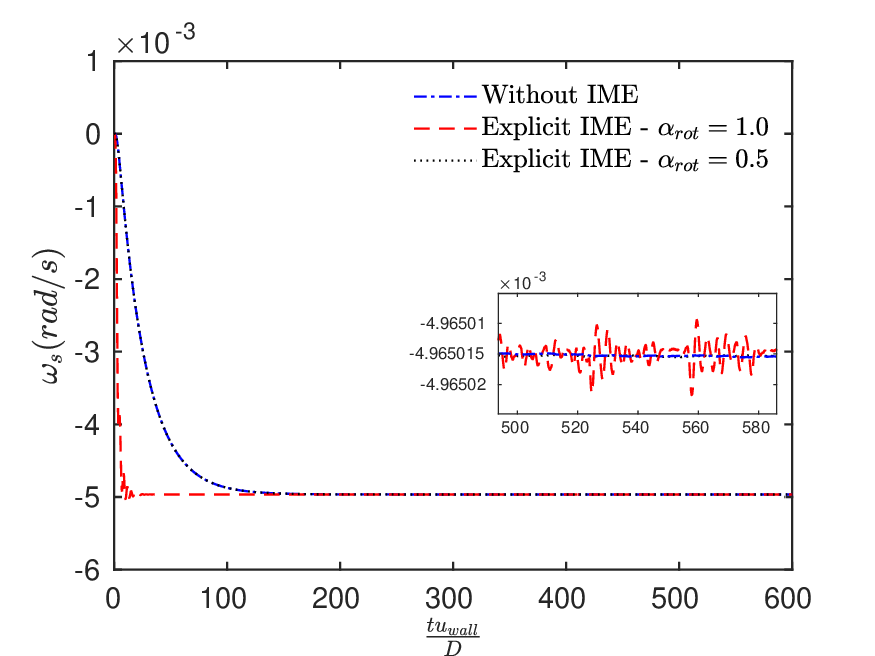}
\caption{Non-dimensional time history of the angular velocity $\omega_s$ for different IME treatment for the shear flow-induced rotation of a 2D disk with diameter $D = 0.004~m$.}
\label{fig: Non-dimensional time history of the disk angular velocity for different FSI coupling strategies (explicit/implicit) and IME (without/explicit/explicit with relaxation) for a circular disk with diameter D = 0.004.}
\end{figure}

The first study focuses on testing various IME treatments. Both the exclusion of the IME and the explicit treatment, with and without relaxation, are tested. The results for the disk with a diameter $D = 0.004~m$ are shown in Fig.~(\ref{fig: Non-dimensional time history of the disk angular velocity for different FSI coupling strategies (explicit/implicit) and IME (without/explicit/explicit with relaxation) for a circular disk with diameter D = 0.004.}). All simulations were able to achieve a steady-state solution as the shear flow developed.

Instabilities arising from the rigid body approximation in the rotational dynamics are clearly observed when the explicit IME is used without relaxation. However, the simulation was able to recover and reach convergence. These instabilities are resolved once a relaxation parameter is introduced, as indicated in Fig.~(\ref{fig: Non-dimensional time history of the disk angular velocity for different FSI coupling strategies (explicit/implicit) and IME (without/explicit/explicit with relaxation) for a circular disk with diameter D = 0.004.}). No significant difference is seen once steady-state solution is reached. This is consistent with the findings in~\cite{suzuki_IME}, which state that for $Re_{solid}^{rot} < 1$, both the exclusion and explicit treatment yield similar results, as the IME is relatively small in this case. The minor oscillations that appeared with the explicit IME without the relaxation, during the steady-state portion of the solution, are avoided by using the relaxation parameter.

The next step in this problem is to validate the implementation of rotational dynamics for different disk diameters. Fig.~(\ref{fig: Non-dimensional time history of the disk angular velocity for different circular disk diameters, using the implicit FSI coupling and the explicit IME with a relaxation parameter.}) reveals the evolution of the angular velocity $\omega_s$ for several disk diameters. As expected, the steady-state angular velocity decreases with increasing disk diameter. Additionally, the time required to reach the steady-state solution is directly proportional to the disk diameter.

\begin{figure}[!h]
\centering
\includegraphics[width=0.55\linewidth]{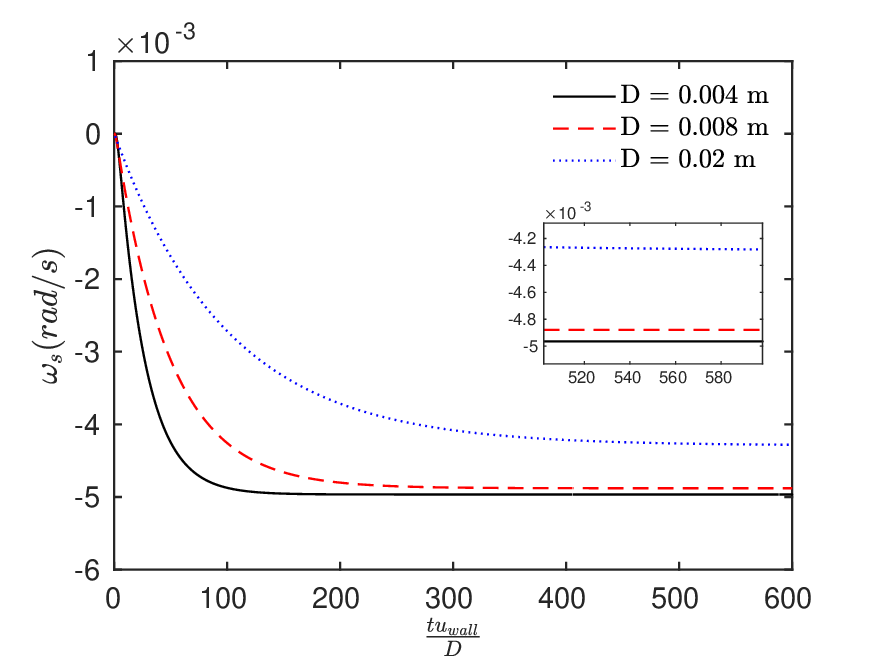}
\caption{Non-dimensional time history of the angular velocity $\omega_s$ for the shear flow-induced rotation of a 2D disk for different diameters $D = 0.004,~0.008,~\text{and}~0.02~m$, using the explicit IME with a relaxation parameter $\alpha_r = 0.5$.}
\label{fig: Non-dimensional time history of the disk angular velocity for different circular disk diameters, using the implicit FSI coupling and the explicit IME with a relaxation parameter.}
\end{figure}

The steady-state angular velocity, also referred to as the terminal angular velocity $\omega_t$, is compared with data from the literature in Table~\ref{tab: Comparison of the terminal angular velocity of the circular disk for different disk diameters.}. Close agreement is observed with the literature for all disk diameters.

\begin{table}[!h]
\centering
\caption{Comparison of the terminal angular velocity $\omega_t$ of the disk for different diameters.}
\label{tab: Comparison of the terminal angular velocity of the circular disk for different disk diameters.}
\begin{tabular}{lccc}
\hline
References & $D = 0.004~m$ & $D = 0.008~m$ & $D = 0.02~m$ \\ 
\hline
Wan and Turek{~\cite{WanAndTurek2006}} & $-0.0049584$ & $-0.0048697$ & $-0.0043148$ \\
Cai{~\cite{shangguiPhd2016}}           & $-0.0049818$ & $-0.0049167$ & $-0.0043878$ \\
Present                                & $-0.0049650$ & $-0.0048795$ & $-0.0042814$ \\
\hline
\end{tabular}
\end{table}

\subsection{Flow-Induced Rotation of a 2D NACA0012 Airfoil}
\begin{figure}[!h]
\centering
\includegraphics[width=0.5\linewidth]{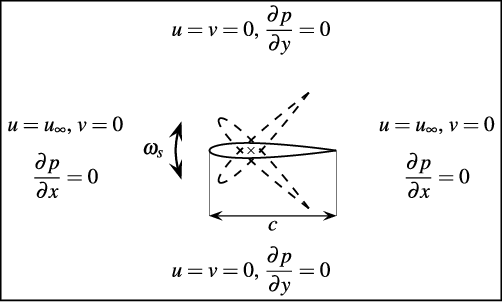}
\caption{Computational domain and boundary conditions for the freely rotating 2D NACA0012 airfoil subjected to a freestream and fixed center of mass.}
\label{fig: Computational domain and boundary conditions for the freely rotating 2D NACA0012 airfoil subjected to a freestream and fixed center of mass.}
\end{figure}

The flow-induced rotation of a 2D NACA0012 airfoil subjected to a freestream is considered for a density ratio $\rho_s / \rho_f = 1.1$. The symmetrical airfoil is positioned within a fluid Eulerian domain, $\Omega_f = [-0.04,~0.16] \times [-0.02,~0.02]~m^2$. Two Eulerian grid resolutions, $h = 1/4800~m$ and $h = 1/9600~m$, and two different time-steps, $\Delta t = 0.002~s$ and $\Delta t = 0.001~s$, are simulated to analyze their influence on the coupling. It is noteworthy that the simulation diverged when the IME was omitted therefore only the explicit IME results will be evaluated. The airfoil has a chord length $c = 1.008 \times 10^{-2}~m$ and is fixed at its center of mass, located at $(X_s,~Y_s) = (4.205 \times 10^{-3}~m,~0~m)$. The airfoil density is set to $\rho_s = 1100~kg/m^3$, with a volume $V = 8.222 \times 10^{-6}~m^3$ and a mass moment of inertia $I_s = 5.071 \times 10^{-8}~kg \cdot m^2$. The initial angle of attack, $\theta_s$, and angular velocity, $\omega_s$, are both set to zero. The boundary conditions for this problem are illustrated in Fig.~(\ref{fig: Computational domain and boundary conditions for the freely rotating 2D NACA0012 airfoil subjected to a freestream and fixed center of mass.}). A freestream velocity $u_{\infty} = 0.01~m/s$, is applied at the inflow and outflow boundaries. The fluid density is set to $\rho_f = 1000~kg/m^3$, with a kinematic viscosity $\nu_f = 10^{-6}~m^2/s$, resulting in a fluid Reynolds number $Re_{fluid} = u_{\infty} c / \nu_f \approx 100.9$, based on the airfoil chord length, consistent with the values used by Glowinski et al.~\cite{glowinski2001} to allow for comparison. The airfoil is allowed to rotate solely around $\mathbf{X}_s$ due to its interaction with the fluid flow (hydrodynamic torque). To stabilize the simulation, a relaxation parameter for the Euler equation is set to $\alpha_r = 0.5$, as the simulation failed to converge when no relaxation was applied $(\alpha_r = 1.0)$ due to the IME approximation in rotational dynamics. 

\begin{figure}[H]
\centering
\begin{subfigure}[b]{.5\textwidth}
\centering
\includegraphics[width=0.95\linewidth]{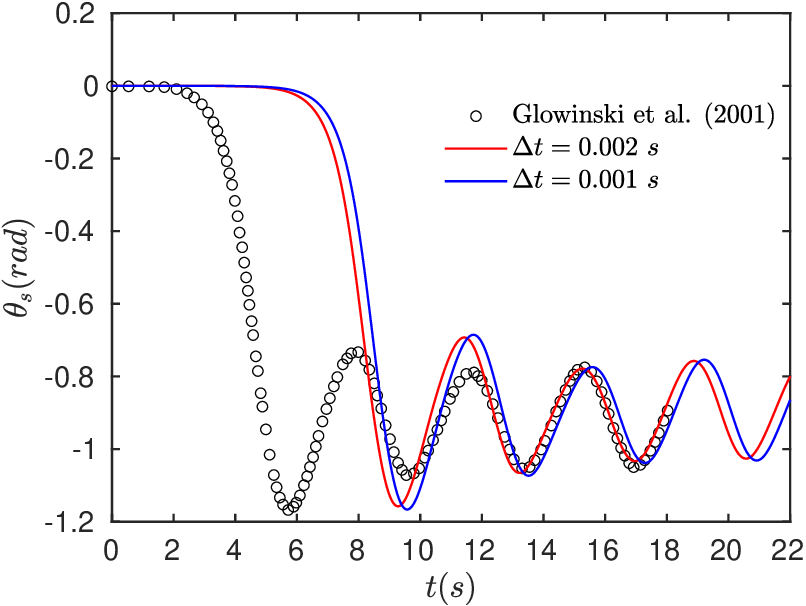}
\subcaption{}
\end{subfigure}%
\begin{subfigure}[b]{.5\textwidth}
\centering
\includegraphics[width=0.95\linewidth]{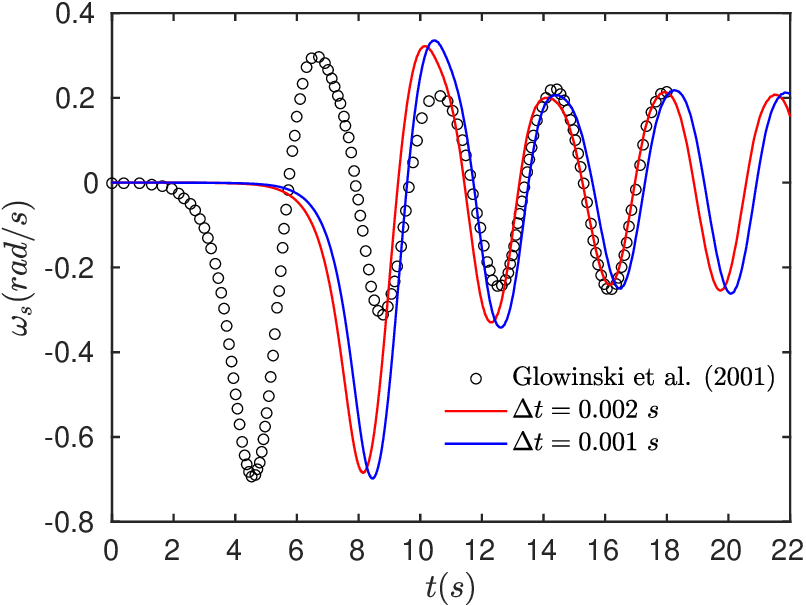}
\subcaption{}
\end{subfigure}
\caption{Time history of the angle of attack $\theta_s$ (a) and angular velocity $\omega_s$ (b) at two different time-steps for the freely rotating 2D NACA0012 airfoil for $h = 1/9600~m$.}
\label{fig: Time history of the rotation angle (a) and angular velocity (b) at two different time-steps for the freely rotating 2D NACA0012 airfoil for h = 1/9600.}
\end{figure}

The first study involves varying the time-step $\Delta t$ to assess its impact on the computed solution using the finer grid resolution, $h = 1/9600~m$. Fig.~(\ref{fig: Time history of the rotation angle (a) and angular velocity (b) at two different time-steps for the freely rotating 2D NACA0012 airfoil for h = 1/9600.}) presents the time evolution of the angle of attack $\theta_s$ and the angular velocity $\omega_s$ of the rotating airfoil. 

A phase shift is observed in the computed results compared to those obtained by Glowinski et al.~\cite{glowinski2001}. In the present results, the airfoil's rotation begins at approximately $t \approx 6.0~s$, whereas for Glowinski et al., the rotation was initiated at $t \approx 3.0~s$. This phase discrepancy aligns with findings from~\cite{suzuki_IME}, which state that an explicit IME treatment introduces a phase shift in rotational dynamics quantities when the solid rotational Reynolds number $Re_{solid}^{rot}$ exceeds 10. For the current FSI scenario, $Re_{solid}^{rot} = c^2 \max\{|\omega_s|\} / 2 \nu_f \approx 35$. However, when the plots are compared with data from the literature, identical patterns emerge, with the maxima and minima of all oscillations captured correctly. This suggests that, despite the delayed interactions caused by the specific IME treatment, the essential flow physics are still effectively captured. From Fig.~(\ref{fig: Time history of the rotation angle (a) and angular velocity (b) at two different time-steps for the freely rotating 2D NACA0012 airfoil for h = 1/9600.}), it can be concluded that refining the time-step further delayed the rotational dynamics of the airfoil while preserving nearly identical oscillation peaks.

\begin{figure}[!h]
\centering
\begin{subfigure}[b]{.5\textwidth}
\centering
\includegraphics[width=0.95\linewidth]{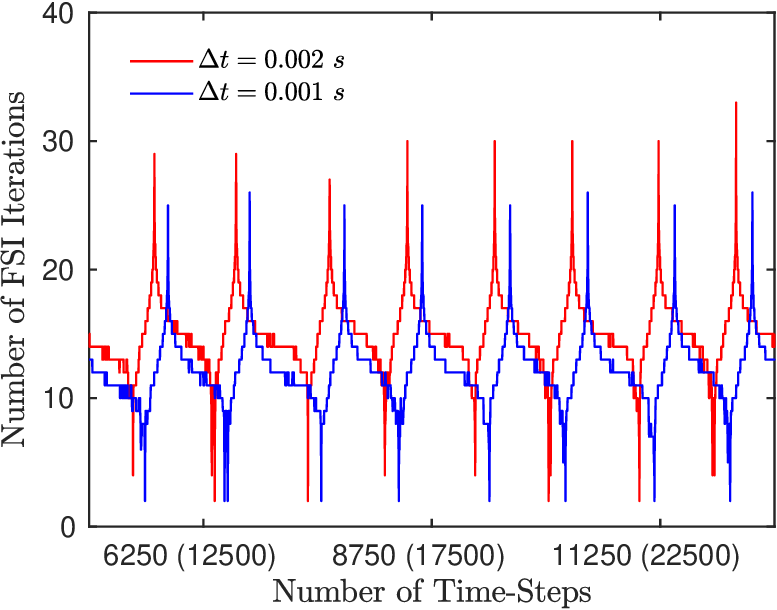}
\subcaption{}
\label{fig: Number of implicit FSI iterations for the freely rotating 2D NACA0012 airfoil for h = 1/9600.}
\end{subfigure}%
\begin{subfigure}[b]{.5\textwidth}
\centering
\includegraphics[width=0.95\linewidth]{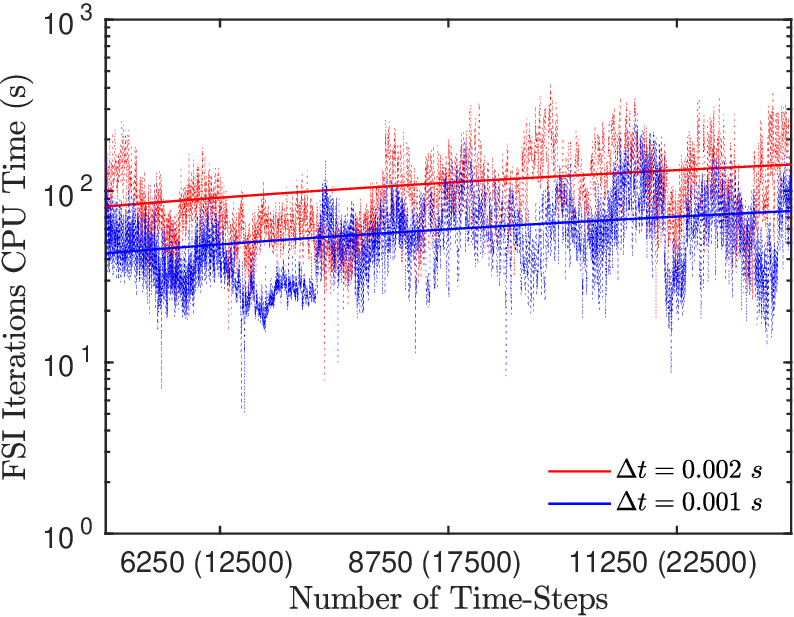}
\subcaption{}
\label{fig: CPU time per FSI iteration for the freely rotating 2D NACA0012 airfoil for h = 1/9600.}
\end{subfigure}
\caption{Number of implicit FSI iterations (a) and CPU time per time-step (b) at two different time-steps for the freely rotating 2D NACA0012 airfoil for $h = 1/9600~m$. The number of time-steps corresponding to $\Delta t = 0.001~s$ are shown in brackets; the rest corresponds to $\Delta t = 0.002~s$. Dashed lines correspond to the actual CPU time; solid lines correspond to a first-degree (linear) polynomial fit of the CPU time data.}
\label{fig: Number of implicit FSI iterations (a) and CPU time per FSI iteration (b) for the freely rotating 2D NACA0012 airfoil for h = 1/9600.}
\end{figure}

Next, to evaluate the effect of the time-step on the implicit coupling, Fig.~(\ref{fig: Number of implicit FSI iterations (a) and CPU time per FSI iteration (b) for the freely rotating 2D NACA0012 airfoil for h = 1/9600.}) is presented. The number of FSI iterations decreases with a smaller time-step, as revealed in Fig.~(\ref{fig: Number of implicit FSI iterations for the freely rotating 2D NACA0012 airfoil for h = 1/9600.}). This reduction in implicit FSI iterations directly translates to a decrease in the CPU time required per time-step in this case, as demonstrated in Fig.~(\ref{fig: CPU time per FSI iteration for the freely rotating 2D NACA0012 airfoil for h = 1/9600.}). The total CPU time for the simulations shows a relatively small difference, with a $10\%$ increase observed for the smaller $\Delta t$ compared to the larger $\Delta t$, despite requiring twice the number of time-steps for the refined $\Delta t$. This is attributed to the lower computational cost per time-step due to the reduced number of FSI iterations when the time-step is refined, leading to overall small increase in the total simulation time.

\begin{figure}[!h]
\centering
\begin{subfigure}[b]{.5\textwidth}
\centering
\includegraphics[width=0.95\linewidth]{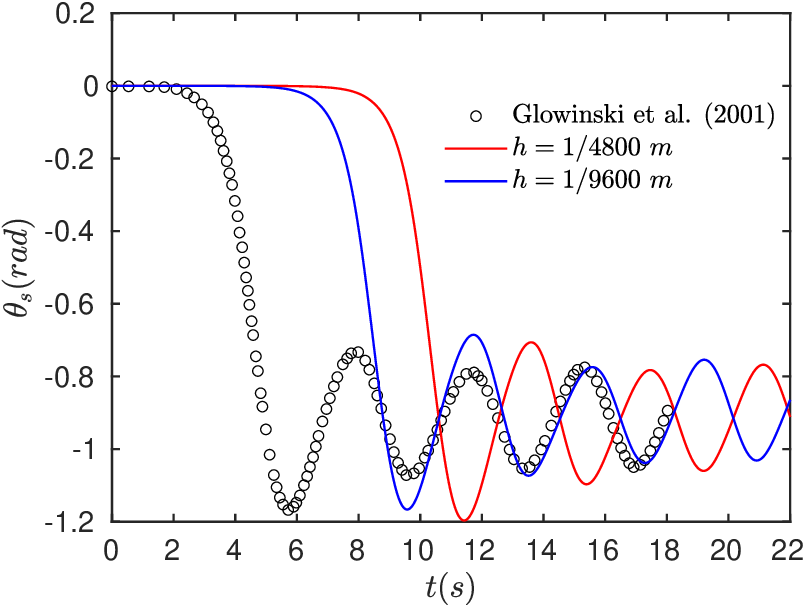}
\subcaption{}
\end{subfigure}%
\begin{subfigure}[b]{.5\textwidth}
\centering
\includegraphics[width=0.95\linewidth]{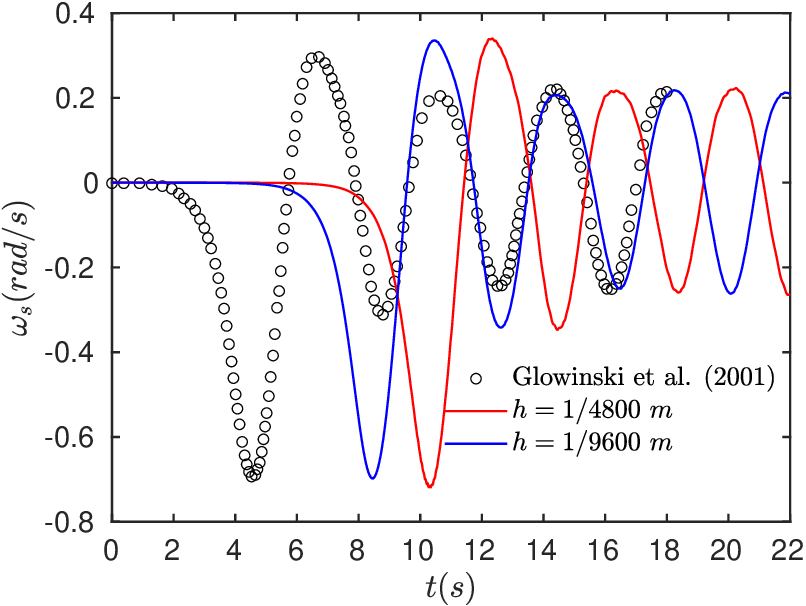}
\subcaption{}
\end{subfigure}
\caption{Time history of the angle of attack $\theta_s$ (a) and angular velocity $\omega_s$ (b) at two different mesh sizes for the freely rotating 2D NACA0012 airfoil for $\Delta t = 0.001~s$.}
\label{fig: Time history of the rotation angle (a) and angular velocity (b) at two different mesh sizes for the freely rotating 2D NACA0012 airfoil for Delta t = 0.001.}
\end{figure}

The sensitivity of the FSI solution to mesh resolution is also examined for this problem. A time-step of $\Delta t = 0.001~s$ was selected for this comparison. The results for $\theta_s$ and $\omega_s$ of the rotating NACA airfoil are presented in Fig.~(\ref{fig: Time history of the rotation angle (a) and angular velocity (b) at two different mesh sizes for the freely rotating 2D NACA0012 airfoil for Delta t = 0.001.}). A similar conclusion to the time-step analysis can be drawn for mesh resolution. Although the phase difference significantly increases when the mesh resolution is doubled (i.e., for the coarser grid), compared to the smaller phase difference observed when the time-step was halved. Nevertheless, both simulations recorded the same oscillation peaks, aligning with the data reported by Glowinski et al.~\cite{glowinski2001}.

\begin{figure}[H]
\centering
\begin{subfigure}[b]{.5\textwidth}
\centering
\includegraphics[width=0.95\linewidth]{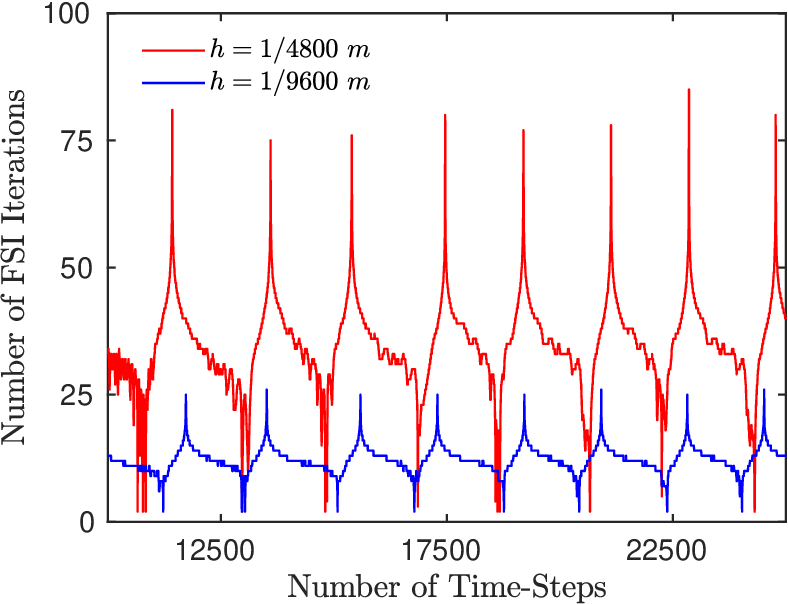}
\subcaption{}
\label{fig: Number of implicit FSI iterations for the freely rotating 2D NACA0012 airfoil for Delta t = 0.001.}
\end{subfigure}%
\begin{subfigure}[b]{.5\textwidth}
\centering
\includegraphics[width=0.95\linewidth]{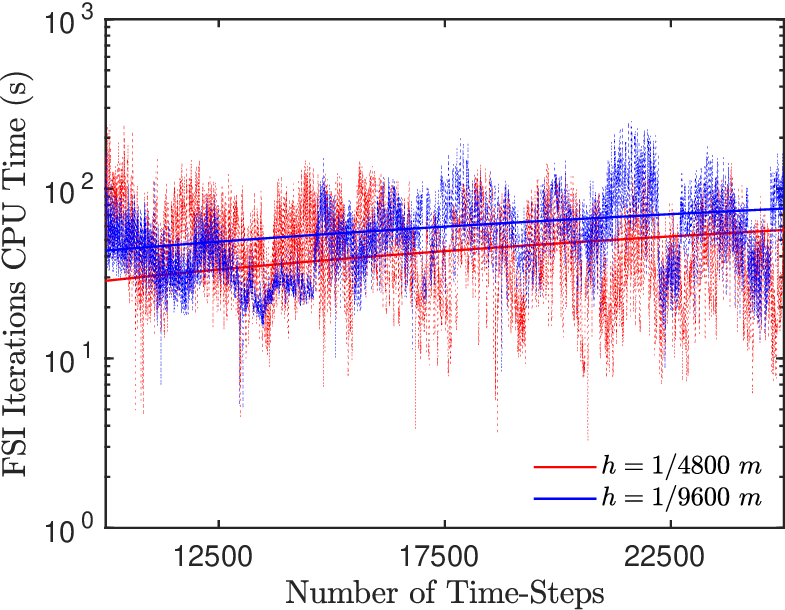}
\subcaption{}
\label{fig: CPU time per FSI iteration (b) for the freely rotating 2D NACA0012 airfoil for Delta t = 0.001.}
\end{subfigure}
\caption{Number of implicit FSI iterations (a) and CPU time per time-step (b) at two different mesh sizes for the freely rotating 2D NACA0012 airfoil for $\Delta t = 0.001~s$. Dashed lines correspond to the actual CPU time; solid lines correspond to a first-degree (linear) polynomial fit of the CPU time data.}
\label{fig: Number of implicit FSI iterations (a) and CPU time per FSI iteration (b) for the freely rotating 2D NACA0012 airfoil for Delta t = 0.001.}
\end{figure}

The effect of mesh resolution on the implicit coupling is depicted in Fig.~(\ref{fig: Number of implicit FSI iterations (a) and CPU time per FSI iteration (b) for the freely rotating 2D NACA0012 airfoil for Delta t = 0.001.}). A notable increase in the number of implicit iterations is observed, with the number of iterations almost tripling when the mesh resolution is doubled, as seen in Fig.~(\ref{fig: Number of implicit FSI iterations for the freely rotating 2D NACA0012 airfoil for Delta t = 0.001.}). Conversely, the CPU time required per time-step is higher for the finer grid compared to the coarser one, as illustrated in Fig.~(\ref{fig: CPU time per FSI iteration (b) for the freely rotating 2D NACA0012 airfoil for Delta t = 0.001.}).

This suggests that, despite the reduced number of iterations per time-step, the size of the fluid grid matrix is the dominant factor affecting the computational cost, rather than the number of iterations. In fact, a smaller number of iterations does not necessarily result in lower computational costs. As a result, the total simulation CPU time for the fine mesh was found to increase by nearly $60\%$ compared to the coarse mesh.

\subsection{Sedimentation of a 2D Ellipse Inside a Confined Channel}
The interaction of a sedimenting ellipse is considered, including coupled translational and rotational motions. A comparison is conducted with the results obtained by Xia et al.~\cite{ellipseSedimentation}, who employed the conventional ALE method and a finite element solver. To facilitate a meaningful comparison with the literature, four non-dimensional parameters are respected. The density ratio $\rho_s/\rho_f = 1.1$, the ellipse aspect ratio $a/b = 2$ where $a$ and $b$ are the major and minor axes of the ellipse respectively, the domain blockage ratio $L/a = 4$ with $L$ the channel width, and the translational solid Reynolds number $Re_{solid}^{trans} = V_t a / \nu_f \approx 12.96$, where $V_t = 1.296 \times 10^{-2}~m/s$ is the vertical v-component of the ellipse's terminal velocity.

\begin{figure}[!h]
\centering
\includegraphics[width=.4\linewidth]{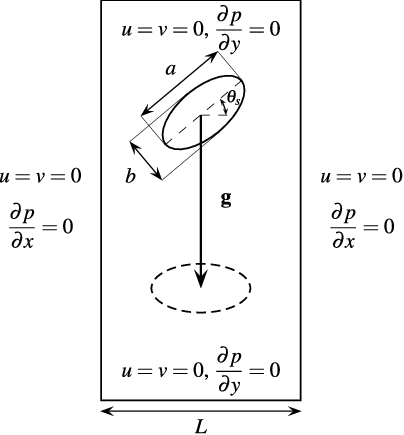}
\caption{Computational domain and boundary conditions for the sedimenting 2D ellipse inside a confined channel.}
\label{fig: Computational domain and boundary conditions for the sedimenting 2D ellipse inside a confined channel.}
\end{figure}

The 2D ellipse is placed inside a confined channel with a narrow width $L$ and a height of $17.5L$. In this simulation, the width is set to $L = 0.004m$. This results in a fluid domain with the following dimensions, $\Omega_f = [0,~L] \times [0,~17.5L] = [0,~0.004] \times [0,~0.07]$, having a uniform grid resolution of $h = 2 \times 10^{-5}~m$. A time-step of $\Delta t = 1 \times 10^{-4}~s$ is used. The boundary conditions are analogous to those considered in the 2D circular disk sedimentation problem, as illustrated in Fig.~(\ref{fig: Computational domain and boundary conditions for the sedimenting 2D ellipse inside a confined channel.}). The quiescent fluid has a density of $\rho_f = 1000~kg/m^3$ and a kinematic viscosity of $\nu_f = 10^{-6}~m^2/s$. For the ellipse, the major and minor axes are $a = 1 \times 10^{-3}~m$ and $b = 5 \times 10^{-4}~m$, respectively. Initially, at $t = 0$, the center of mass of the rigid body is located at $(X_s,~Y_s) = (0.5L~m,~15L~m) = (0.002~m,~0.06~m)$, with an orientation angle $\theta_s = \pi/4$. The solid density is set to $\rho_s = 1100~kg/m^3$ to satisfy the density ratio imposed by Xia et al.~\cite{ellipseSedimentation}. Based on the results obtained in the previous problems, the explicit IME is employed in addition to a relaxation parameter of $\alpha_r = 0.5$ for the Euler equation of motion.

\begin{figure}[!h]
\centering
\begin{subfigure}[b]{.5\textwidth}
\centering
\includegraphics[width=0.75\linewidth]{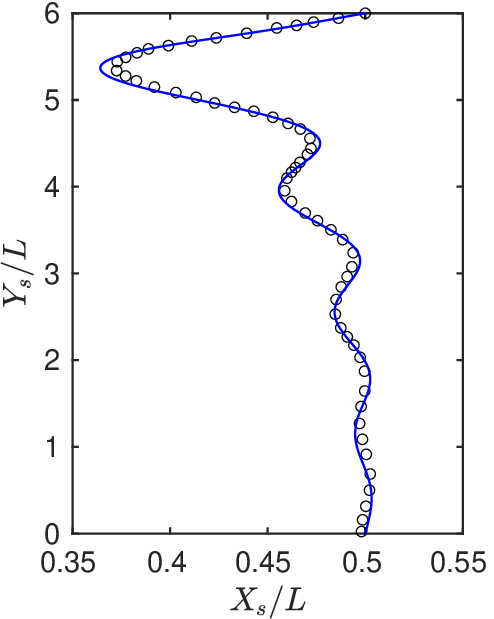}
\subcaption{}
\end{subfigure}%
\begin{subfigure}[b]{.5\textwidth}
\centering
\includegraphics[width=0.75\linewidth]{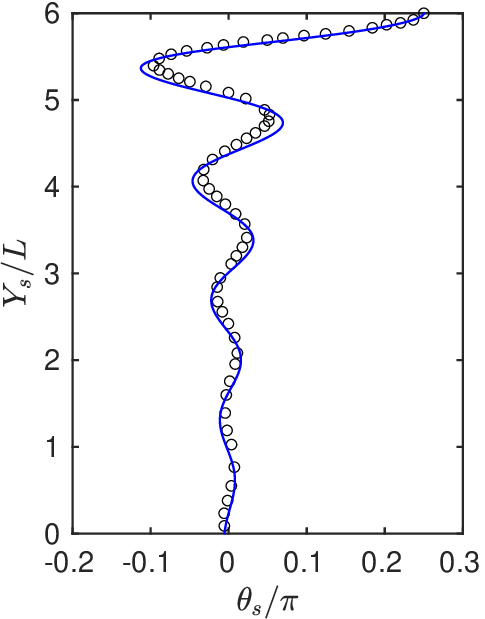}
\subcaption{}
\end{subfigure}
\caption{Trajectory of the center of mass $(X_s,~Y_s)$ (a) and orientation $\theta_s$ (b) of the 2D sedimenting ellipse in a confined channel. The current results are marked with solid lines (\protect\tikz[baseline=-0.1ex] \protect\draw [thick,solid] (0,0.075) -- (0.5,0.075);) and the numerical results extracted from the work of Xia et al.~\cite{ellipseSedimentation} are marked with symbols ($\circ$).}
\label{fig: Trajectory of the center of mass (a) and orientation (b) of a 2D sedimenting ellipse in a confined channel.}
\end{figure}

Fig.~(\ref{fig: Trajectory of the center of mass (a) and orientation (b) of a 2D sedimenting ellipse in a confined channel.}) illustrates the non-dimensional trajectory of the ellipse's center of mass $(X_s,~Y_s)$ and its non-dimensional orientation $\theta_s$ during the sedimentation process. Due to the initial orientation of the ellipse, the symmetry of the flow is disrupted compared to the circular disk case, causing the ellipse to rotate and deviate from its initial horizontal position for $t > 0$. Over time, the ellipse progressively returns to a horizontal position, aligning its major axis perpendicular to the direction of the gravitational force. The simulation results show good agreement with the data extracted from the work of Xia et al.~\cite{ellipseSedimentation}. The solid rotational Reynolds number based on the ellipse's major axis is computed as $Re_{solid}^{rot} = a^2 \max\{|\omega_s|\} / 2 \nu_f \approx 2.291$. This value satisfies the constraints imposed by the IME explicit scheme on $Re_{solid}^{rot}$.

\begin{figure}[!h]
\centering
\centering
\includegraphics[width=0.55\linewidth]{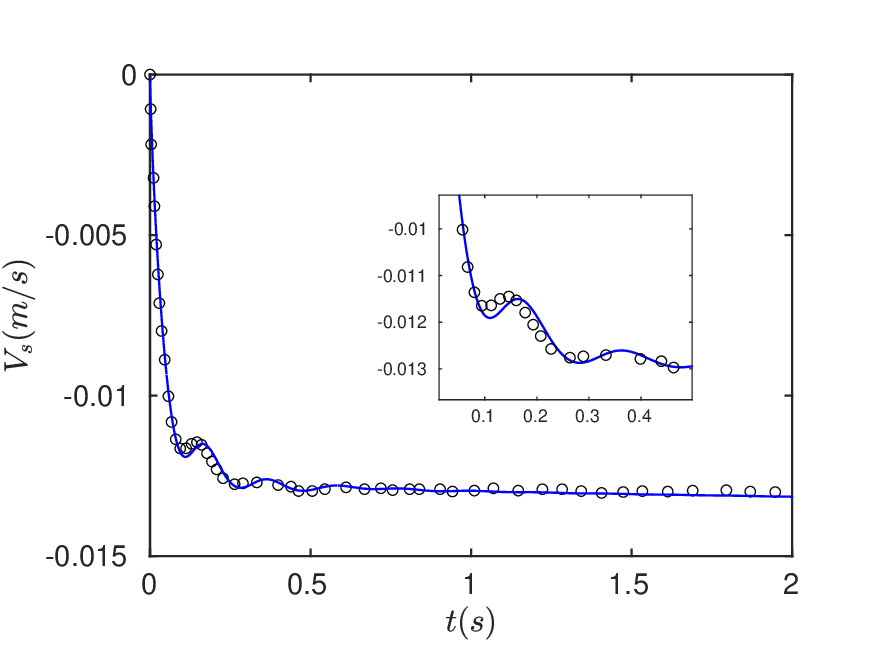}
\caption{Vertical v-component of the velocity $V_s$ of the 2D sedimenting ellipse in a confined channel. The current results are marked with solid lines (\protect\tikz[baseline=-0.1ex] \protect\draw [thick,solid] (0,0.075) -- (0.5,0.075);) and the numerical results extracted from the work of Xia et al.~\cite{ellipseSedimentation} are marked with symbols ($\circ$).}
\label{fig: Vertical v-component of the velocity of the 2D sedimenting ellipse in a confined channel.}
\end{figure}

To compare the current $Re_{solid}^{trans}$ with that obtained by Xia et al., the terminal velocity of the ellipse must be calculated. Fig.~(\ref{fig: Vertical v-component of the velocity of the 2D sedimenting ellipse in a confined channel.}) shows the vertical v-component of the ellipse's velocity. The mean value of the vertical velocity between $t = 1.0~s$ and $2.0~s$ is $V_t = 1.306 \times 10^{-2}~m/s$. Consequently, the computed translational solid Reynolds number for the present simulation is $Re_{solid}^{trans} \approx 13.06$, which is in close agreement with the value reported by Xia et al.~\cite{ellipseSedimentation}. Furthermore, the variation of the vertical velocity aligns well with the data taken from the literature.

\begin{figure}[!h]
\centering
\begin{subfigure}[b]{.16\textwidth}
\centering
\includegraphics[width=.92\linewidth]{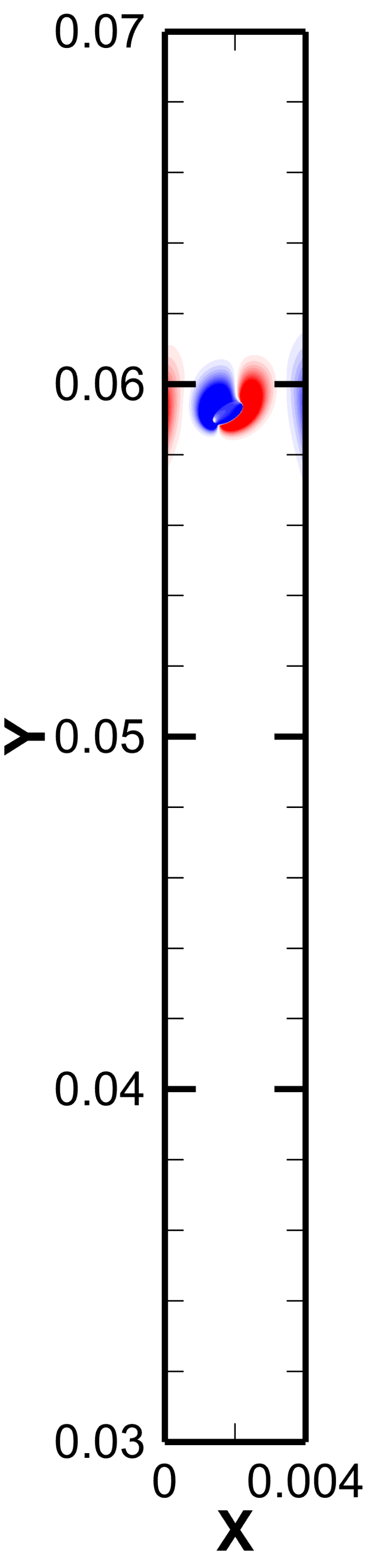}
\subcaption{$t = 0.1~s$}
\end{subfigure}%
\begin{subfigure}[b]{.16\textwidth}
\centering
\includegraphics[width=.92\linewidth]{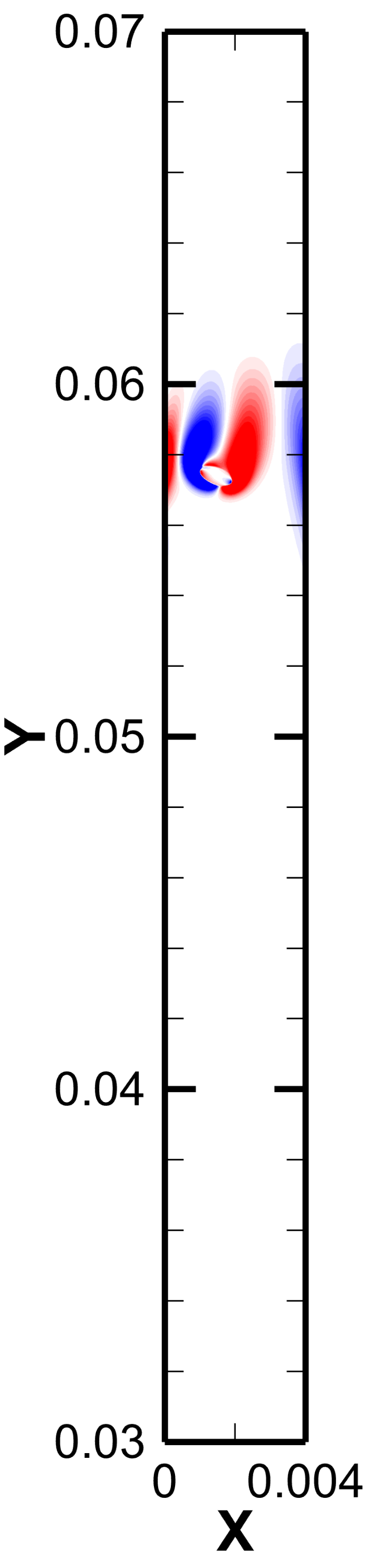}
\subcaption{$t = 0.25~s$}
\end{subfigure}%
\begin{subfigure}[b]{.16\textwidth}
\centering
\includegraphics[width=.92\linewidth]{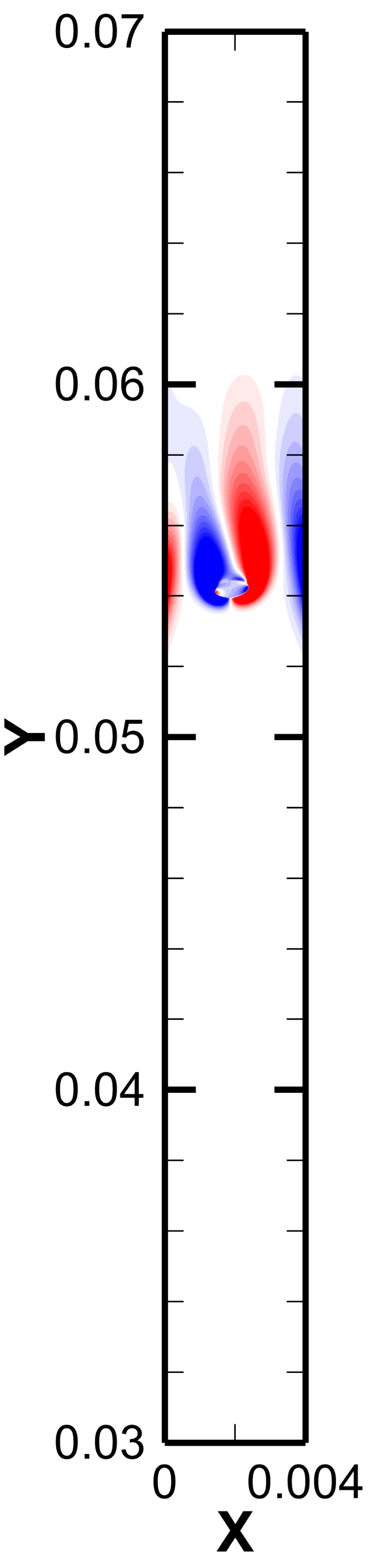}
\subcaption{$t = 0.5~s$}
\end{subfigure}%
\begin{subfigure}[b]{.16\textwidth}
\centering
\includegraphics[width=.92\linewidth]{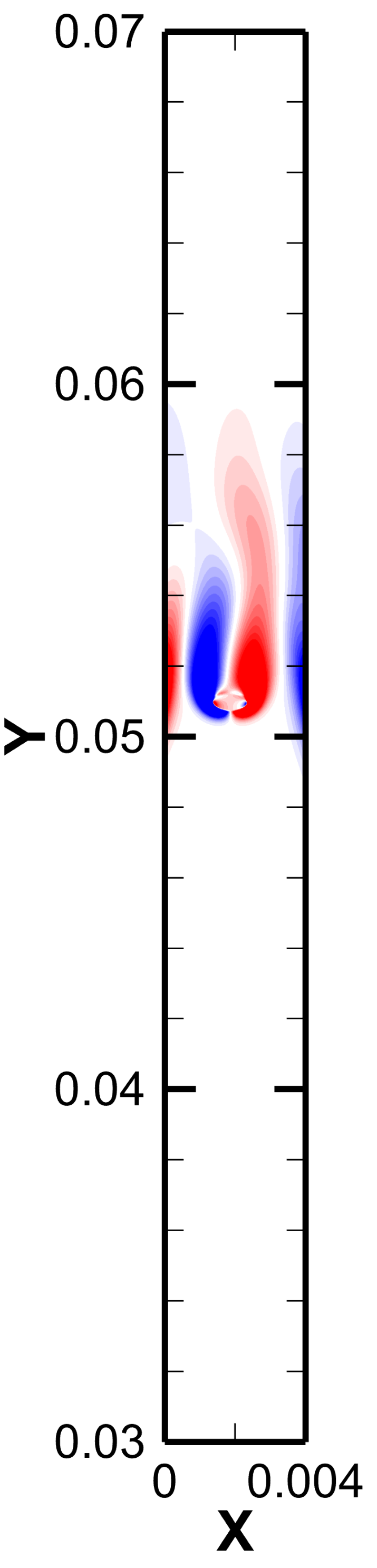}
\subcaption{$t = 0.75~s$}
\end{subfigure}%
\begin{subfigure}[b]{.16\textwidth}
\centering
\includegraphics[width=.92\linewidth]{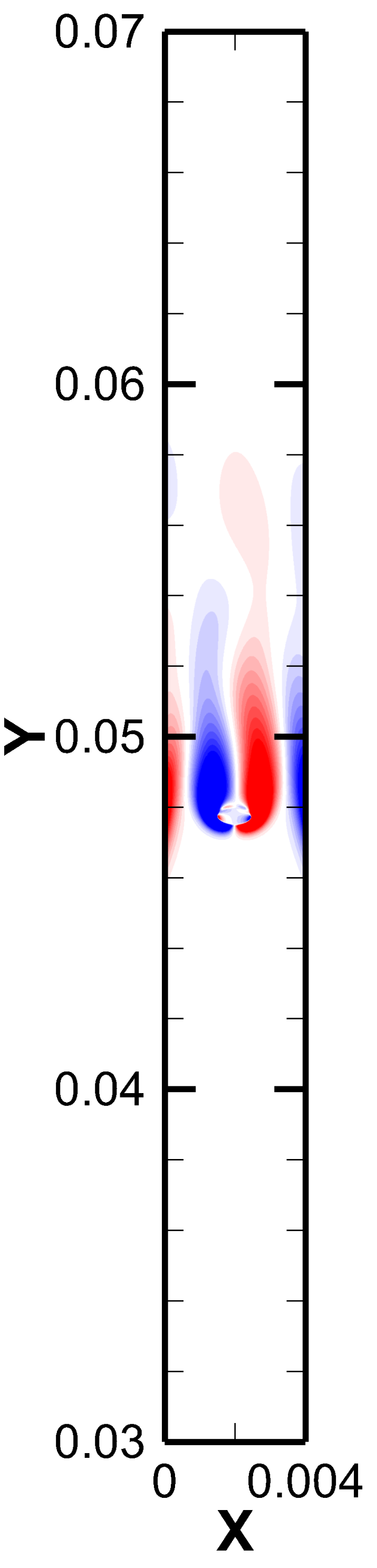}
\subcaption{$t = 1.0~s$}
\end{subfigure}%
\begin{subfigure}[b]{.16\textwidth}
\centering
\includegraphics[width=1.15\linewidth]{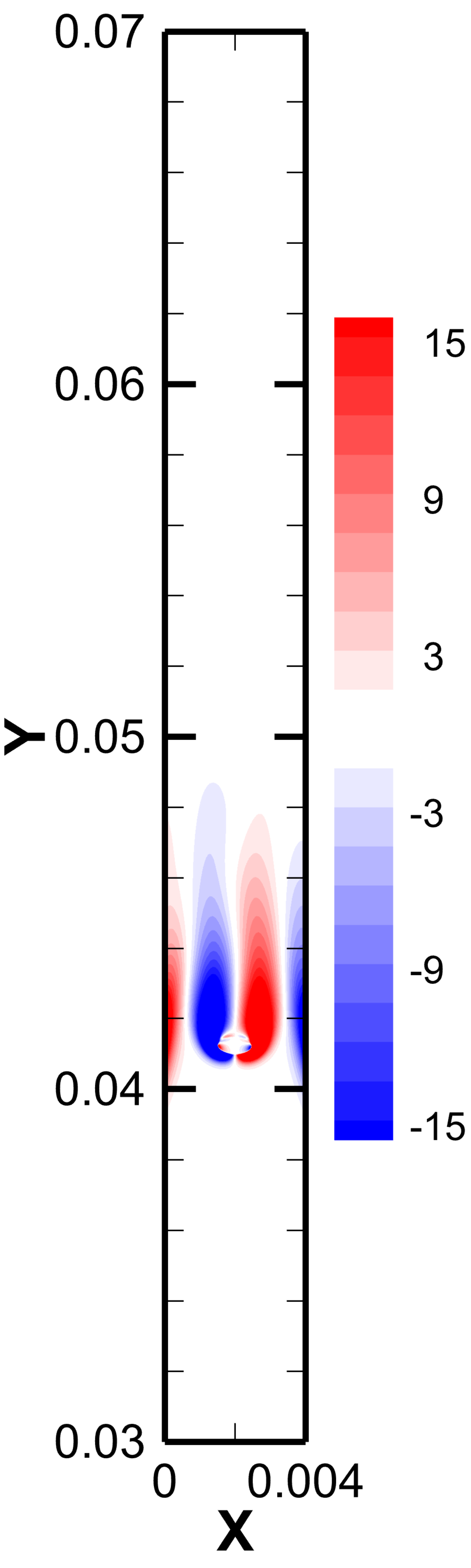}
\subcaption{$t = 1.5~s$}
\end{subfigure}
\caption{Vorticity contours at different time $t$ instances of the sedimenting 2D ellipse.}
\label{fig: Instantaneous vorticity contours at different time t instances of the sedimenting 2D ellipse.}
\end{figure}

Snapshots of the vorticity fields at various time instances, $t = 0.1,~0.25,~0.5,~0.75,~1.0,~\text{and}$ $1.5~s$, are shown in Fig.~(\ref{fig: Instantaneous vorticity contours at different time t instances of the sedimenting 2D ellipse.}). The results demonstrate the capability of the present work to accurately capture the complex flow physics arising from the interaction between the sedimenting ellipse and the initially quiescent flow field.

\subsection{Freely Falling (Dense) and Rising (Light) of a 2D Disk}
\begin{figure}[!h]
\centering
\begin{subfigure}[b]{.5\textwidth}
\centering
\includegraphics[width=.7\linewidth]{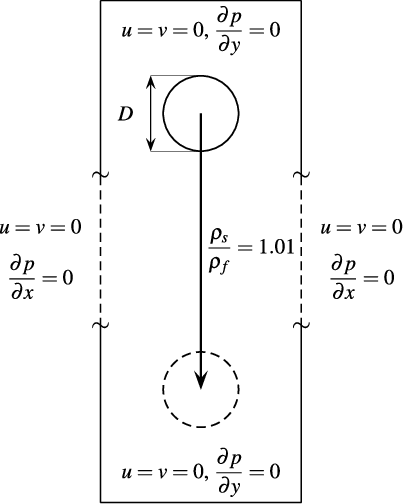}
\subcaption{}
\label{fig: Computational domain and boundary conditions for the freely falling (a) 2D circular disk in an open domain.}
\end{subfigure}%
\begin{subfigure}[b]{.5\textwidth}
\centering
\includegraphics[width=.7\linewidth]{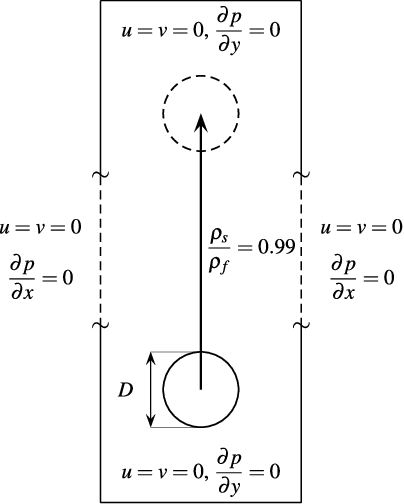}
\subcaption{}
\label{fig: Computational domain and boundary conditions for the freely rising (b) 2D circular disk in an open domain.}
\end{subfigure}
\caption{Computational domain and boundary conditions for the freely falling (a) and freely rising (b) 2D disk in an open domain.}
\label{fig: Computational domain and boundary conditions for the freely falling (a) and freely rising (b) 2D circular disk in an open domain.}
\end{figure}

This study examines cases where the solid-fluid density ratio is near unity, considering both upper and lower bounds. Therefore, the free fall of a dense disk $(\rho_s/\rho_f = 1.01)$ and the free rise of a lighter disk $(\rho_s/\rho_f = 0.99)$ are simulated in a stagnant fluid. Despite the apparent simplicity, this problem is considered challenging due to the critical density ratios chosen, which are known to destabilize the numerical coupling algorithm. The computed results are compared with reference data~\cite{fallingRisingCylinder}, which utilized the ALE method and a finite element solver. For consistency, the following non-dimensional Galileo number, $Ga = 1/\nu_f \sqrt{\left| \rho_s/\rho_f - 1 \right| g D^3} = 138$ and translational solid Reynolds number, $Re_{solid}^{trans} = V_t D/\nu_f = 156$, with $V_t$ the vertical v-component of the terminal velocity, are adhered to during the analysis. To model the effect of an open (infinite) domain, an extensively extended finite domain is employed. The domain dimensions are defined as a function of the disk diameter $D$, having $\Omega_f = [-5D,~5D] \times [-70D,~70D] = [-0.025,~0.025] \times [-0.35,~0.35]$ for a disk diameter $D = 0.005~m$. A uniform mesh size of $h = 2 \times 10^{-4}~m$ is used. The simulations are advanced with a time-step $\Delta t = 2 \times 10^{-4}~s$. The incompressible fluid has the following properties, $\rho_f = 996~kg/m^3$ and $\nu_f = 8.02\times10^{-7}~m^2/s$. Figs.~(\ref{fig: Computational domain and boundary conditions for the freely falling (a) 2D circular disk in an open domain.}) and~(\ref{fig: Computational domain and boundary conditions for the freely rising (b) 2D circular disk in an open domain.}) illustrate the computational domain and boundary conditions for the freely falling and rising cases, respectively.

The initial position of the disk is $(X_s,~Y_s) = (0,~\pm65D) = (0~m,~\pm0.325~m)$ and the densities are found as $\rho_s = 1005.96$ and $986.04~kg/m^3$ for the dense and lighter disk, respectively. Due to the critical density ratios simulated in this problem, a relaxation parameter $\alpha_r = 0.5$ was used for the Newton equation of motion, as the moving disk undergoes only translational motion. The explicit IME is employed as well here.

\begin{figure}[!h]
\centering
\begin{subfigure}[b]{.5\textwidth}
\centering
\includegraphics[width=0.75\linewidth]{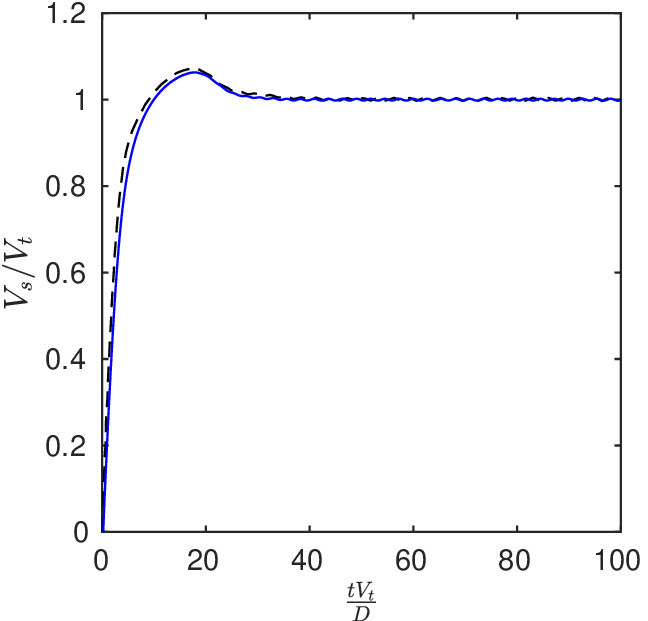}
\subcaption{}
\end{subfigure}%
\begin{subfigure}[b]{.5\textwidth}
\centering
\includegraphics[width=0.77\linewidth]{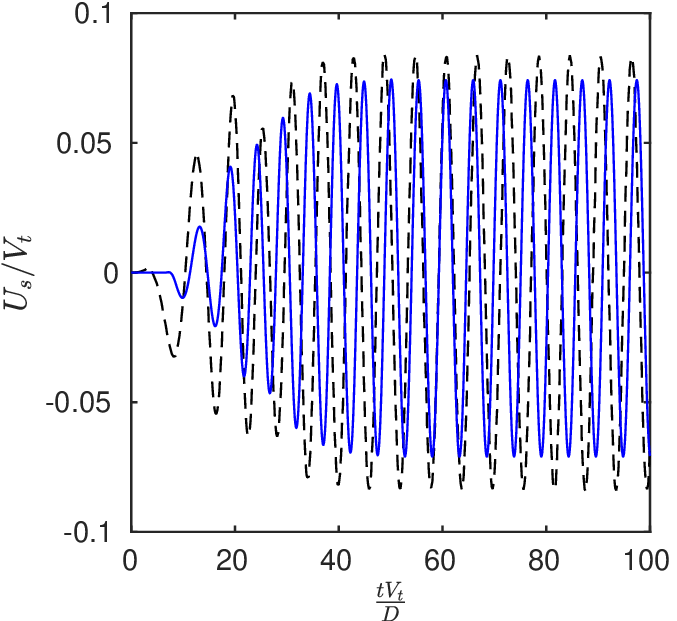}
\subcaption{}
\end{subfigure}
\caption{Non-dimensional vertical v-component (a) and horizontal u-component (b) of the disk velocity for the freely rising disk for $\rho_s/\rho_f = 0.99$ at $Re = 156$ and $Ga = 138$. The current results are marked with solid lines (\protect\tikz[baseline=-0.1ex] \protect\draw [thick,solid] (0,0.075) -- (0.5,0.075);) and the numerical data extracted from~\cite{fallingRisingCylinder} are marked with dashed lines (\protect\tikz[baseline=-0.1ex] \protect\draw [thick,dashed] (0,0.075) -- (0.5,0.075);).}
\label{fig: Non-dimensional vertical v-component (a) and horizontal u-component (b) of the circular disk velocity for the freely rising circular disk.}
\end{figure}

The non-dimensional vertical v-component and horizontal u-component of the velocity for the freely rising disk are shown in Fig.~(\ref{fig: Non-dimensional vertical v-component (a) and horizontal u-component (b) of the circular disk velocity for the freely rising circular disk.}), respectively. The results are in good agreement with the data shown in~\cite{fallingRisingCylinder}. The disk's horizontal velocity component is less than $10\%$ of the vertical velocity component. For $t V_t/D > 50$, the disk attains a periodic motion, with an average vertical terminal velocity $V_t = 2.494 \times 10^{-2}~m/s$ for the falling case, compared to $V_t = 2.501 \times 10^{-2}~m/s$ reported in~\cite{fallingRisingCylinder}. 

A comparison of the force coefficients and Strouhal number is provided in Table~\ref{tab: Comparison of the mean drag, amplitude of lift coefficient, and Strouhal number for the freely falling and rising circular disk.}. The mean drag, lift coefficient amplitude, and Strouhal number are computed using the terminal velocity $V_t$. The results show satisfactory agreement with values reported in the literature.

\begin{table}[!h]
\centering
\caption{Comparison of the mean drag, lift coefficient amplitude, and Strouhal number for the freely falling and rising disk.}
\label{tab: Comparison of the mean drag, amplitude of lift coefficient, and Strouhal number for the freely falling and rising circular disk.}
\begin{tabular}{lcccccccc}
\hline
\multirow{2}{*}{References} & \multicolumn{3}{c}{$\rho_s / \rho_f = 1.01$ -- Falling} & & & \multicolumn{3}{c}{$\rho_s / \rho_f = 0.99$ -- Rising} \\ \cline{2-4} \cline{7-9} 
 & $\overline{C_D}$ & $\max|C_L|$ & $\text{Str}$ & ~ & ~ & $\overline{C_D}$ & $\max|C_L|$ & $\text{Str}$ \\ \hline
Namkoong et al.~\cite{fallingRisingCylinder} & 1.23 & 0.15 & 0.16840 &  &  & $-$  & $-$  & 0.16870 \\
Lācis et al.~\cite{timeLaggedIBPM_1}         & 1.29 & 0.14 & 0.17185 &  &  & 1.29 & 0.14 & 0.17188 \\
Cai~\cite{shangguiPhd2016}                   & 1.35 & 0.10 & 0.18900 &  &  & 1.35 & 0.10 & 0.18900 \\
Present                                      & 1.24 & 0.10 & 0.16252 &  &  & 1.24 & 0.10 & 0.16473 \\ \hline
\end{tabular}%
\end{table}

\begin{figure}[!h]
\centering
\begin{subfigure}[b]{.4\textwidth}
\centering
\includegraphics[width=.5\linewidth]{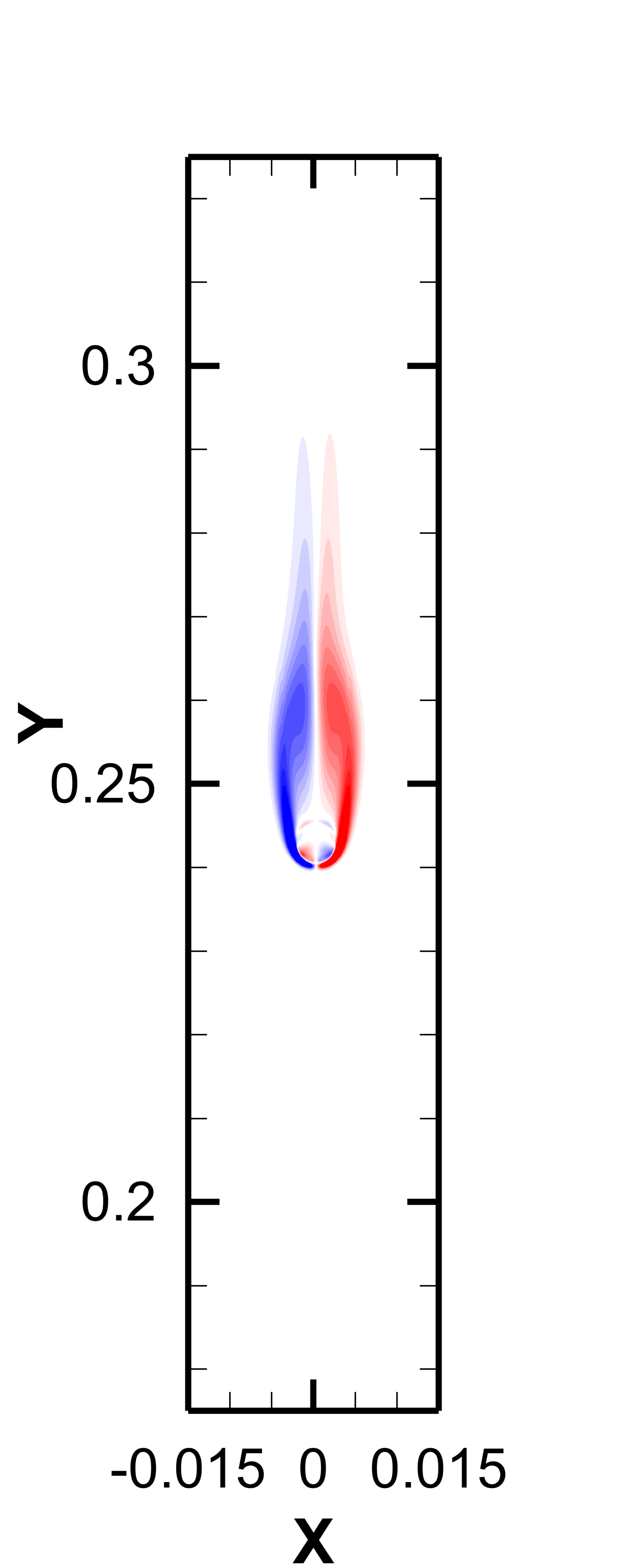}%
\hfill
\includegraphics[width=.5\linewidth]{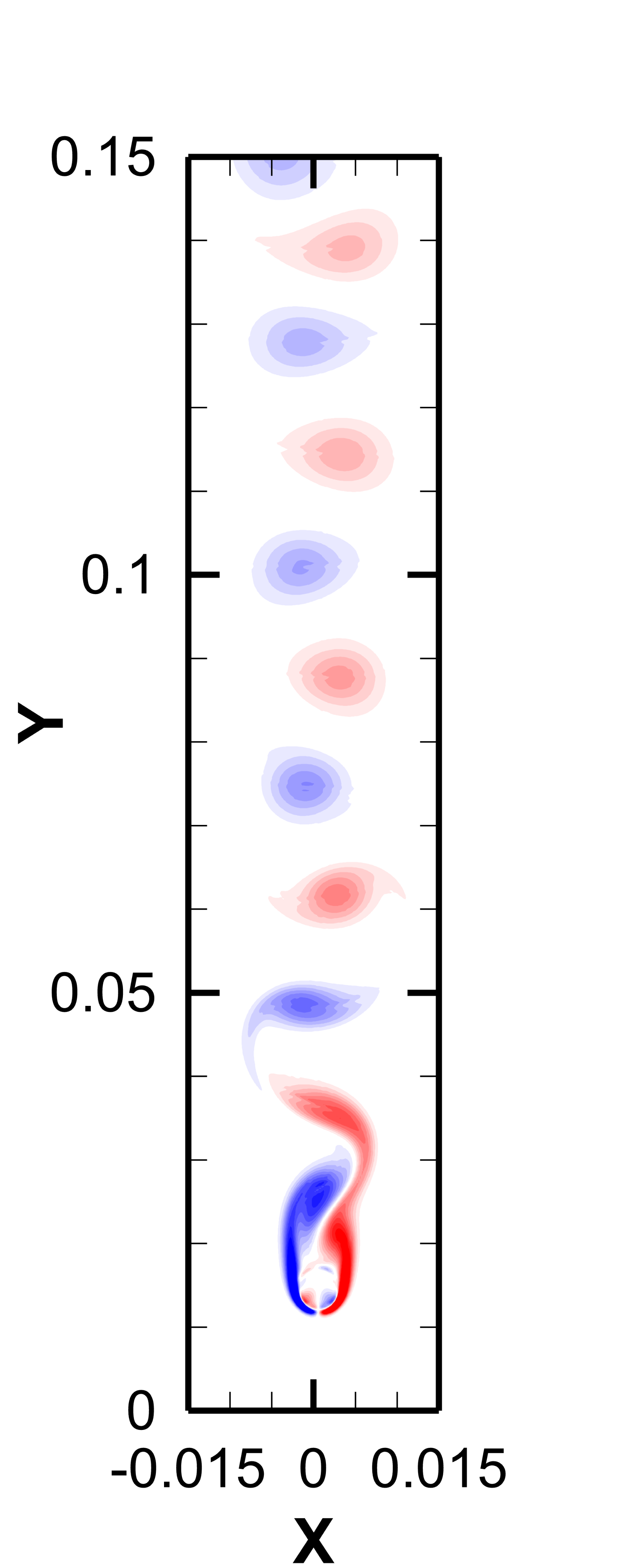}
\subcaption{$\rho_s/\rho_f = 1.01$}
\end{subfigure}%
\begin{subfigure}[b]{.4\textwidth}
\centering
\includegraphics[width=.5\linewidth]{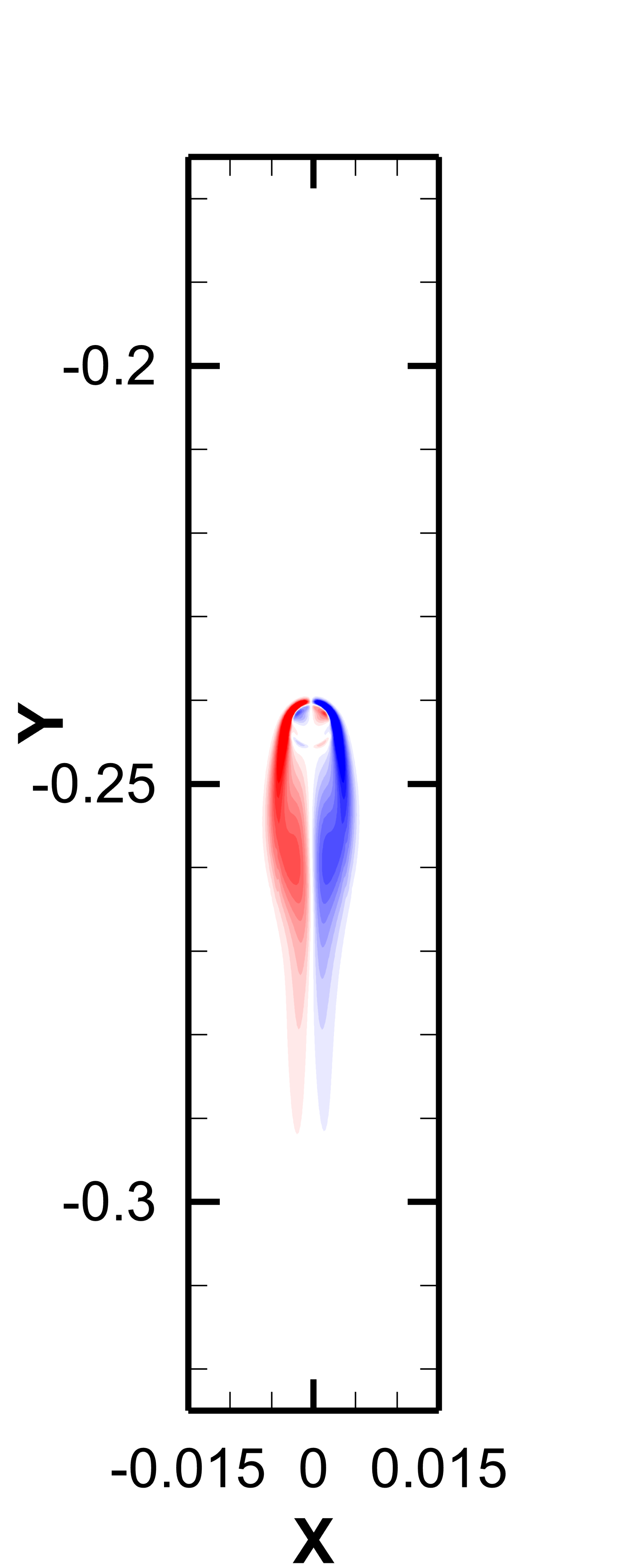}%
\hfill
\includegraphics[width=.5\linewidth]{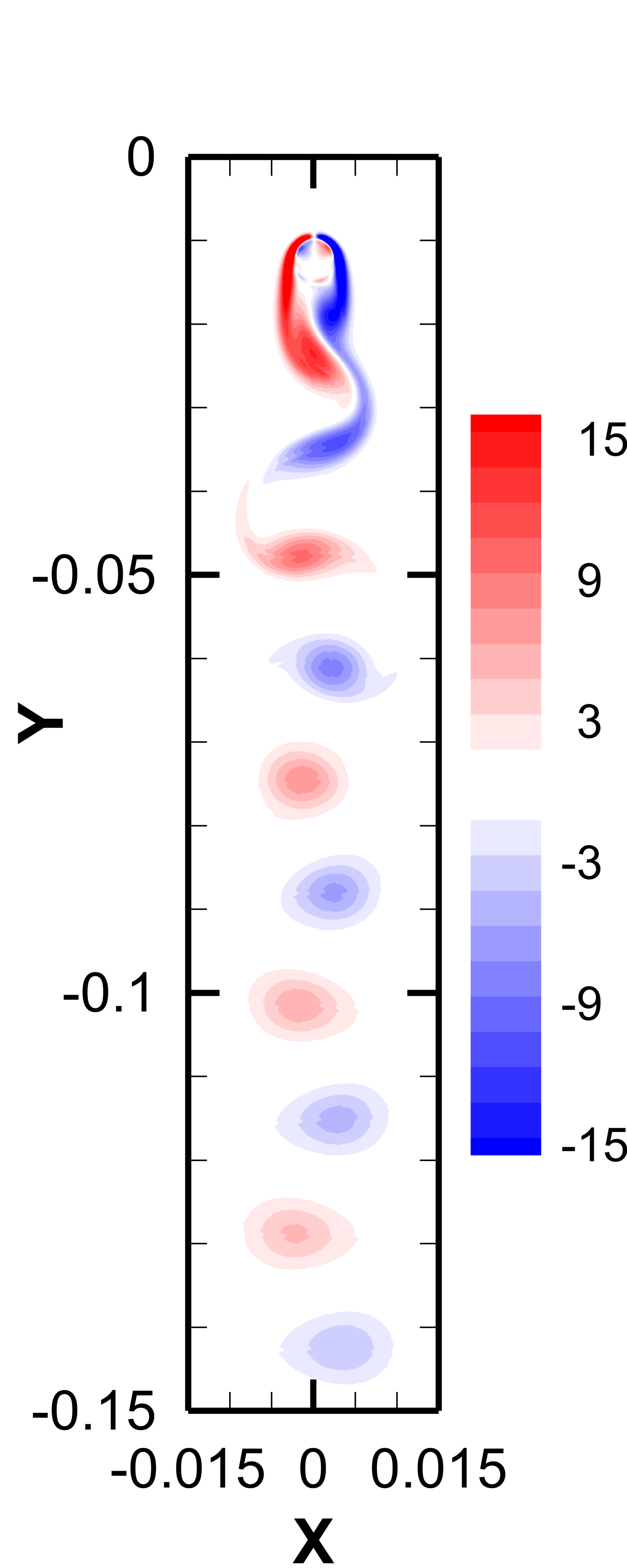}
\subcaption{$\rho_s/\rho_f = 0.99$}
\end{subfigure}
\caption{Vorticity contours at different time instances for the freely falling (a) and rising (b) disk.}
\label{fig: Instantaneous vorticity contours at different time instances for the freely falling circular disk.}
\end{figure}

Instantaneous vorticity contours for the freely falling and rising disk are shown in Fig.~(\ref{fig: Instantaneous vorticity contours at different time instances for the freely falling circular disk.}). The flow fields exhibit strong similarities due to the similar Reynolds numbers and closely matched density ratios. Initially, the flow remains symmetric, with a symmetric vortex pair forming downstream of the disk around $t V_t/D = 20$. However, symmetry begins to break at approximately $t V_t/D = 30$, and by $t V_t/D = 55$, the wake becomes fully asymmetric, leading to unsteady periodic vortex shedding phenomena, causing the horizontal oscillations of the disk.

\subsection{Rising of a Highly Buoyant 2D Disk Inside a Quiescent Flow}
After validating the present method against benchmark problems from the literature, we consider a more demanding test case to further assess the robustness of the proposed algorithm in the low density ratio regime, corresponding to highly buoyant rigid bodies. Specifically, we investigate the rising motion of a light rigid body in a quiescent fluid by extending the 2D disk sedimentation case introduced in Section~\ref{subsec: Sedimentation of a 2D Disk Inside a Confined Channel} to density ratios below unity. This configuration is computationally efficient and widely used to evaluate the stability of fluid-structure interaction coupling strategies, making it well suited for a parametric study of challenging FSI regimes. In addition, the present case is proposed as a benchmark problem for fluid-rigid body interaction at low density ratios, for which, to the best of the authors’ knowledge, no reference data are currently available for quantitative comparison.

All physical and numerical parameters are kept identical to those used in Section~\ref{subsec: Sedimentation of a 2D Disk Inside a Confined Channel}, except for the initial position of the disk center of mass, which is set to $(X_s,~Y_s) = (0.01~m,~0.02~m)$ instead of $(0.01~m,~0.04~m)$ to allow sufficient vertical space for the upward motion of the buoyant disk. The fluid density is set to $\rho_f = 996~kg/m^3$, and the simulation is terminated when the disk reaches a vertical position of $Y_s = 0.05~m$.

\begin{figure}[!h]
\centering
\centering
\includegraphics[width=0.6\linewidth]{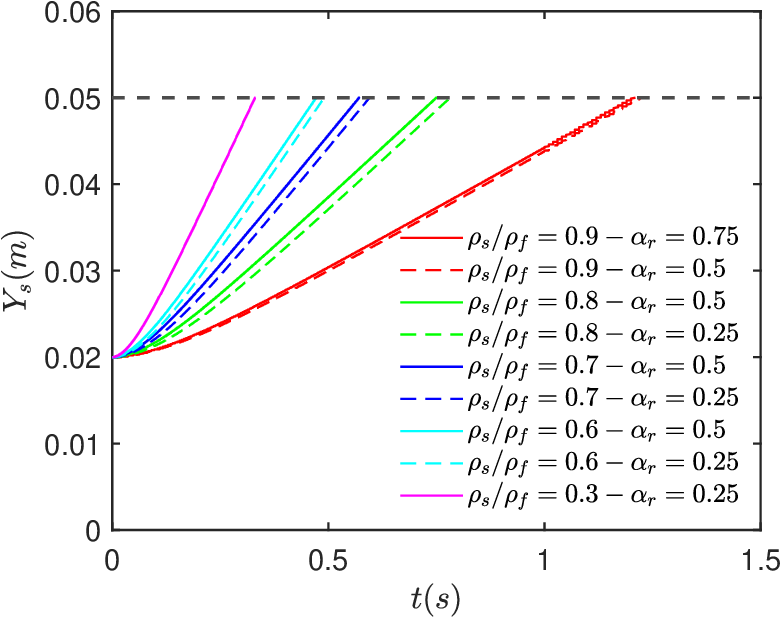}
\caption{Time history of the vertical position $Y_s$ of the rising 2D disk for different solid to fluid density ratios $\rho_s/\rho_f$.}
\label{fig: Time history of the vertical position of the rising 2D disk for different solid to fluid density ratios.}
\end{figure}

The objective of this study is not to determine the lower stability limit of the method, but to demonstrate its robustness over a representative range of density ratios below unity. Simulations are performed for $\rho_s/\rho_f = 0.9,~0.8,~0.7,~0.6$, and $0.3$. For each case, the time evolution of the disk center of mass position is shown in Fig.~(\ref{fig: Time history of the vertical position of the rising 2D disk for different solid to fluid density ratios.}). As the density ratio decreases, the time required for the disk to reach the desired vertical position decreases, as expected, due to the increasing density contrast between the solid and the fluid, which results in a proportionally larger net buoyant force acting on the disk.

\begin{figure}[!h]
\centering
\centering
\includegraphics[width=0.6\linewidth]{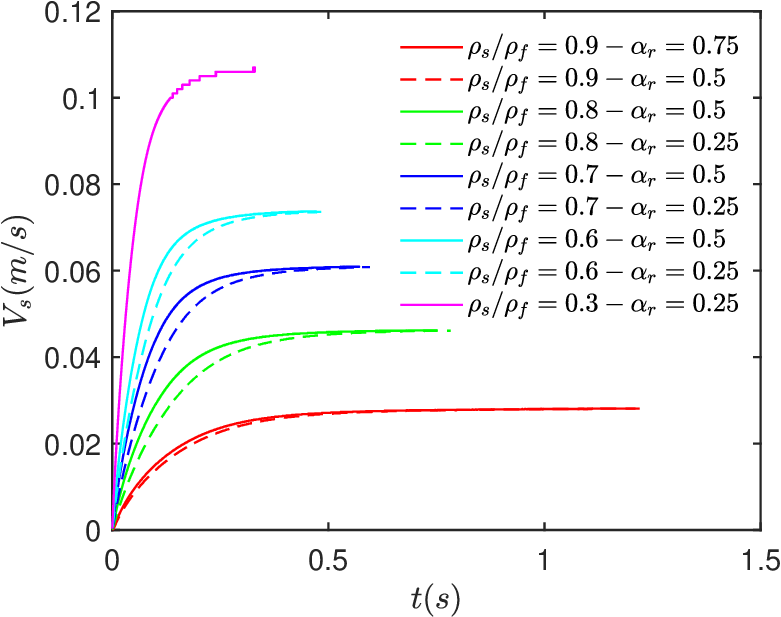}
\caption{Time history of the velocity v-component $V_s$ of the rising 2D disk for different solid to fluid density ratios $\rho_s/\rho_f$.}
\label{fig: Time history of the v-component of the rising 2D disk for different solid to fluid density ratios.}
\end{figure}

We emphasize that for these low $\rho_s/\rho_f$ cases, simulations performed without relaxation diverged from the very first iterations, highlighting the intrinsic numerical instability of this regime. In contrast, the proposed method remained stable and yielded physically consistent results for all tested density ratios. Two relaxation parameters, $\alpha_r = 0.5$ and $0.75$, were initially examined; however, for density ratios $\rho_s/\rho_f \leq 0.8$, a value of $\alpha_r = 0.75$ proved insufficient to ensure convergence. A smaller relaxation factor, $\alpha_r = 0.25$, was therefore adopted to guarantee numerical stability and convergence in the low density ratio regime. For the lowest density ratio considered, $\rho_s/\rho_f = 0.3$, even a relaxation of $\alpha_r = 0.5$ led to large velocity fluctuations and divergence of the solution. Consequently, the velocity was over-relaxed using the smaller relaxation factor, $\alpha_r = 0.25$, to maintain stable convergence for the lowest density ratio investigated. Fig.~(\ref{fig: Time history of the v-component of the rising 2D disk for different solid to fluid density ratios.}) shows the variation of the disk velocity for different solid to fluid density ratios. It is evident that for highly buoyant rigid bodies ($\rho_s/\rho_f = 0.3$), velocity fluctuations are inevitable, even with strong relaxation ($\alpha_r = 0.25$). Nevertheless, the algorithm was able to maintain stability and convergence throughout the implicit iterations.

Additional data are provided for future benchmark comparisons, specifically related to the kinematics of the rising disk. Table~\ref{tab: Data obtained for the terminal velocity and the time taken to reach the desired vertical position for the rising 2D disk at different solid to fluid density ratios.} summarizes the terminal velocity and the time required to reach the desired vertical position for the different solid to fluid density ratios.

\begin{table}[!h]
\centering
\caption{Data obtained for the terminal velocity $V_t$ and the time taken to reach the desired vertical position $t_{Y_s=0.05}$ for the rising 2D disk at different solid to fluid density ratios $\rho_s/\rho_f$.}
\label{tab: Data obtained for the terminal velocity and the time taken to reach the desired vertical position for the rising 2D disk at different solid to fluid density ratios.}
\begin{tabular}{lccc}
\hline
$\rho_s/\rho_f$       & $\alpha_r$ & $V_t$ & $t_{Y_s=0.05}$ \\
\hline
\multirow{2}{*}{0.9} & 0.75       & 0.0281 & 1.210     \\
                     & 0.50       & 0.0281 & 1.220     \\
\multirow{2}{*}{0.8} & 0.50       & 0.0462 & 0.751     \\
                     & 0.25       & 0.0461 & 0.783     \\
\multirow{2}{*}{0.7} & 0.50       & 0.0609 & 0.572     \\
                     & 0.25       & 0.0608 & 0.596     \\
\multirow{2}{*}{0.6} & 0.50       & 0.0737 & 0.472     \\
                     & 0.25       & 0.0736 & 0.491     \\
0.3                  & 0.25       & 0.1070 & 0.331     \\
\hline
\end{tabular}
\end{table}

These additional results provide clear numerical evidence that the proposed method can robustly handle low density ratio problems, including highly buoyant rigid bodies, thereby strengthening its overall applicability. The results are presented to demonstrate the stability and robustness of the method in a regime that is well known to be numerically challenging.

\section{Conclusions} \label{sec: conclusions}
In this work, an extension to the previously developed implicit DF-IBM algorithm~\cite{eliasIBM} was proposed to enable simulations of unrestricted flow-induced rigid body motion and broaden the applicability of DF-IBM. The fluid governing equations are coupled with the rigid body dynamics Newton-Euler equations within the DF-IBM framework. The predictor-corrector PISO algorithm was leveraged to handle fluid-rigid body interactions, structured into three main steps, (1) momentum predictor, (2) fluid-rigid body coupling, and (3) PISO corrector loops. An implicit coupling strategy was developed for strongly coupled interface condition following the partitioned coupling approach. Due to the nature of the fluid-rigid body coupling in the DF-IBM, steps (1) and (3) are executed outside the implicit coupling iterations, while only step (2) is included within the implicit loop. Knowing that step (3) includes solving the PPE, its exclusion significantly improves computational efficiency compared to implicit coupling under the ALE formulation.

Additionally, the proposed algorithm accurately enforces the dual constraints of the coupled system sequentially in the solution process, ensuring a divergence-free condition in the fluid domain, an interface matching and a no-slip velocity boundary condition at the immersed interface. This study also identifies key challenges associated with the IME. The first arises from critical density ratios, particularly for neutrally buoyant and buoyant rigid bodies. The second stems from the rigid body approximation of the IME in the Euler equation. To address these issues, a fixed relaxation technique is introduced for rigid body kinematics, improving convergence, stabilizing computations, and ensuring robust implicit coupling.

Overall, the proposed algorithm has proven to effectively handle a broad range of density ratios, achieving stable and converged solutions. Good agreement was found with reference data from the literature across all simulated cases, validating the algorithm's capability and robustness for a broad range of challenging fluid-rigid body interaction problems. IBM significantly accelerates the simulation process, especially in complex fluid-rigid body interaction flows, by eliminating the need for complex body-fitted meshes and enabling a straightforward handling of moving rigid bodies. Direct applications of the method include, for instance, the flow-induced rotation of vertical axis turbines. The developed method’s efficiency and flexibility could become a standard tool for both academic research and industrial practice, whenever quick information on fluid-rigid body interaction problems is required.

\section*{Acknowledgements}  \label{sec: acknowledgements}
This research was supported by the UK Engineering and Physical Sciences Research Council (EPSRC) (Grant No. EP/T518104/1, Project Reference No. 2676291). The computations were performed on the platform PILCAM2 from the Université de Technologie de Compiègne, Alliance Sorbonne Université. The data that support the findings of this study are openly available in CERES Research at https://doi.org/10.57996/cran.ceres-2804.

\appendix
\section{Hydrodynamic Force and Torque}
\label{sec: Appendix A}
Due to the interaction between the viscous fluid and the rigid body immersed boundary, $\Gamma_s$ termed as the fluid-solid interface, a hydrodynamic force and torque are exerted on the solid by the fluid stress tensor, resulting in the drag and lift forces as well as the torque on the rigid body, having the following forms:
\begin{subequations}
\begin{equation}
\mathbf{F}_{h} = 
\begin{pmatrix}
F_D \\
F_L
\end{pmatrix}
= \rho_f \oint_{\Gamma_s} \boldsymbol{\sigma}_f \cdot \mathbf{n} ~dS,
\label{eq: surface_integral_force}
\end{equation}
\begin{equation}
\mathbf{T}_{h} = \rho_f \oint_{\Gamma_s} \mathbf{r}_{i,j} \times (\boldsymbol{\sigma}_f \cdot \mathbf{n}) ~dS,
\label{eq: surface_integral_torque}
\end{equation}
\end{subequations}
where $\mathbf{r}_{i,j} = \mathbf{x}_{i,j} - \mathbf{X}_s$ is the relative position vector of any fluid Eulerian grid point $\mathbf{x}_{i,j}$ with respect to the rigid body center of mass $\mathbf{X}_s$, $\mathbf{n}$ is the outward-pointing unit normal vector on the immersed boundary surface $\Gamma_s$, $dS$ is an infinitesimal surface element over $\Gamma_s$. The constitutive relation for a Newtonian incompressible fluid is given by the total stress tensor equation:
\begin{equation}
\boldsymbol{\sigma}_f = -p\mathbf{I} + \nu_f \left( \nabla \mathbf{u}_f + (\nabla \mathbf{u}_f)^{\mathrm{T}} \right),
\label{eq: total_stress_tensor}
\end{equation}
having $\nu_f = \mu_f / \rho_f$ the effective kinematic viscosity, $p = p_s / \rho_f$ a normalized pressure by the fluid density $\rho_f$, and $\mathbf{I}$ the identity matrix, with $\mu_f$ being the effective dynamic viscosity and $p_s$ is the static pressure.

A direct evaluation of Eqs.~(\ref{eq: surface_integral_force}) and~(\ref{eq: surface_integral_torque}) are considered a cumbersome task and numerically exigent. Therefore, Uhlmann~\cite{uhlmann2003} proposed an accurate and efficient method using the Cauchy principle~\cite{cauchyPrinciple} to solve the surface integrals using the linear and angular momentum balance by taking advantage of the specific volume forces already computed in the DF-IBM framework. In other words, the rate of change of the linear and angular momentum over $\Omega_s$, is set equal to the sum of the forces and torques applied on the volume $\Omega_s$ and the surface $\Gamma_s$ of that rigid body subjected to motion:
\begin{subequations}
\begin{equation}
\rho_f \dfrac{d}{dt} \int_{\Omega_s} \mathbf{u}_f ~dV = \rho_f \int_{\Omega_s} \mathbf{f} ~dV + \rho_f \oint_{\Gamma_s} \boldsymbol{\sigma}_f \cdot \mathbf{n} ~dS,
\label{eq: cauchy_principale_force}
\end{equation}
\begin{equation}
\rho_f \dfrac{d}{dt} \int_{\Omega_s} \mathbf{r}_{i,j} \times \mathbf{u}_f ~dV = \rho_f \int_{\Omega_s} \mathbf{r}_{i,j} \times \mathbf{f} ~dV + \rho_f \oint_{\Gamma_s} \mathbf{r}_{i,j} \times (\boldsymbol{\sigma}_f \cdot \mathbf{n}) ~dS.
\label{eq: cauchy_principale_torque}
\end{equation}
\end{subequations}

Therefore, the hydrodynamic force and torque acting upon a solid body fully immersed in a fluid are computed by substituting Eqs.~(\ref{eq: cauchy_principale_force}) and~(\ref{eq: cauchy_principale_torque}) into Eqs.~(\ref{eq: surface_integral_force}) and~(\ref{eq: surface_integral_torque}):
\begin{subequations}
\begin{equation}
\mathbf{F}_h = -\rho_f \int_{\Omega_s} \mathbf{f} ~dV + \rho_f \dfrac{d}{dt} \int_{\Omega_s} \mathbf{u}_f ~dV,
\label{eq: hydrodynamic_force}
\end{equation}
\begin{equation}
\mathbf{T}_h = -\rho_f \int_{\Omega_s} \mathbf{r}_{i,j} \times \mathbf{f} ~dV + \rho_f \dfrac{d}{dt} \int_{\Omega_s} \mathbf{r}_{i,j} \times \mathbf{u}_f ~dV.
\label{eq: hydrodynamic_torque}
\end{equation}
\end{subequations}

The first term on the right-hand side of Eqs.~(\ref{eq: hydrodynamic_force}) and~(\ref{eq: hydrodynamic_torque}), is evaluated over grid points from both the external and internal fluids to the rigid body. This implies that the internal fluid has basically moved. While this internal flow does not affect the flow outside $\Gamma_s$, the moving or accelerating internal fluid has a major influence on the hydrodynamic force and torque exerted on $\Gamma_s$. This influence of internal mass is referred to as the internal mass effect (IME), denoted here by $\mathbf{F}_{IME}$ and $\mathbf{T}_{IME}$, and is equal to the second term on the right-hand side of Eqs.~(\ref{eq: hydrodynamic_force}) and~(\ref{eq: hydrodynamic_torque}). Since only the external contribution of the fluid acting on the rigid body is of interest, i.e., the hydrodynamic force and torque acting on $\Gamma_s$ from the external fluid, the total force and torque are compensated with the internal force and torque used to move the internal fluid~\cite{tenCate_IME,uhlmann2005,suzuki_IME}.

In the DF-IBM framework, the total force and torque shown in Eqs.~(\ref{eq: hydrodynamic_force}) and~(\ref{eq: hydrodynamic_torque}) are computed in a discrete form, and by employing the total force and total torque identities, properties of the IBM~\cite{peskin2002}, they are re-formulated as:
\begin{subequations}
\begin{equation}
-\rho_f \int_{\Omega_s} \mathbf{f} ~dV = -\rho_f \sum_{\forall \mathbf{x}_{i,j} \in \Omega_f} \mathbf{f}(\mathbf{x}_{i,j}, t) \Delta V_{i,j} = -\rho_f \sum_{\forall \mathbf{X}_{n} \in \Gamma_s} \mathbf{F}_{n}(\mathbf{X}_{n}, t) W_{n},
\label{eq: total_hydrodynamic_force_discrete_eulerian}
\end{equation}
\begin{equation}
-\rho_f \int_{\Omega_s} \mathbf{r}_{i,j} \times \mathbf{f} ~dV = -\rho_f \sum_{\forall \mathbf{x}_{i,j} \in \Omega_f} \mathbf{r}_{i,j}(\mathbf{x}_{i,j}, t) \times \mathbf{f} (\mathbf{x}_{i,j}, t) \Delta V_{i,j} = -\rho_f \sum_{\forall \mathbf{X}_{n} \in \Gamma_s} \mathbf{r}_{n}(\mathbf{X}_{n}, t) \times \mathbf{F}_{n}(\mathbf{X}_{n}, t) W_{n}.
\label{eq: total_hydrodynamic_torque_discrete_eulerian}
\end{equation}
\end{subequations}

Subsequently, the rate of change of the linear and angular momentum of the internal fluid occupying the rigid body, designated here as the internal force and torque, needs to be computed. It was proved theoretically by Uhlmann~\cite{uhlmann2003}, that the rate of change of the linear momentum of the internal fluid is exactly equal to the rate of change of the linear momentum of the rigid body, despite actual internal flow development. Therefore $\mathbf{F}_{IME}$ is re-written as:
\begin{equation}
\mathbf{F}_{IME} = \rho_f \dfrac{d}{dt} \int_{\Omega_s} \mathbf{u}_f ~dV = \rho_f \dfrac{m_s}{\rho_s} \dfrac{d \mathbf{U}_{s}}{dt} = \rho_f V \dfrac{d \mathbf{U}_{s}}{dt}.
\label{eq: internal_hydrodynamic_force_rigidbodymotion}
\end{equation}

This equality signifies that the force generated by the internal fluid within the volume of the rigid body $\Omega_s$, whose boundary $\Gamma_s$ is undergoing a rigid body motion, is unaffected by the specific nature of the motion inside the rigid body. Whether the internal fluid follows the rigid body motion or an arbitrary motion, Eq.~(\ref{eq: internal_hydrodynamic_force_rigidbodymotion}) always holds.

However, the same principle does not apply analogously to the rate of change of angular momentum, unless the internal fluid follows the rigid body motion, as demonstrated by Uhlmann~\cite{uhlmann2003}. Hence, it becomes necessary to impose an additional approximation by assuming that the internal fluid is forced to follow the rigid body motion, i.e., by setting:
\begin{equation}
\mathbf{u}_f(\mathbf{x}_{i,j}, t) = \mathbf{U}_{s}(t) + \omega_s (t) \times \mathbf{r}_{i,j}(\mathbf{x}_{i,j}, t), \quad \forall \mathbf{x}_{i,j} \in \Omega_s,
\end{equation}
therefore $\mathbf{T}_{IME}$ is reformulated as:
\begin{equation}
\mathbf{T}_{IME} = \rho_f \dfrac{d}{dt} \int_{\Omega_s} \mathbf{r}_{i,j} \times \mathbf{u}_f ~dV \approx \rho_f \dfrac{I_s}{\rho_s} \dfrac{d \omega_s}{dt}.
\label{eq: internal_hydrodynamic_torque_rigidbodymotion}
\end{equation}

This approximation forces the torque generated by the internal fluid, enclosed by $\Omega_s$, to follow the rigid body motion regardless of the true nature of the internal fluid motion.

\section{Collision Force}
\label{sec: Appendix B}
Two critical issues arise in the DF-IBM framework when a rigid body approaches a wall within the fluid Eulerian domain. The first problem is related to the rigid body penetrating the domain's wall, and the second problem primarily stems from the violation of the zeroth moment condition also known as the partition of unity~\cite{kempe&frohlich2012}, inherent to the kernel function used. 

\begin{figure}[!h]
\centering
\includegraphics[width=.35\linewidth]{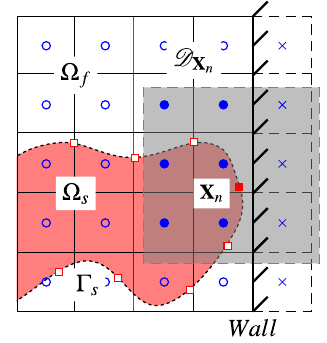}
\caption{Schematic of the solid rigid body-wall collision in the DF-IBM framework.}
\label{fig: Schematic of the solid-wall collision in the DF-IBM framework.}
\end{figure}

The second problem is illustrated in Fig.~(\ref{fig: Schematic of the solid-wall collision in the DF-IBM framework.}). It is evident that the Lagrangian support domain $\mathcal{D}_{\mathbf{X}_{n}}$ only grouped six Eulerian grid nodes that lie inside $\Omega_f$, and three additional grid nodes, called ghost nodes, that fall outside of $\Omega_f$. These additional ghost nodes are actually not recognized by the computational domain, implying that the IBM-related linear operators are executed solely between the six Eulerian grid nodes and the Lagrangian marker $\mathbf{X}_{n}$. This incomplete interaction leads to a violation of the zeroth moment condition, property of the kernel function, causing its value to fall below one (or unity). This loss of partition of unity will introduce errors per time-step. Consequently, decreasing the time-step will only exacerbate the accumulation of these errors over time.

One easy approach to mitigate these issues is by using specialized models designed to prevent a rigid body from colliding with a wall of the fluid domain, when the distance or the gap between the rigid body and the wall exceeds a certain pre-defined safe zone threshold $\xi$. In this study, the simplest collision model adopted here is the repulsive potential model (RPM) proposed by Glowinski et al.~\cite{glowinski1999,glowinski2000_1}. This collision model generates an artificial repulsive force that operates only in a short distance range. The artificial repulsive force is computed as follows:
\begin{equation}
\mathbf{F}_c = 
\begin{cases}
0, & d_{s-s'} > 2 d_{n-s} + \xi, \\
\dfrac{1}{\epsilon_w} (\mathbf{X}_s - \mathbf{X}_{s}') (2d_{n-s} + \xi - d_{s-s'})^2, & d_{s-s'} \leq 2 d_{n-s} + \xi,
\end{cases}
\label{eq: collision_force}
\end{equation}
where $\mathbf{X}_s$ is the center of mass of the rigid body, $\mathbf{X}_{s}'$ is the center of mass of the imaginary rigid body on the outer side of the wall, $d_{s-s'} = \| \mathbf{X}_s - \mathbf{X}_{s}' \|$ is the distance between the center of masses of the rigid body and its imaginary counterpart, $d_{n-s} = \| \mathbf{X}_{n} - \mathbf{X}_s \| = \| \mathbf{r}_{n} \|$ is the distance between the center of mass and a Lagrangian marker $\mathbf{X}_{n}$ on the rigid body, $\xi$ is the safe zone threshold and is based on the support radius $r_s$ of the kernel function, $\epsilon_w$ is a model constant called the stiffness parameter. A value of $\epsilon_w = h^2/2$ according to~\cite{WanAndTurek2006} was found to produce satisfactory results in the current simulations.

\bibliographystyle{elsarticle-num} 
\bibliography{references}






\end{document}